\documentclass[english,12pt, draftclsnofoot, onecolumn]{IEEEtran}
\usepackage[T1]{fontenc}
\usepackage[latin9]{inputenc}
\usepackage{color}
\usepackage{babel}
\usepackage{array}
\usepackage{verbatim}
\usepackage{varioref}
\usepackage{float}
\usepackage{url}
\usepackage{multirow}
\usepackage{amsmath}
\usepackage{amsthm}
\usepackage{amssymb}
\usepackage{graphicx}
\usepackage[unicode=true,pdfusetitle,
 bookmarks=true,bookmarksnumbered=false,bookmarksopen=false,
 breaklinks=false,pdfborder={0 0 1},backref=false,colorlinks=false]
 {hyperref}

\makeatletter

\providecommand{\tabularnewline}{\\}

\theoremstyle{plain}
\newtheorem{thm}{\protect\theoremname}
\theoremstyle{plain}
\newtheorem{cor}[thm]{\protect\corollaryname}
\theoremstyle{plain}
\newtheorem{lem}[thm]{\protect\lemmaname}

\makeatother

\providecommand{\corollaryname}{Corollary}
\providecommand{\lemmaname}{Lemma}
\providecommand{\theoremname}{Theorem}

\begin{document}
\global\long\def\expect#1{\mathbb{E}\left[#1\right]}%

\global\long\def\abs#1{\left\lvert #1\right\lvert }%

\global\long\def\twonorm#1{\left\Vert #1\right\Vert }%

\global\long\def\brac#1{\left(#1\right)}%

\global\long\def\lgbrac#1{\log\left(#1\right)}%

\global\long\def\lnbrac#1{\ln\left(#1\right)}%

\global\long\def\lghbrac#1{\log\left(2\pi e\brac{#1}\right)}%

\global\long\def\cbrac#1{\left\{  #1\right\}  }%

\global\long\def\rline#1{\left.#1\right| }%

\global\long\def\sbrac#1{\left[#1\right] }%

\global\long\def\Det#1{\left|#1\right|}%

\global\long\def\prob#1{\mathbb{P}\brac{#1}}%

\global\long\def\union{\bigcup}%

\global\long\def\inter{\bigcap}%

\global\long\def\real{\mathbb{R}}%

\global\long\def\eqdof{\doteq}%

\global\long\def\leqdof{\overset{.}{\leq}}%

\global\long\def\geqdof{\overset{.}{\geq}}%

\global\long\def\tran{\mathsf{Tran}}%

\global\long\def\snr{\mathsf{SNR}}%

\global\long\def\inr{\mathsf{INR}}%

\global\long\def\pderiv#1#2{\frac{\partial#1}{\partial#2}}%

\newcommand{\etal}{{\it et al.}~}

\newcommand*{\xdash}[1][3em]{\rule[0.5ex]{#1}{0.55pt}}

\allowdisplaybreaks

\newif\ifarxiv

\arxivtrue

\title{Generalized Degrees Freedom of Noncoherent MIMO Channels with Asymmetric
Link Strengths\thanks{Shorter version of this work appeared in \cite{Joyson_2x2MIMO_isit}
with an outline of proofs. This version has complete proofs. This
work was supported in part by NSF grants 1514531 and 1314937.}}
\author{Joyson Sebastian, Suhas N. Diggavi}
\maketitle
\begin{abstract}
We study the generalized degrees of freedom (gDoF) of block-fading
noncoherent multiple input multiple output (MIMO) channels with asymmetric
distributions of link strengths and a coherence time of $T$ symbol
durations. We derive the optimal signaling structure for communication
for the asymmetric MIMO channel, which is distinct from that for the
MIMO channel with independent and identically distributed (i.i.d.)
links. We extend the existing results for the single input multiple
output (SIMO) channel with i.i.d. links to the asymmetric case, proving
that selecting the statistically best antenna is gDoF-optimal. Using
the gDoF result for the SIMO channel, we prove that for $T=1$, the
gDoF is zero for MIMO channels with arbitrary link strengths.%
{} We show that selecting the statistically best antenna is gDoF-optimal
for the multiple input single output (MISO) channel. We also derive
the gDoF for the $2\times2$ MIMO channel with different exponents
in the direct and cross links. In this setting, we show that it is
always necessary to use both the antennas to achieve the gDoF, in
contrast to the results for the $2\times2$ MIMO channel with i.i.d.
links. We show that having  weaker crosslinks, gives gDoF gain compared
to the case with i.i.d. links. For the noncoherent MIMO channel with
i.i.d. links, the traditional method of training each transmit antenna
independently is degrees of freedom (DoF) optimal, whereas we observe
that for the asymmetric $2\times2$ MIMO channel, the traditional
training is not gDoF-optimal. We extend this observation to a larger
$M\times M$ MIMO channel by demonstrating a strategy that can achieve
larger gDoF than a traditional training-based method.
\end{abstract}

\section{Introduction}

The capacity of fading multiple input multiple output (MIMO) channels
when neither the receiver nor the transmitter knows the fading coefficients
was first studied by Marzetta and Hochwald \cite{marzetta1999capacity}.
They considered a block fading channel model where the fading gains
are independent and identically distributed (i.i.d.) Rayleigh random
variables and remain constant for $T$ symbol periods. In \cite{Zheng_Tse_Grassmann_MIMO},
Zheng and Tse introduced the idea of communication over a Grassmanian
manifold for the noncoherent MIMO channel and derived the capacity
behavior when the links are i.i.d. and the signal-to-noise-ratio ($\snr$)
is high. Their characterization was tight for the capacity at large
$\snr$, when the coherence time was large compared to the number
of antennas. In \cite{Wang_2013_noncoh_large_MIMO}, this tight characterization
was extended to the case when the number of antennas was large compared
to the coherence time.

Some works have especially considered the case with coherence time
$T=1$. The noncoherent single input single output (SISO) channel
with $T=1$ was considered by Taricco and Elia \cite{Taricco_Elia_97}.
They obtained the capacity behavior in asymptotically low and high
$\snr$ regimes. The noncoherent SISO channel with $T=1$ was further
studied by Abou-Faycal \etal \cite{Abou_Faycal_noncoherent} and
they showed that for any given $\mathsf{SNR}$, the capacity is achieved
by an input distribution with a finite number of mass points. For
the noncoherent MIMO channel with $T=1$ and high $\snr$, Lapidoth
and Moser \cite{lapidoth2003capacity} showed that the capacity behaves
double logarithmically with $\snr$.

When a complete capacity characterization of a wireless network is
difficult to obtain, the notions of degrees of freedom (DoF) and generalized
degrees of freedom (gDoF) can be used to understand the asymptotic
behavior of the capacity. For example, for a point-to-point network
parameterized by channel strengths\footnote{For ease of analysis, we absorb the transmit SNR into the channel
strengths, hence the capacity characterization does not include the
transmit SNR explicitly. The noise at the receivers are assumed to
be of unit variance.} $\rho_{1}^{2},\rho_{2}^{2},\ldots,\rho_{L}^{2}$ on its links, the
complete capacity characterization obtains the capacity for all values
of $\rho_{1}^{2},\rho_{2}^{2},\ldots,\rho_{L}^{2}$. The DoF characterization
finds the asymptotic behavior of the prelog of capacity along the
line $\lgbrac{\rho_{1}^{2}}=\lgbrac{\rho_{2}^{2}}=\cdots=\lgbrac{\rho_{L}^{2}}$
in the $L-$dimensional space of link strengths in dBm. The gDoF characterization
is more general; it finds the asymptotic behavior of the prelog of
capacity along the line $\brac{\lgbrac{\rho_{1}^{2}}/\gamma_{1}}=\brac{\lgbrac{\rho_{2}^{2}}/\gamma_{2}}=\cdots=\brac{\lgbrac{\rho_{L}^{2}}/\gamma_{L}}$
with constants $\gamma_{1},\ldots,\gamma_{L}$. Equivalently, for
the gDoF characterization, we can set $\lgbrac{\rho_{1}^{2}}/\gamma_{1}=\lgbrac{\rho_{2}^{2}}/\gamma_{2}=\cdots=\lgbrac{\rho_{L}^{2}}/\gamma_{L}=\lgbrac{\snr}$
and let $\snr\rightarrow\infty$. The gDoF characterization was first
used in \cite{etkin_tse_no_fb_IC} to characterize the asymptotic
behavior of the capacity region of a 2-user symmetric interference
channel (IC) for high SNR. There the link strengths were set to scale
as $\snr,\snr^{\alpha},\snr^{\alpha},\snr$ for the 4 links of the
IC. This method of scaling the channel strengths with different SNR
exponents to obtain the gDoF region is also done in other works like
\cite{gDoF_K_user_IC,gDoF_MIMO_IC}.

The DoF characterization has been used to study the noncoherent MIMO
channel with temporal correlation within each fading block. In \cite{Morgenshtern_SIMO_correl_2013},
Morgenshtern \etal studied the single input multiple output (SIMO)
channel with temporally correlated Rayleigh block-fading and showed
that the SIMO channel can have a larger DoF than the SISO channel,
under some mild assumptions on the temporal correlation. The noncoherent
MIMO channel with temporally correlated block fading was studied in
\cite{Koliander_MIMO_correl_2014}. There it was shown that the noncoherent
MIMO channel with temporally correlated block fading can have a larger
DoF than the noncoherent MIMO channel with constant block fading.

Some works have studied noncoherent networks (with more than two nodes)
for the capacity behavior at high $\snr$. In \cite{Lapidoth_network},
it was shown that for noncoherent networks with $T=1$, the gDoF is
zero; this was an extension of the result for the noncoherent MIMO
channel with $T=1$ from \cite{lapidoth2003capacity}. Koch and Kramer
studied the noncoherent single relay network \cite{Koch2013} and
showed that under certain conditions on the fading statistics, the
relay does not increase the capacity at high $\snr$. In \cite{Gohary_non_coherent_2014},
the noncoherent MIMO full-duplex single relay channel with block-fading
was studied, and it was shown that Grassmanian signaling can achieve
the DoF without using the relay. Also, the results in \cite{Gohary_non_coherent_2014}
show that for certain regimes, decode-and-forward with Grassmanian
signaling can approximately achieve the capacity at high $\snr$.

To the best of our knowledge, the existing works consider a DoF framework
for studying the noncoherent channels, \emph{i.e., }the links in the
network scale with the same $\snr$ exponent. However, in networks,
the links could have asymmetry in the channel strengths. In this case,
a gDoF framework could better capture the system behavior. We consider
the noncoherent MIMO channel with asymmetric link strengths as a first
step in the direction of studying the asymmetric noncoherent networks.

\subsection{Contributions and Outline}

In this paper, we consider a noncoherent channel model with coherence
time of $T$ symbol periods and asymmetric link distributions, where
the link strengths are scaled with different exponents of $\mathsf{SNR}$.
In essence, we are moving from the DoF-framework in \cite{marzetta1999capacity,Zheng_Tse_Grassmann_MIMO}
to the generalized DoF of noncoherent MIMO channels.

\begin{figure}[H]
\centering{}\includegraphics{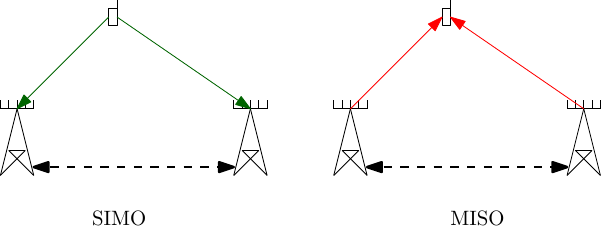}\caption{\label{fig:SIMO_and_MISO_from_COMP}Noncoherent SIMO channels and
MISO channels with asymmetric statistics can arise in COMP architecture
where multiple basestations can cooperate through the backhaul.}
\end{figure}

\begin{figure}[H]
\centering{}\includegraphics{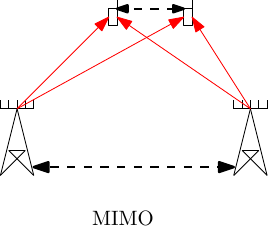}\caption{\label{fig:comp_toMIMO}Noncoherent MIMO channel with asymmetric statistics
can arise with COMP architecture and device-to-device cooperation.}
\end{figure}

\begin{figure}[H]
\centering{}\includegraphics[scale=0.6]{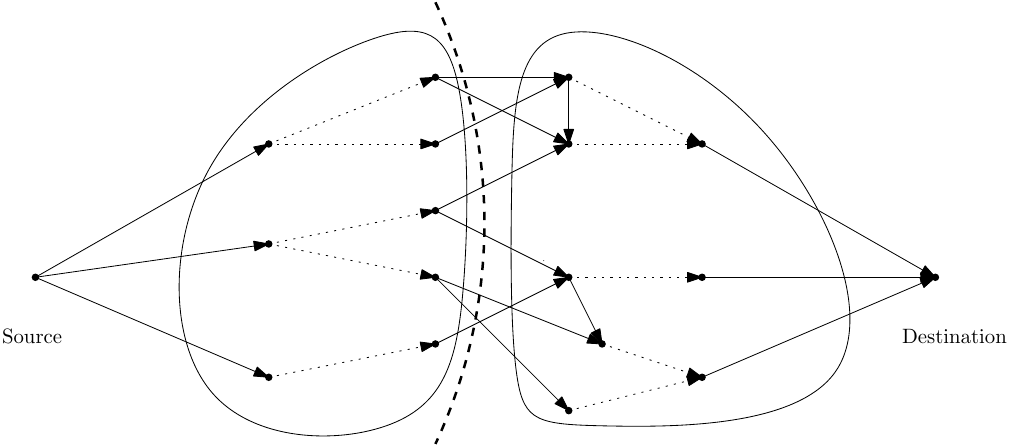}\caption{\label{fig:mimo_from_cutset}Noncoherent MIMO channel with asymmetric
statistics can arise in the analysis of noncoherent networks.}
\end{figure}

Next generation wireless architecture envisages dense deployment of
access points \cite{bhushan2014network_densification}. Another architectural
proposal is to use cloud radio access networks (CRAN) \cite{wu2015cloud}.
These imply that multiple access points could be connected through
a (reliable) backhaul. The implication of this is that of widely separated
antennas, which form a virtual antenna array. Such widely separated
antennas could be used for coordinated transmission and reception,
\emph{e.g.,} coordinated multipoint COMP \cite{Irmer_comp_2011}.
These widely separated antennas could have disparate average strengths
motivating our model (especially SIMO channels and MISO channels).
This is illustrated in Figure \ref{fig:SIMO_and_MISO_from_COMP}.
The MIMO case arises when the receiver could be widely spread (see
Figure \ref{fig:comp_toMIMO}) as would be the case when users can
cooperate using a separate sidechannel \cite{karakus_d2d_sidechannel_2017}.
Another motivation for this model comes from the study of networks.
Here one can think of the cut-set as a distributed MIMO channel (see
Figure \ref{fig:mimo_from_cutset}) where the nodes are widely separated
again resulting in this model.
 The asymmetric case is also motivated by a fundamental question about
the robustness of the results in \cite{marzetta1999capacity,Zheng_Tse_Grassmann_MIMO}
to the changes in the i.i.d. channel model.

For our channel model with arbitrary (fading) link strengths, we show
in Theorem \ref{thm:structure_optimal_distr} that the capacity achieving
input distribution is of the form $LQ$ where $L$ is a lower triangular
matrix and $Q$ is an isotropically distributed unitary matrix independent
of $L$. This is in contrast to the result for the i.i.d. setting,
where the capacity achieving distribution has the form $DQ$ with
a diagonal matrix $D$ \cite{marzetta1999capacity}. In Theorem \ref{thm:simo},
we demonstrate that the gDoF of a SIMO channel can be achieved by
retaining only the signal received by the best receive antenna. The
gDoF result for the SIMO channel is used in Theorem \ref{thm:zero_dof_T=00003D1}
to show that for $T=1$, the gDoF is always zero for a MIMO channel
of any size. In Theorem \ref{thm:MISO_dof}, we show that the gDoF
of the MISO channel can be achieved by signaling only over the statistically
best transmit antenna.

In a setting with $N$ receive antennas, when the exponents in the
$\snr$-scaling are same for all the links (i.i.d. setting), the number
of transmit antennas $M$, required to attain the optimal DoF was
shown to be $\min\brac{\left\lfloor T/2\right\rfloor ,N}$ \cite{Zheng_Tse_Grassmann_MIMO}.
It was shown that increasing the number of transmit antennas beyond
this value reduces the DoF. In this paper, we provide evidence that
this is not the case when the $\snr$ exponents are different: in
Theorem \ref{thm:2x2-sym-mimo}, we show that for a $2\times2$ MIMO
channel with different $\snr$ exponents in the direct and cross links,
and $T=2$, both the transmit antennas are required to achieve the
gDoF. We also show that having smaller exponents in crosslinks lead
to a gDoF gain of $\brac{2/T}\gamma_{\text{diff}}$ compared to the
case with same $\snr$ exponents in all the links, where $\gamma_{\text{diff}}$
is the difference in the $\snr$ exponents. In showing this, several
novel techniques were needed. In particular, we would like to highlight
the technique used in Lemma \ref{lem:discretization}, where in the
outer bound optimization problem, we show that using a discrete probability
distribution with a single mass point is gDoF optimal. To obtain this
result, we discretized the input distribution without a loss in gDoF,
and subsequently used linear programming arguments to show that there
exists an optimal distribution with just one mass point. We believe
that our techniques for the $2\times2$ MIMO channel provide intuitions
for studying larger noncoherent networks, especially in analyzing
the cut-sets.

Traditional training-based schemes for MIMO systems allocate a training
symbol to train each transmit antenna independently. Our results for
the $2\times2$ MIMO channel also demonstrate that a traditional training-based
scheme is not gDoF-optimal. Our scheme has a gDoF gain of $\brac{2/T}\gamma_{\text{diff}}$
compared to a training-based scheme. We also numerically evaluate
the rates achievable using our scheme in some specific scenarios and
compare it to the traditional training-based schemes. For a $2\times2$
system with coherence time $T=2$, transmit $\snr=23$ dB, direct
links with average strength $0.1$ and crosslinks with average strength
$0.025$, our noncoherent scheme can obtain a $7\%$ gain in the rate\footnote{The channel strengths in this example are given without absorbing
the transmit SNR into them. The noise is assumed to be of unit variance.} compared to the schemes that use a symbol for training. We demonstrate
more rate points and the gains in Table \ref{tab:Comparison-of-rates}
on page \pageref{tab:Comparison-of-rates}. In Theorem \ref{thm:gaussian_codebooks},
we extend our observation on the nonoptimality of traditional training-based
schemes to larger $M\times M$ MIMO channels with given $\snr$ exponents
on the direct and cross links, where we demonstrate a strategy that
can achieve larger gDoF than a training-based scheme.

 Extending our outer bounds to the general MIMO channel seems a difficult
task at the moment; the LQ transformation process used for deriving
the outer bound for the $2\times2$ MIMO channel as done in  (\ref{eq:2x2_outer_lq}),
(\ref{eq:xi_11_defn}), (\ref{eq:xi_21_defn}), (\ref{eq:xi_22_defn})
and the subsequent Lemmas (Lemma \ref{lem:dof_equivalence_abs_lin_comb_gaussian_vector},
Lemma \ref{lem:dof_equivalence_h(xi22)} and Lemma \ref{lem:dof_equivalence_h(xi21)})
for bounding the terms in those equations do not easily extend to
$3\times3$ or higher MIMO channels.

\noindent\textbf{Outline:} The rest of this paper is organized as
follows: in Section \ref{sec:System-model-and-notation}, we give
the notations and set up the system model; Section \ref{sec:Main-results}
presents our main results, and Section \ref{sec:Analysis} provides
analysis and proofs for the results in Section \ref{sec:Main-results}.
Some details of the proofs are deferred to the Appendixes. In Section
\ref{sec:Conclusion}, we give our conclusions and final remarks.

\section{Notation and system model\label{sec:System-model-and-notation}}

\subsection{Notational Conventions}

We use the notation $\mathcal{CN}\brac{\mu,\sigma^{2}}$ for circularly
symmetric complex Gaussian distribution with mean $\mu$ and variance
$\sigma^{2}$. We use the symbol $\sim$ with overloaded meanings:
one to indicate that a random variable has a given distribution and
second to indicate that two random variables have the same distribution.
The logarithm to base 2 is denoted by $\lgbrac{}$. The notation $A^{\dagger}$
indicates the Hermitian conjugate of a matrix $A$ and $\tran\brac A$
indicates the transpose of $A$.

The gDoF characterization for a point-to-point network with different
link strengths $\rho_{1}^{2},\rho_{2}^{2},\ldots,\rho_{L}^{2}$ captures
the asymptotic behavior of the capacity along the curve $\lgbrac{\rho_{1}^{2}}/\gamma_{1}=\lgbrac{\rho_{2}^{2}}/\gamma_{2}=\cdots=\lgbrac{\rho_{L}^{2}}/\gamma_{L}$
for any given constants $\gamma_{1},\ldots,\gamma_{L}$ as
\[
\text{gDoF}_{\gamma_{1},\ldots,\gamma_{L}}=\underset{\footnotesize{\lgbrac{\rho_{1}^{2}}/\gamma_{1}=\lgbrac{\rho_{2}^{2}}/\gamma_{2}=\cdots=\lgbrac{\rho_{L}^{2}}/\gamma_{L}=\lgbrac{\snr},\mathsf{SNR}\rightarrow\infty}}{\lim}\frac{C\brac{\rho_{1}^{2},\rho_{2}^{2},\ldots,\rho_{L}^{2}}}{\lgbrac{\mathsf{SNR}}},
\]
where $C\brac{\rho_{1}^{2},\rho_{2}^{2},\ldots,\rho_{L}^{2}}$ is
the capacity of the network for a given value of channel strengths
$\rho_{1}^{2},\rho_{2}^{2},\ldots,\rho_{L}^{2}$. We use the notation
$\doteq$ for relative equality, \emph{i.e.,} we say
\begin{equation}
f_{1}\brac{\mathsf{SNR}}\doteq f_{2}\brac{\mathsf{SNR}}
\end{equation}
 if
\begin{equation}
\text{lim}_{\snr\rightarrow\infty}\frac{f_{1}\brac{\mathsf{SNR}}}{\lgbrac{\mathsf{SNR}}}=\text{lim}_{\snr\rightarrow\infty}\frac{f_{2}\brac{\mathsf{SNR}}}{\lgbrac{\mathsf{SNR}}}.
\end{equation}
The notations $\leqdof,\geqdof$ are defined analogously. The script
$\mathcal{P}$ is used to indicate an optimization problem and $\brac{\mathcal{P}}$
is used to denote the optimal value of the objective function. We
use the \emph{overloaded} notation
\[
\text{gDoF}\brac{\mathcal{P}}=\text{lim}_{\snr\rightarrow\infty}\frac{\brac{\mathcal{P}}}{\lgbrac{\mathsf{SNR}}}
\]
to indicate the scaling of the optimal value of $\mathcal{P}$ when
the optimization problem depends on the $\snr$.

\subsection{System Model}

We consider a block-fading MIMO channel with $M$ transmit and $N$
receive antennas, and a coherence time of $T$ symbol durations. The
signal flow (over a blocklength $T$) is given by:
\begin{equation}
Y=GX+W
\end{equation}
where $X$ is the $M\times T$ matrix of transmitted symbols with
rows $\underline{X_{i}}$ corresponding to each transmit antenna%
; $G$ represents the $N\times M$ channel matrix (which is independently
generated every $T$ symbols), and its elements $g_{ij}$ are independent
with $g_{ij}\sim\mathcal{CN}\brac{0,\rho_{ij}^{2}}=\mathcal{CN}\brac{0,\mathsf{SNR}^{\gamma_{ij}}}$,
where the exponents $\gamma_{ij}$ are (constant) parameters of the
MIMO channel. For convenience, we also use the notation $\underline{\rho}^{2}\brac n$
to denote the row vector of channel strengths to $n^{\text{th}}$
receiver antenna. The columns of $G$ are denoted by $\overline{g_{i}}$,
and these correspond to channels from each transmit antenna%
. The variable $Y$ represents the $N\times T$ matrix of received
symbols, with rows corresponding to each receive antenna and $W$
is an $N\times T$ noise matrix with i.i.d. elements $w_{ij}\sim\mathcal{CN}\brac{0,1}$.
The transmit signals have the average power constraint:
\begin{equation}
\frac{1}{MT}\sum_{m=1}^{M}\sum_{t=1}^{T}\expect{\abs{x_{mt}}^{2}}=1.\label{eq:power_constraint}
\end{equation}

\section{Main results\label{sec:Main-results}}

In this section, we go through the main results of our paper. We first
look at the general results for the noncoherent MIMO channel with
asymmetric link strengths. In Theorem \ref{thm:structure_optimal_distr},
we prove a structural result for the optimizing distribution for the
noncoherent MIMO channel. This result has some similarities to that
for the noncoherent MIMO channel with i.i.d. links in the sense that
part of the structure is similar. We then consider a noncoherent MIMO
channel that can be decomposed into smaller disjoint channels. In
this case, the channel matrix is a block diagonal matrix. We prove
similar to the coherent case, that the power can be allocated across
the disjoint parts and coding can be done separately among the disjoint
parts to achieve the capacity. This result is proved in Theorem \ref{thm:decompose_graph}.
This result can be used to derive the gDoF of noncoherent parallel
channels. This is stated as Corollary \ref{cor:parallel_channels}.

Then we look at noncoherent MIMO channels with specific structures.
In Theorem \ref{thm:simo}, we consider the noncoherent SIMO channel
and derive its gDoF. For this case, we prove that the gDoF is achieved
by using the statistically best antenna. The gDoF result for the SIMO
channel can be used to prove that the gDoF is zero for any MIMO channel
when $T=1$. We obtain this by decomposing the MIMO channel into different
SIMO channels. We obtain this result in Theorem \ref{thm:zero_dof_T=00003D1}.
Next, we consider the noncoherent MISO channel and prove a similar
result, that its gDoF can be achieved using the statistically best
antenna. This is proved in Theorem \ref{thm:MISO_dof}.

The next specific structure we look at is the noncoherent $2\times2$
MIMO channel with a given $\snr$ exponent in the direct links and
another $\snr$ exponent in the crosslinks. We handle this case in
Theorem \ref{thm:2x2-sym-mimo}. We observe that standard training-based
schemes are not gDoF-optimal for $2\times2$ MIMO channels in general.
In Theorem \ref{thm:gaussian_codebooks}, we extend this observation
to larger $M\times M$ MIMO channels.

\subsection{Results for General Noncoherent MIMO Channels\label{subsec:general_results}}
\begin{figure}[H]
\centering{}\includegraphics{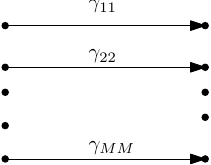}\caption{\label{fig:Parallel-channels}Parallel channels with given $\protect\snr$
exponents.}
\end{figure}

\begin{thm}
The capacity of a noncoherent MIMO system can be achieved with input
signal $X$ of the form $X=LQ$ with $L$  being a lower triangular
matrix and $Q$ being an isotropically distributed unitary matrix
independent of $L$.\label{thm:structure_optimal_distr}
\end{thm}
\begin{IEEEproof}
The proof is given in Section \ref{subsec:properties of capacity achieving distribution}.
\end{IEEEproof}
This theorem is in contrast with the result for the case when the
elements of $G$ and $W$ are i.i.d. Gaussian. In that case, the structure
of an optimal $X$ could be written as $X=DQ$ where $D$ is diagonal
\cite{marzetta1999capacity}. In our system model, only $W$ has i.i.d.
elements which ends up restricting the structure to the form $LQ$.
\begin{thm}
\label{thm:decompose_graph}Let the channel matrix $G$ of the MIMO
system be block diagonal as $G=\text{diag}\brac{G_{1},\ldots,G_{K}}$
where $G_{i}$ are the diagonal blocks of $G$, then the capacity
$C\brac{P,\text{diag}\brac{G_{1},\ldots,G_{K}}}$ of the channel for
a power $P$ can be achieved by splitting the power across the blocks:
$C\brac{P,\text{diag}\brac{G_{1},\ldots,G_{K}}}=\max_{P_{1}+\cdots+P_{K}\leq P}\brac{C\brac{P_{1},G_{1}}+\cdots+C\brac{P_{K},G_{K}}}$.
\end{thm}
\begin{IEEEproof}
[Proof idea]This result holds for the coherent MIMO channel and the
proof for noncoherent case is similar. We just need to show $C\brac{P,\text{diag}\brac{G_{1},G_{2}}}=\max_{P_{1}+P_{2}\leq P}\brac{C\brac{P_{1},G_{1}}+C\brac{P_{2},G_{2}}}$
because of induction. Let $X_{G1},X_{G2}$ be the transmitted symbols
in the parts $G_{1}$ and $G_{2}$ of the channel. Similarly $Y_{G1},Y_{G2}$
be the corresponding received symbols. Now $I\brac{X;Y}\leq I\brac{X_{G1};Y_{G1}}+I\brac{X_{G2};Y_{G2}}$
because $\brac{X_{G2},Y_{G2}}-X_{G1}-Y_{G1}$ , $\brac{X_{G1},Y_{G1}}-X_{G2}-Y_{G2}$
are Markov chains and the desired result easily follows. The detailed
steps are given in \ifarxiv  Appendix \ref{app:decompose_graph}\else  \cite[Appendix C]{Joyson_2x2_mimov5}\fi.
\end{IEEEproof}
Now we have the following corollary from the above theorem.
\begin{cor}
The gDoF of the parallel channel system (Figure \ref{fig:Parallel-channels})
with $G=\text{diag}\brac{\begin{array}{cccc}
g_{11} & . & . & g_{MM}\end{array}}$ and links $g_{ii}\sim\mathcal{CN}\brac{0,\rho_{ii}^{2}}=\mathcal{CN}\brac{0,\snr^{\gamma_{ii}}}$
is $\sum_{i}\brac{1-\frac{1}{T}}\gamma_{ii}$. \label{cor:parallel_channels}
\end{cor}

\begin{IEEEproof}
For a SISO channel with link $g_{ii}$ distributed according to $\mathcal{CN}\brac{0,\rho_{ii}^{2}}=\mathcal{CN}\brac{0,\snr^{\gamma_{ii}}}$,
the gDoF is $\brac{1-\frac{1}{T}}\gamma_{ii}$ \cite{Zheng_Tse_Grassmann_MIMO}.
The result for the parallel channel system follows by decomposing
the parallel channel into individual SISO channels and using Theorem
\ref{thm:decompose_graph}.
\end{IEEEproof}

\subsection{SIMO Channels and MISO Channels\label{subsec:SIMO-and-MISO}}

In this subsection, we consider noncoherent SIMO channels and MISO
channels with asymmetric link strengths. The gDoF result for the SIMO
channel can be easily derived by extending the results for the case
with i.i.d. links. For $T=1$, the gDoF result for the SIMO channel
can be extended to the arbitrary MIMO case. For the MISO case, the
existing techniques are not sufficient for computing the outer bound.
We develop new techniques, manipulating entropy expressions using
linear algebra techniques to derive the gDoF of the MISO channel.

\begin{figure}
\begin{centering}
\includegraphics{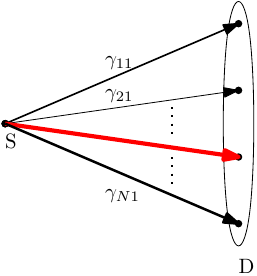}\caption{\label{fig:Noncoherent-SIMO}Selecting the statistically best antenna
is gDoF-optimal for the noncoherent SIMO channel with given $\protect\snr$
exponents.}
\par\end{centering}
\end{figure}

\begin{thm}
For the noncoherent SIMO channel (Figure \ref{fig:Noncoherent-SIMO})
with $G=\tran\brac{\left[\begin{array}{cccc}
g_{11} & . & . & g_{N1}\end{array}\right]}$, where $g_{i1}\sim\mathcal{CN}\brac{0,\rho_{i1}^{2}}=\mathcal{CN}\brac{0,\mathsf{SNR}^{\gamma_{i1}}}$,
the gDoF is $\brac{1-\frac{1}{T}}\max_{i}\gamma_{i1}$, i.e., the
gDoF can be achieved by using only the statistically best receive
antenna. \label{thm:simo}
\end{thm}
\begin{IEEEproof}
We only need to prove the outer bound, since the achievability follows
by using the statistically best receive antenna. The outer bound can
be proved as an extension of results for the SIMO channel with i.i.d.
links. We construct another SIMO channel with a larger capacity than
the given asymmetric SIMO channel. Let $\rho_{*}^{2}=\max_{i}\rho_{i1}^{2}$.
Now with $W$ being a $T\times1$ noise vector with i.i.d. $\mathcal{CN}\brac{0,1}$
elements, $G'$ being a $1\times N$ channel matrix with i.i.d. $\mathcal{CN}\brac{0,\rho_{*}^{2}}$
elements, $W_{1}$ being a noise vector with independent (but not
identical) Gaussian elements $w_{1i}\sim\mathcal{CN}\brac{0,\rho_{*}^{2}/\rho_{1i}^{2}-1},\ i\in\cbrac{1,2,\ldots,T}$
and $K$ being a constant diagonal matrix with elements $k_{ii}=\rho_{1i}/\rho_{*},\ i\in\cbrac{1,2,\ldots,T}$,
we observe that $K\brac{G'X+W+W_{1}}$ has the same distribution as
\[
Y=GX+W.
\]
Hence by the data processing inequality $I\brac{X;GX+W}\leq I\brac{X;G'X+W}$.
Now due to the results for i.i.d. noncoherent MIMO channels \cite{Zheng_Tse_Grassmann_MIMO},
we have $I\brac{X;G'X+W}\leqdof\brac{T-1}\lgbrac{\rho_{*}^{2}}$.
Hence the required result follows.

Using the above result for the SIMO channel, we can now prove that
the gDoF is zero for any MIMO channel for $T=1$.
\end{IEEEproof}
\begin{thm}
(gDoF of arbitrary MIMO channel for $T=1$) For any $G$ with $T=1$,
the gDoF is zero. \label{thm:zero_dof_T=00003D1}
\end{thm}
\begin{IEEEproof}
This can be shown by separately examining the SIMO channels constructed
using $\overline{g_{i}}\ i\in\cbrac{1,2,\ldots,N}$ from $G=\left[\begin{array}{ccccc}
\overline{g_{1}} & \overline{g_{2}} & . & . & \overline{g_{N}}\end{array}\right]$ and $\underline{X_{i}}$ from $X=\tran\sbrac{\begin{array}{ccc}
\tran\brac{\underline{X_{1}}} & \ldots & \tran\brac{\underline{X_{N}}}\end{array}}$, $G$ being the channel and $X$ being the symbols for the whole
MIMO channel. Consider  $N$ SIMO channels $Y_{i}=\overline{g_{i}}\underline{X_{i}}+\frac{W_{i}}{\sqrt{N}}$,
where $W_{i}$ and $W$ have the same distribution but are independent.
Now
\begin{align}
I\brac{X;GX+W} & \quad\leq I\brac{X;\overline{g_{1}}\underline{X_{1}}+\frac{W_{1}}{\sqrt{N}},\ldots,\overline{g_{N}}\underline{X_{N}}+\frac{W_{N}}{\sqrt{N}}}
\end{align}
 using the data processing inequality since $\sum_{i=1}^{N}\brac{W_{i}/\sqrt{N}}\sim W$
and $\sum_{i=1}^{N}\overline{g_{i}}\underline{X_{i}}=GX$. This creates
a new channel which is decomposable into $N$ SIMO channels, and the
new channel has a larger capacity than the original channel. Hence
the required result follows due to Theorem \ref{thm:decompose_graph}
by decomposing the new channel into $N$ SIMO channels and using the
fact that each SIMO channel has zero gDoF for $T=1$ (due to Theorem
\ref{thm:simo}).
\end{IEEEproof}
Note that the above result is a generalization of the zero DoF result
for MIMO channels by Lapidoth and Moser \cite{lapidoth2003capacity}.
In their model, the channel statistics is fixed and the power of the
i.i.d. noise is scaled. However, our result is more general in the
sense that we allow the fading channel strengths to be scaled with
different exponents.

\begin{figure}[h]
\centering{}\includegraphics{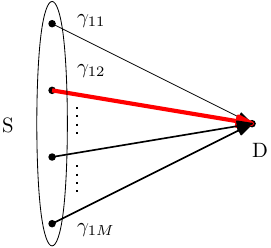}\caption{\label{fig:Noncoherent-MISO}Selecting the statistically best antenna
is gDoF-optimal for a noncoherent MISO channel with given $\protect\snr$
exponents.}
\end{figure}

\begin{thm}
\label{thm:MISO_dof}For the noncoherent MISO channel (Figure \ref{fig:Noncoherent-MISO})
with $G=\left[\begin{array}{cccc}
g_{11} & . & . & g_{1M}\end{array}\right]$, the gDoF is $\brac{1-\frac{1}{T}}\max_{i}\gamma_{1i}$, i.e., the
gDoF can be achieved by only using the statistically best transmit
antenna.
\end{thm}
\begin{IEEEproof}
[Proof idea]We only need to prove the outer bound. In this case,
$Y$ is a column vector and $h\brac Y$ can be evaluated using Lemma
\ref{lem:isotropic_entropy_to_radial}. Also, we prove that
\[
h\brac{Y|X}\geqdof\expect{\lgbrac{1+\sum_{i=1}^{M}\rho_{1i}^{2}\twonorm{\underline{X_{i}}}^{2}}}
\]
 using linear algebra techniques. With these two results, the gDoF
result follows. See Section \ref{subsec:MISO} for details.
\end{IEEEproof}

\subsection{The $2\times2$ MIMO Channel}

\begin{figure}[h]

\begin{centering}
\includegraphics[scale=0.75]{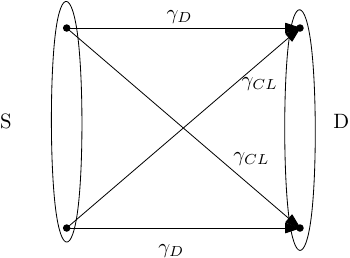}\caption{The $2\times2$ MIMO channel with $\protect\snr$ exponents $\gamma_{D}$
in the direct links and $\gamma_{CL}$ in the crosslinks.\label{fig:2x2MIMO}}
\par\end{centering}
\end{figure}

In this subsection, we describe the results for the $2\times2$ MIMO
channel with $\snr$ exponents $\gamma_{D}$ in the direct links and
$\gamma_{CL}$ in the crosslinks (Figure \ref{fig:2x2MIMO}). This
is one of the simple extensions starting from the MIMO channel with
i.i.d. links and this extension demonstrates different properties
than the i.i.d. case. We describe our outer bound and obtain a signaling
distribution to solve the outer bound optimization problem in terms
of gDoF. The signaling distribution for achievability uses the structure
of our solution to the outer bound optimization problem.

\begin{thm}
\label{thm:2x2-sym-mimo}For the $2\times2$ noncoherent MIMO channel
with
\[
G=\left[\begin{array}{cc}
g_{11} & g_{12}\\
g_{21} & g_{22}
\end{array}\right],
\]
where $g_{11}\sim g_{22}\sim\mathcal{CN}\brac{0,\mathsf{SNR^{\gamma_{D}}}}$
, $g_{12}\sim g_{21}\sim\mathcal{CN}\brac{0,\mathsf{SNR^{\gamma_{CL}}}}$
and $\gamma_{D}\geq\gamma_{CL}$, the gDoF is given in Table \ref{tab:Sol_outerbd_2x2_sym_mimo},
and can be achieved by
\[
X=\left[\begin{array}{cccccc}
a & 0 & 0 & . & . & 0\\
\eta & c & 0 & . & . & 0
\end{array}\right]Q,
\]
where $\eta\sim\mathcal{CN}\brac{0,\abs b^{2}}$, $\abs a^{2}=\mathsf{SNR}^{-\gamma_{a}},\ \abs b^{2}=\mathsf{SNR^{-\gamma_{b}}},\ \abs c^{2}=\mathsf{SNR^{-\gamma_{c}}}$
with the values of $\brac{\gamma_{a},\gamma_{b},\gamma_{c}}$ taken
from Table \ref{tab:Sol_outerbd_2x2_sym_mimo} and Q is an isotropically
distributed unitary matrix independent of $\eta$.

\begin{table}[h]
\centering{}\caption{gDoF of the $2\times2$ MIMO channel with $\gamma_{11}=\gamma_{22}=\gamma_{D}\protect\geq\gamma_{CL}=\gamma_{12}=\gamma_{21}$\label{tab:Sol_outerbd_2x2_sym_mimo}}
\vspace{2 mm}%
\begin{tabular}{|l|l|l|}
\hline
Regime & Solution $\brac{\gamma_{a},\gamma_{b},\gamma_{c}}$ & gDoF\tabularnewline
\hline
\hline
$T=2$ & $\brac{0,0,\gamma_{CL}}$ & $\gamma_{D}-\frac{1}{2}\gamma_{CL}$\tabularnewline
\hline
$T\geq3$ & $\brac{0,0,0}$ & $2\brac{\brac{1-\frac{1}{T}}\gamma_{D}-\frac{1}{T}\gamma_{CL}}$\tabularnewline
\hline
\end{tabular}
\end{table}

\end{thm}
\begin{IEEEproof}
[Proof idea]From Theorem \ref{thm:structure_optimal_distr}, we have
an optimal distribution of the form
\begin{equation}
X=\left[\begin{array}{cccccc}
a & 0 & 0 & . & . & 0\\
b & c & 0 & . & . & 0
\end{array}\right]Q,
\end{equation}
where $Q$ is an isotropically distributed unitary matrix independent
of $a,b,c$. We first obtain a capacity outer bound as the maximum
of the expected value of a function $f\brac{\abs a^{2},\abs b^{2},\abs c^{2}}$.
This is using Lemma \ref{lem:dof_equivalence_abs_lin_comb_gaussian_vector},
Lemma \ref{lem:dof_equivalence_h(xi22)} and Lemma \ref{lem:dof_equivalence_h(xi21)}
which help to convert the entropy terms $h\brac{}$ into expected
values. Then, in Lemma \ref{lem:discretization} we prove that the
maximization of $\expect{f\brac{\abs a^{2},\abs b^{2},\abs c^{2}}}$
can have a solution with a single mass point of $\brac{\abs a^{2},\abs b^{2},\abs c^{2}}$
for achieving the gDoF. Then the gDoF outer bound can be expressed
as the solution of a piecewise linear optimization problem, which
yields the solution as above. The detailed proof of the outer bound
is in Section \ref{subsec:2x2MIMO}. The inner bound can be verified
by using the distribution stated in the Theorem to evaluate the mutual
information; the calculation is given in \ifarxiv  Appendix \ref{app:Inner-bound-for2x2mimo}\else  \cite[Appendix D]{Joyson_2x2_mimov5}\fi.
Also note that $\gamma_{D}\geq\gamma_{CL}$ is without loss of generality,
since we can relabel the links to satisfy this condition.
\end{IEEEproof}
Note that the above result shows that we need to use both the antennas
for achieving the gDoF for $T=2$, since with only one antenna we
can only achieve $\brac{1/2}\gamma_{D}$ due to Theorem \ref{thm:simo}.
This is in contrast to the result for the $2\times2$ MIMO channel
with i.i.d. links, where the gDoF could be achieved using a single
transmit antenna for $T=2$; also, using both the antennas was sub-optimal
\cite{Zheng_Tse_Grassmann_MIMO}. For $T\geq3$, for a $2\times2$
MIMO channel with the value $\gamma_{D}$ for $\snr$ exponents in
all the links, the gDoF is $2\brac{1-2/T}\gamma_{D}$ \cite{Zheng_Tse_Grassmann_MIMO},
whereas in our model with direct link exponents $\gamma_{D}$ and
cross link exponents $\gamma_{CL}$, the gDoF is $2\brac{\brac{1-1/T}\gamma_{D}-\brac{1/T}\gamma_{CL}}$.
Thus having weaker crosslinks gives a gDoF gain of $\brac{2/T}\brac{\gamma_{D}-\gamma_{CL}}$.
Also as $T\rightarrow\infty$, the gDoF achieved is $2\gamma_{D}$,
which agrees with the gDoF results for the coherent MIMO channel \cite[Theorem 2]{Martina_gdof_MIMO}.

Also, it is clear that training-based schemes are suboptimal for the
$2\times2$ MIMO channel. For $T=2$, if one were to train the links,
one has to use two time slots, which leaves no time for communicating.
For $T=2$, if one were to use only one transmit antenna and use one
training slot, the gDoF achievable is $\gamma_{D}/2$ which is smaller
than what we achieve. For $T\geq3$, the gDoF achievable after using
two time slots to communicate is $2\brac{1-2/T}\gamma_{D}$ which
is less than the gDoF $2\brac{\brac{1-1/T}\gamma_{D}-\brac{1/T}\gamma_{CL}}$
that we achieve. The gain in gDoF that we have is $\brac{2/T}\brac{\gamma_{D}-\gamma_{CL}}$.

\begin{table}[H]
\centering{}\caption{\label{tab:Comparison-of-rates}Comparison of rates achievable for
the $2\times2$ MIMO channel with different schemes for $T=2$, $\protect\expect{\protect\abs{g_{11}}^{2}}=\protect\expect{\protect\abs{g_{22}}^{2}}=0.1$}
\begin{tabular}{|c|c|c|c|c|c|}
\hline
\multirow{2}{*}{Transmit $\snr$ per antenna (dB)} & \multirow{2}{*}{$\expect{\abs{g_{21}}^{2}}$} & \multicolumn{3}{c|}{Rates for different schemes} & \multirow{2}{*}{Gain using noncoherent scheme}\tabularnewline
\cline{3-5} \cline{4-5} \cline{5-5}
 &  & Noncoherent  & SISO & Parallel & \tabularnewline
\hline
22 & .025 & 1.364 & 1.305 & 1.063 & 0.059\tabularnewline
\hline
23 & .025 & 1.536 & 1.438 & 1.095 & 0.098\tabularnewline
\hline
23 & .016 & 1.657 & 1.438 & 1.396 & 0.220\tabularnewline
\hline
23 & .040 & 1.454 & 1.438 & 0.807 & 0.017\tabularnewline
\hline
\end{tabular}
\end{table}

Although Theorem \ref{thm:2x2-sym-mimo} is for the gDoF of the system,
our results can provide design guidelines for specific scenarios.
For example, for a $2\times2$ system with coherence time $T=2$,
transmit $\snr=23$ dB, direct links with average strength $0.1$
and crosslinks with average strength $0.025$ (which corresponds to
$\gamma_{D}=0.56,\ \gamma_{CL}=0.30$), our noncoherent scheme can
obtain a $7\%$ gain in the rate compared to the schemes that use
a symbol for training. We illustrate more examples\footnote{The link strengths in the examples are given without absorbing the
transmit SNR into them.} in Table \ref{tab:Comparison-of-rates} where our noncoherent scheme
can obtain gain in the rates compared to the schemes that use a symbol
for training. One possible training-based scheme is to use only one
antenna (reducing the MIMO channel to a SISO channel) and using one
symbol to train the channel. Another possible scheme is to use both
antennas and treat the system as a parallel antenna system, treating
the crosslinks as noise. For the parallel case also, the training-based
scheme uses one symbol to train the channel. The rate points in Table
\ref{tab:Comparison-of-rates} are just a few examples of some specific
scenarios, but we believe that there would be many other cases where
this approach is useful when we have short coherence time and asymmetry
in channel gains. The details of the expressions used for the numerics
are given in \ifarxiv  Appendix \ref{app:simulation_calculation}\else  \cite[Appendix J]{Joyson_2x2_mimov5}\fi.

\subsection{Nonoptimality of Training\label{subsec:Nonoptimality-of-training-MxM}}

\begin{figure}
\centering{}\includegraphics{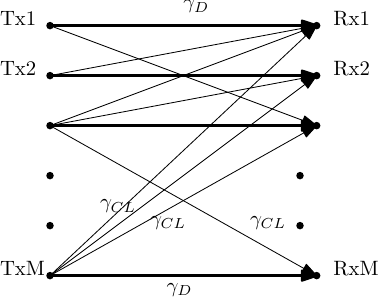}\caption{\label{fig:MIMO-with-direct-and-cross} The $M\times M$ MIMO channel
with exponents $\gamma_{D}$ in the direct links and $\gamma_{CL}$
in crosslinks.}
\end{figure}

We observed in the previous subsection that training-based schemes
cannot achieve the gDoF for $2\times2$ MIMO channels in general.
We can extend this observation to larger MIMO channels. We specifically
consider the $M\times M$ MIMO channel with exponents $\gamma_{D}$
in the direct links and $\gamma_{CL}$ in the crosslinks ($\gamma_{D}>\gamma_{CL}$).
Using the following theorem, we prove that training-based schemes
are suboptimal for this case.
\begin{thm}
A gDoF of $M\brac{\brac{1-1/T}\gamma_{D}-\brac{\brac{M-1}/T}\gamma_{CL}}$
can be achieved for an $M\times M$ MIMO channel with coherence time
$T>M$ and with exponents $\gamma_{D}$ in the direct links and $\gamma_{CL}$
in the crosslinks ($\gamma_{D}>\gamma_{CL}$) (Figure \ref{fig:MIMO-with-direct-and-cross}),
by using i.i.d. Gaussian codebooks across the antennas and time periods.
\label{thm:gaussian_codebooks}
\end{thm}

\begin{IEEEproof}
In this case, the channel matrix $G$ has diagonal elements $g_{ii}$
distributed according to $\mathcal{CN}\brac{0,\snr^{\gamma_{D}}}$
and the rest of the elements are distributed according to $\mathcal{CN}\brac{0,\snr^{\gamma_{CL}}}$.
Using Gaussian codebooks, the rate $R\geq I\brac{GX+W;X}$ is achievable
with $X$ being an $M\times T$ matrix with i.i.d. $\mathcal{CN}\brac{0,1}$
elements. Analyzing this mutual information yields an achievable gDoF
of\newline $M\brac{\brac{1-1/T}\gamma_{D}-\brac{\brac{M-1}/T}\gamma_{CL}}$
per symbol. The calculations are given in \ifarxiv  Appendix \ref{app:Gaussian-code-noncohmimo}\else  \cite[Appendix E]{Joyson_2x2_mimov5}\fi.
\end{IEEEproof}
Note that the gDoF $M\brac{\brac{1-1/T}\gamma_{D}-\brac{\brac{M-1}/T}\gamma_{CL}}$
cannot be achieved by a conventional training scheme where all transmitters
train independently. This is clear since it requires $M$ symbols
in every coherence period for training and the maximum gDoF achievable
using the rest of the symbols is $M\brac{1-M/T}\gamma_{D}$ \cite[Theorem 2]{Martina_gdof_MIMO},
assuming that the channels are available perfectly due to training.
This is smaller than $M\brac{\brac{1-1/T}\gamma_{D}-\brac{\brac{M-1}/T}\gamma_{CL}}$.
Thus using Gaussian codebooks and not using training give a gDoF gain
of $\brac{M\brac{M-1}/T}\brac{\gamma_{D}-\gamma_{CL}}$. This result
also suggests that for noncoherent networks with multiple nodes, a
training-based scheme may not be gDoF-optimal, by viewing the cut-sets
of the networks as MIMO channels.

\section{Analysis\label{sec:Analysis}}

In this section, we provide a more detailed analysis for the results
stated in the previous section. We first state some mathematical preliminaries
required for the analysis. Then in Section \ref{subsec:properties of capacity achieving distribution},
we derive the structure of the capacity achieving distribution for
the noncoherent MIMO channel. In Section \ref{subsec:MISO}, we prove
the gDoF outer bounds for the noncoherent MISO channel and in Section
\ref{subsec:2x2MIMO}, we derive the gDoF outer bounds for the $2\times2$
MIMO system.

\subsection{Mathematical Preliminaries}

\begin{lem}
For an exponentially distributed random variable $\xi$ with mean
$\mu_{\xi}$ and for given constants $a\geq0,b>0$,
\[
\lgbrac{a+b\mu_{\xi}}-\gamma_{E}\lgbrac e\leq\expect{\lgbrac{a+b\xi}}\leq\lgbrac{a+b\mu_{\xi}},
\]
 where $\gamma_{E}$ is Euler's constant.\label{fact:Jensens_gap}
\end{lem}
\begin{IEEEproof}
This is given \textcolor{black}{in \cite[Section VI-B]{Joyson_fading}.}
\end{IEEEproof}

\begin{lem}
\label{fact:Jensens_gap_chi_squared}For a chi-squared random variable
$\chi^{2}\brac k$ and for given constants $a\geq0,b>0$,
\begin{equation}
\lgbrac{a+bk}-\frac{\lgbrac e2}{k}+\lgbrac{1+\frac{1}{k}}\leq\expect{\lgbrac{a+b\chi^{2}\brac k}}\leq\lgbrac{a+bk}.
\end{equation}
\end{lem}
\begin{IEEEproof}
The result is prov\textcolor{black}{ed in \cite[Section VI-A]{Joyson_fading}
for t}he Gamma distribution and the result for the chi-squared distribution
follows as a special case.
\end{IEEEproof}
%

\begin{lem}
For an exponential random variable $\xi$ with mean $\mu_{\xi}$ and
for a given constant $b>0$, we have
\begin{align}
\expect{\frac{b}{b+\xi}} & =\frac{b}{\mu_{\xi}}e^{\frac{b}{\mu_{\xi}}}E_{1}\brac{\frac{b}{\mu_{\xi}}}
\end{align}
and
\begin{align}
1>\frac{b}{\mu_{\xi}}\lnbrac{1+\frac{\mu_{\xi}}{b}} & \geq\frac{b}{\mu_{\xi}}e^{\frac{b}{\mu_{\xi}}}E_{1}\brac{\frac{b}{\mu_{\xi}}}\geq\frac{b}{2\mu_{\xi}}\lnbrac{1+\frac{2\mu_{\xi}}{b}},
\end{align}
where $E_{1}\brac{\cdot}$ is the exponential integral function. \label{fact:expectation_recipr_exponential_distr}
\end{lem}
\begin{IEEEproof}
We have
\begin{align}
\expect{\frac{b}{b+\xi}} & =\expect{\frac{1}{1+\frac{\xi}{b}}}\\
 & \overset{\brac i}{=}\int_{0}^{\infty}\frac{b}{\mu_{\xi}}e^{-\frac{bx}{\mu_{\xi}}}\frac{1}{1+x}dx\\
 & \overset{\brac{ii}}{=}\frac{b}{\mu_{\xi}}e^{\frac{b}{\mu_{\xi}}}\int_{1}^{\infty}e^{-\frac{bx}{\mu_{\xi}}}\frac{1}{x}dx\\
 & \overset{\brac{iii}}{=}\frac{b}{\mu_{\xi}}e^{\frac{b}{\mu_{\xi}}}\int_{\frac{b}{\mu_{\xi}}}^{\infty}e^{-t}\frac{1}{t}dt\\
 & \overset{\brac{iv}}{=}\frac{b}{\mu_{\xi}}e^{\frac{b}{\mu_{\xi}}}E_{1}\brac{\frac{b}{\mu_{\xi}}}.
\end{align}
where the step $\brac i$ is because $\xi/b$ is exponentially distributed
with mean $\mu_{\xi}/b$, the steps $\brac{ii}$, $\brac{iii}$ are
by change of variables, and the step $\brac{iv}$ is by the definition
of the exponential integral. We have
\[
\frac{1}{2}\lnbrac{1+\frac{1}{x}}\leq e^{x}E_{1}\brac x\leq\lnbrac{1+\frac{1}{x}}
\]
 from \cite{abramowitz1964handbook}. Also $\brac{b/\mu_{\xi}}\lnbrac{1+\mu_{\xi}/b}<1$,
because $0<x\lnbrac{1+1/x}<1$ for $x>0$. Thus the desired result
follows.
\end{IEEEproof}
\begin{lem}
\label{fact:HtimesPhi_distribution}Let $H$ be an isotropically distributed
random matrix and $\Phi$ be a random unitary matrix distributed according
to any distribution independent of $H$, then $H,\Phi H,H\Phi$ all
have the same distribution. Moreover, $\Phi H$ and $H\Phi$ are independent
of $\Phi$. See \cite{marzetta1999capacity} for details.
\end{lem}
\begin{lem}
Let $\sbrac{\xi_{1},\xi_{2},\ldots,\xi_{n}}$ be an arbitrary complex
random vector and $Q$ be an $n\times n$ isotropically distributed
random unitary matrix independent of $\xi_{i}$, then
\[
h\brac{\sbrac{\xi_{1},\xi_{2},\ldots,\xi_{n}}Q}=h\brac{\sum\abs{\xi_{i}}^{2}}+\brac{n-1}\expect{\lgbrac{\sum\abs{\xi_{i}}^{2}}}+\lgbrac{\frac{\pi^{n}}{\Gamma\brac n}}.
\]
\label{lem:isotropic_entropy_to_radial}
\end{lem}
\begin{IEEEproof}
[Proof idea]This is proved by using the fact that in radial coordinates,
the distribution of \sloppy $\sbrac{\xi_{1},\xi_{2},\ldots,\xi_{n}}Q$
is dependent only on the radius. See Appendix \ref{app:proof_lemma_isotropic_entropy_to_radial}
 for more details.
\end{IEEEproof}
Note that we can use the above Lemma also on $h\brac{\xi_{1}\overline{q_{1}}^{\brac T}}$
\textcolor{black}{with an isotropically distributed random unit vector
$\overline{q_{1}}^{\brac T}$ by considering the equality $h\brac{\xi_{1}\overline{q_{1}}^{\brac T}}=h\brac{\sbrac{\xi_{1},0,..,0}Q}$,
where the isotropically distributed random unit vector $\overline{q_{1}}^{\brac T}$
can be taken as the first row of an isotropically distributed random
unitary matrix $Q$.}
\begin{cor}
Let $\sbrac{\xi_{1},\xi_{2},\ldots,\xi_{n}}$ be an arbitrary complex
random vector, $\xi$ be an arbitrary complex random variable and
$Q$ be an $n\times n$ \textcolor{black}{isotropically distributed
}random\textcolor{black}{{} unitary matrix} independent of $\xi,\xi_{i}$,
then
\[
h\brac{\rline{\sbrac{\xi_{1},\xi_{2},\ldots,\xi_{n}}Q}\xi}=h\brac{\rline{\sum\abs{\xi_{i}}^{2}}\xi}+\brac{n-1}\expect{\lgbrac{\sum\abs{\xi_{i}}^{2}}}+\lgbrac{\frac{\pi^{n}}{\Gamma\brac n}}.
\]
\label{cor:isotropic_entropy_to_radial_with_conditioning}
\end{cor}
\begin{IEEEproof}
This can be proved similar to the previous lemma since the distribution
\sloppy of $h\brac{\rline{\sbrac{\xi_{1},\xi_{2},\ldots,\xi_{n}}Q}\xi}$
will be dependent only on the radius.
\end{IEEEproof}
We can use this corollary also on $h\brac{\rline{\xi_{1}\overline{q_{1}}^{\brac T}}\xi}$,
similar to the previous Lemma.
\begin{lem}
Among positive continuous random variables with a given mean, the
exponential random variable has the maximum entropy.\label{lem:max_entropy}
\end{lem}
\begin{IEEEproof}
See \textcolor{black}{\cite[Chapter 12]{cover2012elements}.}
\end{IEEEproof}

\subsection{Properties of Transmitted Signals that Achieve Capacity \label{subsec:properties of capacity achieving distribution}}

We now establish the properties of capacity achieving distribution
for the noncoherent MIMO channel with asymmetric statistics. We have
our channel model $Y=GX+W$. Now for any $T\times T$ unitary matrix
$\Phi$ we have $Y\Phi^{\dagger}=GX\Phi^{\dagger}+W\Phi^{\dagger}.$
Since $w_{ij}$ are i.i.d. $\mathcal{CN}\brac{0,1}$, $W\Phi^{\dagger}$
and $W$ have the same distribution, and hence
\begin{equation}
p\brac{Y\Phi^{\dagger}|X\Phi^{\dagger}}=p\brac{Y|X}.\label{eq:p(Y|X)rotation}
\end{equation}
Now
\begin{equation}
C=\sup_{p\brac X}I\brac{X;Y}
\end{equation}
subject to the average power constraint (\ref{eq:power_constraint})
and we have
\begin{align}
I\brac{X;Y} & =\expect{\lgbrac{\frac{p\brac{Y|X}}{p\brac Y}}}\nonumber \\
 & =\int dXp\brac X\int dYp\brac{Y|X}\lgbrac{\frac{p\brac{Y|X}}{\int d\tilde{X}p\brac{\tilde{X}}p\brac{Y|\tilde{X}}}}.\label{eq:mutual_info_formula}
\end{align}

\begin{lem}
(Invariance of $I\brac{X;Y}$ to post-rotations of $X$\label{lem:post_rotation}):
Suppose that $X$ has a probability density $p_{0}\brac X$ that generates
some mutual information $I_{0}$. Then, for any unitary matrix $\Phi$,
the \textquotedblleft post-rotated\textquotedblright{} probability
density, $p_{1}\brac X=p_{0}\brac{X\Phi^{\dagger}}$ also generates
$I_{0}$.
\end{lem}
\begin{IEEEproof}
[Proof idea] This is an adaptation of the existing results for MIMO
channels from \cite[Lemma 1]{marzetta1999capacity}. The proof proceeds
by substituting the post-rotated density $p_{1}\brac X$ into  $\brac{\ref{eq:mutual_info_formula}}$,
changing the variables of integration and using $p\brac{Y\Phi|X\Phi}=p\brac{Y|X}$
from  (\ref{eq:p(Y|X)rotation}). %
\end{IEEEproof}
\begin{lem}
The signal of the form $X=LQ$ with $L$ being a lower triangular
random matrix and $Q$ being an isotropically distributed random unitary
matrix independent of $L$ achieves the capacity of the noncoherent
MIMO channel.\label{lem:signal_structure}
\end{lem}
\begin{IEEEproof}
Let $X$ be a capacity achieving random variable and $I_{0}$ be the
corresponding mutual information achieved. Now $X$ can be decomposed
as $X=L\Phi'$ using the LQ decomposition with $L$ upper diagonal
and $\Phi'$ unitary, but they could be jointly distributed and $\Phi'$
may not be isotropically unitary distributed. Let $\Theta$ be an
isotropically distributed random unitary matrix that is independent
of $L$ and $\Phi'$. Now use $X_{1}=X\Theta$ for signaling. Let
$Y$ be the corresponding received signal. Then
\begin{align}
I\brac{X_{1};Y|\Theta} & =I\brac{\rline{X\Theta;Y}\Theta}\\
 & =I_{0},
\end{align}
where the last step was using Lemma \ref{lem:post_rotation}. Now
\begin{align}
I\brac{X_{1};Y}+I\brac{\Theta;Y|X_{1}} & =I\brac{\Theta;Y}+I\brac{X_{1};Y|\Theta}\\
I\brac{X_{1};Y}+0 & \overset{\brac i}{=}I\brac{\Theta;Y}+I\brac{X_{1};Y|\Theta}\\
I\brac{X_{1};Y} & \overset{\brac{ii}}{\geq}I\brac{X_{1};Y|\Theta}\\
 & =I_{0},
\end{align}
where $\brac i$ was because $I\brac{\Theta;Y|X_{1}}=0$ since $\Theta-X_{1}-Y$
is a Markov chain and $\brac{ii}$ was because $I\brac{X_{1};Y|\Theta}\geq0$.
Hence without loss of generality, the signal of the form $LQ=L\Phi'\Theta$
with $Q=\Phi'\Theta$ achieves the capacity. Now $Q=\Phi'\Theta$
is an isotropically distributed unitary matrix and it is independent
of $\Phi'$ using Lemma \ref{fact:HtimesPhi_distribution} on page
\pageref{fact:HtimesPhi_distribution}.
\end{IEEEproof}
Next, we focus our attention on computing $h\brac{Y|X}$, which will
be necessary in future derivations. Let $Y(n)$ be the $n^{\text{th}}$
row of $Y$. Conditioned on $X$, the rows of $Y$ are independent
Gaussian. Hence
\begin{align}
h\brac{Y|X} & =\sum_{n=1}^{N}h\brac{Y\brac n|X}.\label{eq:h(Y|X)}
\end{align}
With $\underline{\rho}^{2}\brac n$ being the vector of channel strengths
arriving at the $n^{\text{th}}$ receiver antenna, we have:
\begin{align*}
K_{Y\brac n|X} & =\mathbb{E}\sbrac{\rline{Q^{\dagger}L^{\dagger}g^{\dagger}\brac ng\brac nLQ}LQ}+I_{T}\\
 & =Q^{\dagger}L^{\dagger}\mathbb{E}\sbrac{g^{\dagger}\brac ng\brac n}LQ+I_{T}\\
 & =Q^{\dagger}L^{\dagger}\text{diag}\brac{\underline{\rho}^{2}\brac n}LQ+I_{T},
\end{align*}
where $I_{T}$ is a $T\times T$ identity matrix and $\text{diag}\brac{\underline{\rho}^{2}\brac n}$
is the diagonal matrix formed from $\underline{\rho}^{2}\brac n$.
Hence
\begin{align}
 & h\brac{Y\brac n|X}\nonumber \\
 & \quad=\expect{\lgbrac{\det\brac{\pi eK_{Y\brac n|X}}}}\\
 & \quad=\expect{\lgbrac{\det\brac{\pi e\brac{Q^{\dagger}L^{\dagger}\text{diag}\brac{\underline{\rho}^{2}\brac n}LQ+I_{T}}}}}\\
 & \quad\overset{\brac i}{=}\expect{\lgbrac{\det\brac{\pi e\brac{L^{\dagger}\text{diag}\brac{\underline{\rho}^{2}\brac n}L+I_{T}}}}},
\end{align}
where $\brac i$ uses the property of determinants to cancel $Q$
and $Q^{\dagger}$. Also, for $T\geq M$, with $L_{M\times M}$ being
the first $M\times M$ submatrix of $L$ (rest of the elements of
$L$ are zero for $T\geq M$), we have:
\begin{align}
 & h\brac{Y\brac n|X}\nonumber \\
 & \overset{}{=}\expect{\lgbrac{\det\brac{L_{M\times M}^{\dagger}\text{diag}\brac{\underline{\rho}^{2}\brac n}L_{M\times M}+I_{M}}}}+T\lgbrac{\pi e}.\label{eq:h(Y(n)|X)}
\end{align}

\subsection{Outer Bound for the $M\times1$ MISO Channel\label{subsec:MISO}}

We now prove the gDoF outer bound given in Theorem \ref{thm:MISO_dof}
for the $M\times1$ MISO system. We assume that $T>1$, since for
$T=1$ we have the desired result using Theorem \ref{thm:zero_dof_T=00003D1}
on page \pageref{thm:zero_dof_T=00003D1}. Also, we assume that $T\geq M$
in the following outer bound computations. The case for $T<M$ can
be derived similarly and the derivation is given in \ifarxiv  Appendix \ref{app:MISO_T<M}\else  \cite[Appendix H]{Joyson_2x2_mimov5}\fi.

For the capacity achieving distribution, we have the structure $X=\left[\begin{array}{cc}
L_{M\times M} & 0_{M\times\brac{T-M}}\end{array}\right]Q$ (from Theorem \ref{thm:structure_optimal_distr}), where
\[
L_{M\times M}=\left[\begin{array}{cccc}
x_{11} & 0 & 0\\
. & . & 0 & 0\\
. &  & . & 0\\
x_{M1} & . & . & x_{MM}
\end{array}\right],
\]
and $0_{M\times\brac{T-M}}$ is an $M\times\brac{T-M}$ matrix with
elements of value zero. Also $Q$ is an isotropically distributed
random unitary matrix. For the MISO channel, we have $Y=GX+W$ with
$G=\left[\begin{array}{cccc}
g_{11} & . & . & g_{1M}\end{array}\right]$ and $W$ is the $1\times T$ noise vector with i.i.d. $\mathcal{CN}\brac{0,1}$
components. We assume $\rho_{11}^{2}\geq\rho_{1i}^{2}$ without loss
of generality. Now note that $WQ$ has also the same distribution
as $W$ and is independent of $Q$ (using the fact that $W$ is isotropically
distributed and Lemma \ref{fact:HtimesPhi_distribution}). Hence
\begin{align*}
h\brac Y & =h\brac{\brac{GX+W}Q}\\
 & =h\left(\left[\vphantom{a^{a^{a^{a^{a}}}}}\begin{array}{cc}
\brac{w_{11}+\sum_{i=1}^{M}x_{i1}g_{1i}}, & \brac{w_{12}+\sum_{i=2}^{M}x_{i2}g_{1i}},\ldots\end{array}\right.\right.\\
 & \qquad\qquad\left.\left.\ldots,\begin{array}{cccc}
\brac{w_{1M}+\sum_{i=M}^{M}x_{i2}g_{1i}}, & w_{1\brac{M+1}} & \ldots, & w_{1T}\end{array}\vphantom{a^{a^{a^{a^{a}}}}}\right]Q\right),
\end{align*}
where $w_{1i}$ are i.i.d. and $\mathcal{CN}\brac{0,1}$ distributed.
Now using Lemma \ref{lem:isotropic_entropy_to_radial} on page \pageref{lem:isotropic_entropy_to_radial},
we have
\begin{align}
h\brac Y & =h\brac{\sum_{j=1}^{M}\abs{w_{1j}+\sum_{i=j}^{M}x_{ij}g_{1i}}^{2}+\sum_{i=M+1}^{T}\abs{w_{1i}}^{2}}\nonumber \\
 & \qquad+\brac{T-1}\expect{\lgbrac{\sum_{j=1}^{M}\abs{w_{1j}+\sum_{i=j}^{M}x_{ij}g_{1i}}^{2}+\sum_{i=M+1}^{T}\abs{w_{1i}}^{2}}}+\lgbrac{\frac{\pi^{T}}{\Gamma\brac T}}\\
 & \overset{\brac i}{\leq}h\brac{\sum_{j=1}^{M}\abs{w_{1j}+\sum_{i=j}^{M}x_{ij}g_{1i}}^{2}+\sum_{i=M+1}^{T}\abs{w_{1i}}^{2}}\nonumber \\
 & \qquad+\brac{T-1}\expect{\lgbrac{\sum_{i=1}^{M}\rho_{1i}^{2}\brac{\sum_{j=1}^{i}\abs{x_{ij}}^{2}}+T-M}}+\lgbrac{\frac{\pi^{T}}{\Gamma\brac T}},
\end{align}
where $\brac i$ is using the Tower property of expectation, Jensen's
inequality and $\sum_{j=1}^{M}\sum_{i=j}^{M}\abs{x_{ij}}^{2}\rho_{1i}^{2}=\sum_{i=1}^{M}\sum_{j=1}^{i}\abs{x_{ij}}^{2}\rho_{1i}^{2}$.
Now using   (\ref{eq:h(Y(n)|X)}) we have
\begin{align}
h\brac{Y|X} & =\expect{\lgbrac{\det\brac{L_{M\times M}^{\dagger}\text{diag }\brac{\rho_{11}^{2},\ldots,\rho_{1M}^{2}}L_{M\times M}+I_{M}}}}\nonumber \\
 & \qquad+T\lgbrac{\pi e}\\
 & =\expect{\lgbrac{\prod_{i=1}^{M}\brac{1+\omega_{i}}}}+T\lgbrac{\pi e},
\end{align}
where $\omega_{i}$ are the eigenvalues of $L_{M\times M}^{\dagger}\text{diag }\brac{\rho_{11}^{2},\ldots,\rho_{1M}^{2}}L_{M\times M}$.
The eigenvalues are non-negative since the matrix is Hermitian. Hence
\begin{align}
h\brac{Y|X} & =\expect{\lgbrac{\prod_{i=1}^{M}\brac{1+\omega_{i}}}}+T\lgbrac{\pi e}\\
 & \geq\expect{\lgbrac{1+\sum\omega_{i}}}+T\lgbrac{\pi e}.
\end{align}
The last step is true because $\omega_{i}\geq0$. Now
\begin{align}
\sum\omega_{i} & =\text{Trace}\brac{L_{M\times M}^{\dagger}\text{diag }\brac{\rho_{11}^{2},\ldots,\rho_{1M}^{2}}L_{M\times M}}\\
 & =\text{Trace}\brac{\text{diag }\brac{\rho_{11}^{2},\ldots,\rho_{1M}^{2}}L_{M\times M}L_{M\times M}^{\dagger}}\\
 & =\sum_{i=1}^{M}\rho_{1i}^{2}\brac{\sum_{j=1}^{i}\abs{x_{ij}}^{2}}.
\end{align}
Hence
\begin{align}
h\brac{Y|X} & \geq\expect{\lgbrac{1+\sum_{i=1}^{M}\rho_{1i}^{2}\brac{\sum_{j=1}^{i}\abs{x_{ij}}^{2}}}}+T\lgbrac{\pi e}.
\end{align}
Hence
\begin{align}
I\brac{X;Y} & \overset{}{\leq}h\brac{\sum_{j=1}^{M}\abs{w_{1j}+\sum_{i=j}^{M}x_{ij}g_{1i}}^{2}+\sum_{i=M+1}^{T}\abs{w_{1i}}^{2}}\nonumber \\
 & \qquad+\brac{T-1}\expect{\lgbrac{\sum_{i=1}^{M}\rho_{1i}^{2}\brac{\sum_{j=1}^{i}\abs{x_{ij}}^{2}}+T-M}}\nonumber \\
 & \qquad-\expect{\lgbrac{1+\sum_{i=1}^{M}\rho_{1i}^{2}\brac{\sum_{j=1}^{i}\abs{x_{ij}}^{2}}}}\nonumber \\
 & \qquad+\lgbrac{\frac{\pi^{T}}{\Gamma\brac T}}-T\lgbrac{\pi e}\\
 & \overset{.}{\leq}\brac{T-1}\lgbrac{\sum_{i=1}^{M}\rho_{1i}^{2}MT+T},
\end{align}
where in the last step, we used Lemma \ref{lem:max_entropy} and Jensen's
inequality. Hence
\begin{align}
\underset{\snr\rightarrow\infty}{\text{limsup}}\ \frac{1}{T}\frac{I\brac{X;Y}}{\lgbrac{\snr}} & \overset{}{\leq}\brac{1-\frac{1}{T}}\gamma_{11}.
\end{align}

\subsection{Outer Bound for the $2\times2$ MIMO Channel\label{subsec:2x2MIMO}}

In this subsection, we prove the gDoF outer bound from Theorem \ref{thm:2x2-sym-mimo}
for the $2\times2$ MIMO channel with exponents $\gamma_{D}$ in the
direct links and $\gamma_{CL}$ in the crosslinks. We have the structure
of the optimal distribution as
\[
X=\left[\begin{array}{cccccc}
a & 0 & 0 & . & . & 0\\
b & c & 0 & . & . & 0
\end{array}\right]Q
\]
 from Theorem \ref{thm:structure_optimal_distr}. We have
\[
G=\left[\begin{array}{cc}
g_{11} & g_{12}\\
g_{21} & g_{22}
\end{array}\right],
\]
and $Y=GX+W$, where $W$ is a $2\times T$ vector with i.i.d. $\mathcal{CN}\brac{0,1}$
components. For $T=1$, the gDoF is zero due to Theorem \ref{thm:zero_dof_T=00003D1},
hence we consider $T\geq2$ in this proof. We have
\begin{align}
h\brac Y & =h\brac{G\left[\begin{array}{cccccc}
a & 0 & 0 & . & . & 0\\
b & c & 0 & . & . & 0
\end{array}\right]Q+W}\\
 & \overset{\brac i}{=}h\brac{\brac{G\left[\begin{array}{cccccc}
a & 0 & 0 & . & . & 0\\
b & c & 0 & . & . & 0
\end{array}\right]+W}Q}\\
 & =h\brac{\left[\begin{array}{ccccc}
ag_{11}+bg_{12}+w_{11} & cg_{12}+w_{12} & w_{13} & . & w_{1T}\\
ag_{21}+bg_{22}+w_{21} & cg_{22}+w_{22} & w_{23} & . & w_{2T}
\end{array}\right]Q}\label{eq:2x2_preLQ}\\
 & \overset{\brac{ii}}{=}h\brac{\left[\begin{array}{cccccc}
\xi_{11} & 0 & . & . & . & 0\\
\xi_{21} & \xi_{22} & 0 & . & . & 0
\end{array}\right]\Phi Q}\\
 & \overset{\brac{iii}}{=}h\brac{\left[\begin{array}{cccccc}
\xi_{11} & 0 & . & . & . & 0\\
\xi_{21} & \xi_{22} & 0 & . & . & 0
\end{array}\right]Q},\label{eq:2x2_outer_lq}
\end{align}
where the step $\brac i$ used the fact that $W$ and $WQ$ have the
same distribution and $WQ$ is independent of $Q$. In step $\brac{ii}$,
$\xi_{ij}$ arise after LQ transformation (using Gram-Schmidt process):
\[
\left[\begin{array}{ccccc}
ag_{11}+bg_{12}+w_{11} & cg_{12}+w_{12} & w_{13} & . & w_{1T}\\
ag_{21}+bg_{22}+w_{21} & cg_{22}+w_{22} & w_{23} & . & w_{2T}
\end{array}\right]=\left[\begin{array}{cccccc}
\xi_{11} & 0 & . & . & . & 0\\
\xi_{21} & \xi_{22} & 0 & . & . & 0
\end{array}\right]\Phi,
\]
\begin{align}
\abs{\xi_{11}}^{2} & =\abs{ag_{11}+bg_{12}+w_{11}}^{2}+\abs{cg_{12}+w_{12}}^{2}+\sum_{i=3}^{T}\abs{w_{1i}}^{2},\label{eq:xi_11_defn}\\
\abs{\xi_{21}}^{2} & =\frac{\abs{\brac{ag_{21}+bg_{22}+w_{21}}\brac{ag_{11}+bg_{12}+w_{11}}^{*}+\brac{cg_{22}+w_{22}}\brac{cg_{12}+w_{12}}^{*}+\sum_{i=3}^{T}w_{2i}w_{1i}^{*}}^{2}}{\abs{ag_{11}+bg_{12}+w_{11}}^{2}+\abs{cg_{12}+w_{12}}^{2}+\sum_{i=3}^{T}\abs{w_{1i}}^{2}},\label{eq:xi_21_defn}\\
\abs{\xi_{22}}^{2} & =\abs{ag_{21}+bg_{22}+w_{21}}^{2}+\abs{cg_{22}+w_{22}}^{2}+\sum_{i=3}^{T}\abs{w_{2i}}^{2}\nonumber \\
 & \quad-\frac{\abs{\brac{ag_{21}+bg_{22}+w_{21}}\brac{ag_{11}+bg_{12}+w_{11}}^{*}+\brac{cg_{22}+w_{22}}\brac{cg_{12}+w_{12}}^{*}+\sum_{i=3}^{T}w_{2i}w_{1i}^{*}}^{2}}{\abs{ag_{11}+bg_{12}+w_{11}}^{2}+\abs{cg_{12}+w_{12}}^{2}+\sum_{i=3}^{T}\abs{w_{1i}}^{2}},\label{eq:xi_22_defn}
\end{align}
where $\Phi$ is unitary. In step $\brac{iii}$, we absorb $\Phi$
onto $Q$ using Lemma \ref{fact:HtimesPhi_distribution}. The Gram-Schmidt
process for LQ transformation yields $\xi_{ij}$ as given in   (\ref{eq:xi_11_defn}),
 (\ref{eq:xi_21_defn}) and   (\ref{eq:xi_22_defn}).

Also, using (\ref{eq:h(Y(n)|X)}), (\ref{eq:h(Y|X)}), we get
\begin{align}
h\brac{Y|X} & =\expect{\lgbrac{\abs a^{2}\rho_{11}^{2}+\abs b^{2}\rho_{12}^{2}+\abs c^{2}\rho_{12}^{2}+\abs a^{2}\abs c^{2}\rho_{11}^{2}\rho_{12}^{2}+1}}\nonumber \\
 & \quad+\expect{\lgbrac{\abs a^{2}\rho_{21}^{2}+\abs b^{2}\rho_{22}^{2}+\abs c^{2}\rho_{22}^{2}+\abs a^{2}\abs c^{2}\rho_{21}^{2}\rho_{22}^{2}+1}}\nonumber \\
 & \quad+2T\lgbrac{\pi e}.\label{eq:2x2MIMO_hy_x}
\end{align}
For computing $h\brac Y$, let $\overline{q}_{1}^{\brac T},\overline{q}_{2}^{\brac T}$
be the first two rows of $Q$. The vectors $\overline{q}_{1}^{\brac T},\overline{q}_{2}^{\brac T}$
are orthogonal since $Q$ is unitary. We have
\begin{align}
h\brac Y & =h\brac{\left[\begin{array}{cccccc}
\xi_{11} & 0 & . & . & . & 0\\
\xi_{21} & \xi_{22} & 0 & . & . & 0
\end{array}\right]Q}\label{eq:2x2LQsimplification_begin}\\
 & =h\brac{\xi_{11}\overline{q}_{1}^{\brac T}}+h\brac{\rline{\xi_{21}\overline{q}_{1}^{\brac T}+\xi_{22}\overline{q}_{2}^{\brac T}}\xi_{11}\overline{q}_{1}^{\brac T}}.
\end{align}
Now consider $h\brac{\rline{\xi_{21}\overline{q}_{1}^{\brac T}+\xi_{22}\overline{q}_{2}^{\brac T}}\xi_{11}\overline{q}_{1}^{\brac T}}.$
Since $\xi_{11}$ is nonnegative and $\xi_{11}\overline{q}_{1}^{\brac T}$
is given in the conditioning, the direction $\overline{q}_{1}^{\brac T}$
is known in the conditioning. Hence considering $\xi_{21}\overline{q}_{1}^{\brac T}+\xi_{22}\overline{q}_{2}^{\brac T}$
in a new orthonormal basis with the first basis vector chosen as $\overline{q}_{1}^{\brac T}$
and the rest of the basis vectors chosen arbitrarily, the projection
of $\xi_{21}\overline{q}_{1}^{\brac T}+\xi_{22}\overline{q}_{2}^{\brac T}$
onto the first basis vector is $\xi_{21}$. The projection onto the
rest of the $T-1$ vectors forms $\xi_{22}\overline{q}_{2}^{\brac{T-1}}$
where $\overline{q}_{2}^{\brac{T-1}}$ is a $T-1$ dimensional isotropically
distributed random unit vector. Hence
\begin{align}
h\brac{\rline{\xi_{21}\overline{q}_{1}^{\brac T}+\xi_{22}\overline{q}_{2}^{\brac T}}\xi_{11}\overline{q}_{1}^{\brac T}} & =h\brac{\rline{\sbrac{\xi_{21},\xi_{22}\overline{q}_{2}^{\brac{T-1}}}}\xi_{11},\overline{q}_{1}^{\brac T}}\\
 & =h\brac{\rline{\sbrac{\xi_{21},\xi_{22}\overline{q}_{2}^{\brac{T-1}}}}\xi_{11}}\label{eq:2x2LQsimplification_end}
\end{align}
and
\begin{align}
h\brac Y & =h\brac{\xi_{11}\overline{q}_{1}^{\brac T}}+h\brac{\rline{\sbrac{\xi_{21},\xi_{22}\overline{q}_{2}^{\brac{T-1}}}}\xi_{11}}\\
 & \overset{\brac i}{=}h\brac{\xi_{11}\overline{q}_{1}^{\brac T}}+h\brac{\rline{\sbrac{\xi_{21},\xi_{22}\overline{q}_{2}^{\brac{T-1}}}}\abs{\xi_{11}}^{2}}\\
 & \leq h\brac{\xi_{11}\overline{q}_{1}^{\brac T}}+h\brac{\rline{\xi_{22}\overline{q}_{2}^{\brac{T-1}}}\abs{\xi_{11}}^{2}}+h\brac{\rline{\xi_{21}}\abs{\xi_{11}}^{2}},
\end{align}
where $\brac i$ is because $\xi_{11}$ is non-negative. Note that
the above equation contains $\xi_{11},\xi_{22},\xi_{21}$ which we
would like to convert to the form $\abs{\xi_{11}}^{2},\abs{\xi_{22}}^{2},\abs{\xi_{21}}^{2}$
which are available from (\ref{eq:xi_11_defn}),  (\ref{eq:xi_21_defn})
and   (\ref{eq:xi_22_defn}). We handle $h\brac{\rline{\xi_{21}}\abs{\xi_{11}}^{2}}$
with the following lemma.
\begin{lem}
$h\brac{\rline{\xi_{21}}\abs{\xi_{11}}^{2}}\leq h\brac{\rline{\abs{\xi_{21}}^{2}}\abs{\xi_{11}}^{2}}+\lgbrac{\pi}$.
\label{lem:xi_to_xi_squared}
\end{lem}
\begin{IEEEproof}
We have
\begin{align}
h\brac{\rline{\xi_{21}}\abs{\xi_{11}}^{2}} & \overset{\brac i}{=}h\brac{\rline{\xi_{21}e^{i\theta}}\abs{\xi_{11}}^{2},\theta}\\
 & \overset{\brac{ii}}{\leq}h\brac{\rline{\xi_{21}e^{i\theta}}\abs{\xi_{11}}^{2}}\\
 & \overset{\brac{iii}}{=}h\brac{\rline{\abs{\xi_{21}}^{2}}\abs{\xi_{11}}^{2}}+\lgbrac{\pi},
\end{align}
where $\brac i$ uses $\theta\sim\text{Unif}\sbrac{0,2\pi}$ independent
of the other random variables, $\brac{ii}$ is because conditioning
reduces entropy, $\brac{iii}$ is using Lemma \ref{lem:isotropic_entropy_to_radial}
since given $\abs{\xi_{11}}^{2}$, $\xi_{21}e^{i\theta}$ is isotropically
distributed.
\end{IEEEproof}
Using the above lemma, we get
\begin{align}
h\brac Y & \overset{}{\leq}h\brac{\xi_{11}\overline{q_{1}}^{\brac T}}+h\brac{\rline{\xi_{22}\overline{q}_{2}^{\brac{T-1}}}\abs{\xi_{11}}^{2}}+h\brac{\rline{\abs{\xi_{21}}^{2}}\abs{\xi_{11}}^{2}}+\lgbrac{\pi}\\
 & \overset{\brac i}{\leq}h\brac{\abs{ag_{11}+bg_{12}+w_{11}}^{2}+\abs{cg_{12}+w_{12}}^{2}+\sum_{i=3}^{T}\abs{w_{1i}}^{2}}\nonumber \\
 & \qquad+\brac{T-1}\expect{\lgbrac{\abs a^{2}\rho_{11}^{2}+\brac{\abs b^{2}+\abs c^{2}}\rho_{12}^{2}+1}}+\lgbrac{\frac{\pi^{T}}{\Gamma\brac T}}\nonumber \\
 & \qquad+h\brac{\rline{\abs{\xi_{21}}^{2}}\abs{\xi_{11}}^{2}}+h\brac{\rline{\abs{\xi_{22}}^{2}}\abs{\xi_{11}}^{2}}+\brac{T-2}\expect{\lgbrac{\abs{\xi_{22}}^{2}}}+\lgbrac{\pi},
\end{align}
where $\brac i$ is by applying Lemma \ref{lem:isotropic_entropy_to_radial}
on $h\brac{\xi_{11}\overline{q_{1}}^{\brac T}}$ and Corollary \ref{cor:isotropic_entropy_to_radial_with_conditioning}
on $h\brac{\rline{\xi_{22}\overline{q}_{2}^{\brac{T-1}}}\abs{\xi_{11}}^{2}}.$
Now we use the following Lemma to simplify $h\brac{\abs{ag_{11}+bg_{12}+w_{11}}^{2}+\abs{cg_{12}+w_{12}}^{2}+\sum_{i=3}^{T}\abs{w_{1i}}^{2}}$
from the previous expression.
\begin{lem}
For any given distribution on $\brac{a,b,c}$, the terms
\[
h\brac{\abs{ag_{11}+bg_{12}+w_{11}}^{2}+\abs{cg_{12}+w_{12}}^{2}+\sum_{i=3}^{T}\abs{w_{1i}}^{2}}
\]
 and
\[
\expect{\lgbrac{\abs a^{2}\rho_{11}^{2}+\brac{\abs b^{2}+\abs c^{2}}\rho_{12}^{2}+1}}
\]
have the same gDoF. Similarly for any given distribution on $\brac{a,b,c}$,
the terms
\[
h\brac{\abs{ag_{21}+bg_{22}+w_{21}}^{2}+\abs{cg_{22}+w_{22}}^{2}+\sum_{i=3}^{T}\abs{w_{2i}}^{2}}
\]
 and
\[
\expect{\lgbrac{\abs a^{2}\rho_{21}^{2}+\brac{\abs b^{2}+\abs c^{2}}\rho_{22}^{2}+1}}
\]
 have the same gDoF. \label{lem:dof_equivalence_abs_lin_comb_gaussian_vector}
\end{lem}
\begin{IEEEproof}
The proof proceeds by constructing a noncoherent channel
\begin{equation}
\mathcal{C}_{1}:V=\abs{ag_{11}+bg_{12}+w_{11}}^{2}+\abs{cg_{12}+w_{12}}^{2}+\sum_{i=3}^{T}\abs{w_{1i}}^{2}
\end{equation}
with inputs $a,b,c$ and output $V$. Then we show that this channel
has zero gDoF. The proof uses outer bounding techniques from \cite{lapidoth2003capacity}.
See \ifarxiv  Appendix \ref{app:dof_equivalence_squared_norm_gaussiann_vector} \else  \cite[Appendix I]{Joyson_2x2_mimov5} \fi
for details.
\end{IEEEproof}

Hence using the previous lemma, we get
\begin{align}
h\brac Y & \leqdof T\expect{\lgbrac{\abs a^{2}\rho_{11}^{2}+\brac{\abs b^{2}+\abs c^{2}}\rho_{12}^{2}+1}}\nonumber \\
 & \qquad+h\brac{\rline{\abs{\xi_{21}}^{2}\abs{\xi_{11}}^{2}\vphantom{a^{a^{a^{a^{a^{a}}}}}}}\abs{\xi_{11}}^{2}}+h\brac{\rline{\abs{\xi_{22}}^{2}\abs{\xi_{11}}^{2}\vphantom{a^{a^{a^{a^{a^{a}}}}}}}\abs{\xi_{11}}^{2}}\nonumber \\
 & \qquad+\brac{T-2}\expect{\lgbrac{\abs{\xi_{22}}^{2}\abs{\xi_{11}}^{2}}}-T\expect{\lgbrac{\abs{\xi_{11}}^{2}}}.
\end{align}
Now we simplify $\expect{\lgbrac{\abs{\xi_{11}}^{2}}}$ from the previous
expression.
\begin{align}
\expect{\lgbrac{\abs{\xi_{11}}^{2}}} & =\expect{\lgbrac{\abs{ag_{11}+bg_{12}+w_{11}}^{2}+\abs{cg_{12}+w_{12}}^{2}+\sum_{i=3}^{T}\abs{w_{1i}}^{2}}}\nonumber \\
 & =\expect{\expect{\rline{\lgbrac{\abs{ag_{11}+bg_{12}+w_{11}}^{2}+\abs{cg_{12}+w_{12}}^{2}+\sum_{i=3}^{T}\abs{w_{1i}}^{2}}}a,b,c}}\nonumber \\
 & \overset{\brac i}{\eqdof}\expect{\lgbrac{\abs a^{2}\rho_{11}^{2}+\brac{\abs b^{2}+\abs c^{2}}\rho_{12}^{2}+1}},
\end{align}
where $\brac i$ was using Lemma \ref{fact:Jensens_gap} on page \pageref{fact:Jensens_gap}
and using the fact that $\abs{ag_{11}+bg_{12}+w_{11}}^{2},\abs{cg_{12}+w_{12}}^{2},\abs{w_{1i}}^{2}$
are exponentially distributed given $a,b,c$. Hence
\begin{align}
h\brac Y & \leqdof h\brac{\rline{\abs{\xi_{21}}^{2}\abs{\xi_{11}}^{2}\vphantom{a^{a^{a^{a}}}}}\abs{\xi_{11}}^{2}}+h\brac{\rline{\abs{\xi_{22}}^{2}\abs{\xi_{11}}^{2}\vphantom{a^{a^{a^{a}}}}}\abs{\xi_{11}}^{2}}\nonumber \\
 & \qquad+\brac{T-2}\expect{\lgbrac{\abs{\xi_{22}}^{2}\abs{\xi_{11}}^{2}}}.\label{eq:2x2MIMO_hy}
\end{align}
Now we use the following lemmas to further simplify the terms in the
above expression for $h\brac Y$.
\begin{lem}
\label{lem:dof_equivalence_h(xi22)}For any given distribution on
$\brac{a,b,c}$,
\[
h\brac{\rline{\abs{\xi_{22}}^{2}\abs{\xi_{11}}^{2}\vphantom{a^{a^{a^{a}}}}}\abs{\xi_{11}}^{2}}\eqdof h\brac{\rline{\abs{\xi_{22}}^{2}\abs{\xi_{11}}^{2}\vphantom{a^{a^{a^{a}}}}}\abs{\xi_{11}}^{2},a,b,c}\leq\expect{\lgbrac{e\expect{\rline{\abs{\xi_{22}}^{2}\abs{\xi_{11}}^{2}\vphantom{a^{a^{a^{a}}}}}a,b,c}}}.
\]
\end{lem}
\begin{IEEEproof}
The proof uses similar techniques like that for Lemma \ref{lem:dof_equivalence_abs_lin_comb_gaussian_vector}.
See \ifarxiv  Appendix \ref{app:Proof-of-Lemma_dof_equivalence_h(xi22)} \else  \cite[Appendix J]{Joyson_2x2_mimov5} \fi
for details.
\end{IEEEproof}
\begin{lem}
\label{lem:dof_equivalence_h(xi21)}For any given distribution on
$\brac{a,b,c}$,
\[
h\brac{\rline{\abs{\xi_{21}}^{2}\abs{\xi_{11}}^{2}\vphantom{a^{a^{a^{a}}}}}\abs{\xi_{11}}^{2}}\eqdof h\brac{\rline{\abs{\xi_{21}}^{2}\abs{\xi_{11}}^{2}\vphantom{a^{a^{a^{a}}}}}\abs{\xi_{11}}^{2},a,b,c}\leq\expect{\lgbrac{e\expect{\rline{\abs{\xi_{21}}^{2}\abs{\xi_{11}}^{2}\vphantom{\vphantom{a^{a^{a^{a}}}}}}a,b,c}}}.
\]
\end{lem}
\begin{IEEEproof}
This can be proved similar to the previous lemma. We omit the proof.
\end{IEEEproof}
We have
\begin{align*}
\abs{\xi_{21}}^{2}\abs{\xi_{11}}^{2} & =\abs{\brac{ag_{21}+bg_{22}+w_{21}}\brac{ag_{11}+bg_{12}+w_{11}}^{*}+\brac{cg_{22}+w_{22}}\brac{cg_{12}+w_{12}}^{*}+\sum_{i=3}^{T}w_{2i}w_{1i}^{*}}^{2}.
\end{align*}
Hence using  (\ref{eq:xi_11_defn}), (\ref{eq:xi_21_defn}) and Lemma
\ref{lem:dof_equivalence_h(xi21)} to bound $h\brac{\rline{\abs{\xi_{21}}^{2}\abs{\xi_{11}}^{2}\vphantom{a^{a^{a^{a}}}}}\abs{\xi_{11}}^{2}}$,
we get
\begin{align}
 & h\brac{\rline{\abs{\xi_{21}}^{2}\abs{\xi_{11}}^{2}\vphantom{a^{a^{a^{a}}}}}\abs{\xi_{11}}^{2}}\nonumber \\
 & \leqdof\mathbb{E}\left[\log\left(\brac{\abs a^{2}\rho_{11}^{2}+\abs b^{2}\rho_{12}^{2}+1}\brac{\abs a^{2}\rho_{21}^{2}+\abs b^{2}\rho_{22}^{2}+1}\vphantom{a^{a^{a^{a}}}}\right.\right.\nonumber \\
 & \qquad\qquad\left.\left.+2\abs c^{2}\abs b^{2}\rho_{22}^{2}\rho_{12}^{2}+\brac{\abs c^{2}\rho_{12}^{2}+1}\brac{\abs c^{2}\rho_{22}^{2}+1}+T-2\vphantom{a^{a^{a^{a}}}}\right)\right]\\
 & \leqdof\mathbb{E}\left[\log\left(\brac{\abs a^{2}\rho_{11}^{2}+\abs b^{2}\rho_{12}^{2}+1}\brac{\abs a^{2}\rho_{21}^{2}+\abs b^{2}\rho_{22}^{2}+1}\vphantom{a^{a^{a^{a}}}}\right.\right.\\
 & \qquad\qquad\left.\left.+\brac{\abs c^{2}\rho_{12}^{2}+1}\brac{\abs c^{2}\rho_{22}^{2}+1}\vphantom{a^{a^{a^{a}}}}\right)\right],\label{eq:2x2MIMO_hy_part1}
\end{align}
where the last step followed due to AM-GM inequality (arithmetic mean
$\leq$ geometric mean). The AM-GM inequality yields $2\abs c^{2}\abs b^{2}\rho_{22}^{2}\rho_{12}^{2}\leq\abs b^{4}\rho_{22}^{2}\rho_{12}^{2}+\abs c^{4}\rho_{22}^{2}\rho_{12}^{2}$.
Similarly, using  (\ref{eq:xi_11_defn}) and (\ref{eq:xi_22_defn}),
we have{\small{}
\begin{align*}
 & \abs{\xi_{22}}^{2}\abs{\xi_{11}}^{2}\\
 & =\brac{\abs{ag_{21}+bg_{22}+w_{21}}^{2}+\abs{cg_{22}+w_{22}}^{2}+\sum_{i=3}^{T}\abs{w_{2i}}^{2}}\brac{\abs{ag_{11}+bg_{12}+w_{11}}^{2}+\abs{cg_{12}+w_{12}}^{2}+\sum_{i=3}^{T}\abs{w_{1i}}^{2}}\\
 & \qquad-\abs{\brac{ag_{21}+bg_{22}+w_{21}}\brac{ag_{11}+bg_{12}+w_{11}}^{*}+\brac{cg_{22}+w_{22}}\brac{cg_{12}+w_{12}}^{*}+\sum_{i=3}^{T}w_{2i}w_{1i}^{*}}^{2}.
\end{align*}
}After some algebraic manipulations, it can be seen that%
\begin{align}
 & \expect{\rline{\abs{\xi_{22}}^{2}\abs{\xi_{11}}^{2}\vphantom{a^{a^{a^{a}}}}}a,b,c}\nonumber \\
 & =\brac{T-2}^{2}-\brac{T-2}+\brac{T-2}\brac{\abs a^{2}\rho_{21}^{2}+\abs b^{2}\rho_{22}^{2}+\abs c^{2}\rho_{22}^{2}+\abs a^{2}\rho_{11}^{2}+\abs b^{2}\rho_{12}^{2}+\abs c^{2}\rho_{12}^{2}+2}\nonumber \\
 & \qquad+\brac{\abs a^{2}\rho_{11}^{2}+1}\brac{\abs c^{2}\rho_{22}^{2}+1}+\abs b^{2}\rho_{12}^{2}+\brac{\abs a^{2}\rho_{21}^{2}+1}\brac{\abs c^{2}\rho_{12}^{2}+1}+\abs b^{2}\rho_{22}^{2}.
\end{align}
After retaining only the terms that contribute to gDoF from the above
equation, we bound $h\brac{\rline{\abs{\xi_{22}}^{2}\abs{\xi_{11}}^{2}\vphantom{a^{a^{a^{a}}}}}\abs{\xi_{11}}^{2}}$
using Lemma \ref{lem:dof_equivalence_h(xi21)} to get
\begin{align}
h\brac{\rline{\abs{\xi_{22}}^{2}\abs{\xi_{11}}^{2}\vphantom{a^{a^{a^{a}}}}}\abs{\xi_{11}}^{2}} & \leqdof\expect{\lgbrac{e\expect{\rline{\abs{\xi_{22}}^{2}\abs{\xi_{11}}^{2}\vphantom{a^{a^{a^{a}}}}}a,b,c}}}\nonumber \\
 & \leqdof\mathbb{E}\left[\log\left(\brac{\abs a^{2}\rho_{11}^{2}+1}\brac{\abs c^{2}\rho_{22}^{2}+1}+\abs b^{2}\brac{\rho_{12}^{2}+\rho_{22}^{2}}\vphantom{a^{a^{a^{a}}}}\right.\right.\nonumber \\
 & \qquad\qquad\left.\left.+\brac{\abs a^{2}\rho_{21}^{2}+1}\brac{\abs c^{2}\rho_{12}^{2}+1}\vphantom{a^{a^{a^{a}}}}\right)\right].\label{eq:2x2MIMO_hy_part2}
\end{align}
Also
\begin{align}
\expect{\lgbrac{\abs{\xi_{22}}^{2}\abs{\xi_{11}}^{2}}} & \leq\expect{\lgbrac{\expect{\rline{\abs{\xi_{22}}^{2}\abs{\xi_{11}}^{2}\vphantom{a^{a^{a^{a}}}}}a,b,c}}}\nonumber \\
 & \leqdof\mathbb{E}\left[\log\left(\brac{\abs a^{2}\rho_{11}^{2}+1}\brac{\abs c^{2}\rho_{22}^{2}+1}+\abs b^{2}\brac{\rho_{12}^{2}+\rho_{22}^{2}}\vphantom{a^{a^{a^{a}}}}\right.\right.\nonumber \\
 & \qquad\qquad\left.\left.+\brac{\abs a^{2}\rho_{21}^{2}+1}\brac{\abs c^{2}\rho_{12}^{2}+1}\vphantom{a^{a^{a^{a}}}}\right)\right].\label{eq:2x2MIMO_hy_par3}
\end{align}
 Hence using  (\ref{eq:2x2MIMO_hy_par3}), (\ref{eq:2x2MIMO_hy_part2}),
(\ref{eq:2x2MIMO_hy_part1}) in (\ref{eq:2x2MIMO_hy}), we get
\begin{align}
 & h\brac Y\nonumber \\
 & \leqdof\expect{\lgbrac{\brac{\abs a^{2}\rho_{11}^{2}+\abs b^{2}\rho_{12}^{2}+1}\brac{\abs a^{2}\rho_{21}^{2}+\abs b^{2}\rho_{22}^{2}+1}+\brac{\abs c^{2}\rho_{12}^{2}+1}\brac{\abs c^{2}\rho_{22}^{2}+1}}}\nonumber \\
 & \qquad+\brac{T-1}\mathbb{E}\left[\log\left(\brac{\abs a^{2}\rho_{11}^{2}+1}\brac{\abs c^{2}\rho_{22}^{2}+1}+\abs b^{2}\brac{\rho_{12}^{2}+\rho_{22}^{2}}\vphantom{a^{a^{a^{a}}}}\right.\right.\nonumber \\
 & \qquad\qquad\qquad\qquad\qquad\left.\left.+\brac{\abs a^{2}\rho_{21}^{2}+1}\brac{\abs c^{2}\rho_{12}^{2}+1}\vphantom{a^{a^{a^{a}}}}\right)\right].
\end{align}
Using the above equation and  (\ref{eq:2x2MIMO_hy_x}), we get
\begin{align}
 & I\brac{X;Y}\nonumber \\
 & \leqdof\expect{\lgbrac{\brac{\abs a^{2}\rho_{11}^{2}+\abs b^{2}\rho_{12}^{2}+1}\brac{\abs a^{2}\rho_{21}^{2}+\abs b^{2}\rho_{22}^{2}+1}+\brac{\abs c^{2}\rho_{12}^{2}+1}\brac{\abs c^{2}\rho_{22}^{2}+1}}}\nonumber \\
 & \qquad+\brac{T-1}\expect{\lgbrac{\brac{\abs a^{2}\rho_{11}^{2}+1}\brac{\abs c^{2}\rho_{22}^{2}+1}+\abs b^{2}\brac{\rho_{12}^{2}+\rho_{22}^{2}}+\brac{\abs a^{2}\rho_{21}^{2}+1}\brac{\abs c^{2}\rho_{12}^{2}+1}}}\nonumber \\
 & \qquad-\expect{\lgbrac{\abs a^{2}\rho_{11}^{2}+\abs b^{2}\rho_{12}^{2}+\abs c^{2}\rho_{12}^{2}+\abs a^{2}\abs c^{2}\rho_{11}^{2}\rho_{12}^{2}+1}}\nonumber \\
 & \qquad-\expect{\lgbrac{\abs a^{2}\rho_{21}^{2}+\abs b^{2}\rho_{22}^{2}+\abs c^{2}\rho_{22}^{2}+\abs a^{2}\abs c^{2}\rho_{21}^{2}\rho_{22}^{2}+1}}.\\
 & =\expect{f\brac{\abs a^{2},\abs b^{2},\abs c^{2}}},
\end{align}
where the last step included a trivial definition for $f\brac{\abs a^{2},\abs b^{2},\abs c^{2}}$,
by collecting all the terms from the previous equation. Hence an outer
bound for the gDoF of the channel can be obtained by solving the following
optimization problem:
\begin{equation}
\mathcal{P}_{1}:\begin{cases}
\underset{\expect{\abs a^{2}+\abs b^{2}+\abs c^{2}}\leq T}{\text{maximize}}\expect{f\brac{\abs a^{2},\abs b^{2},\abs c^{2}}}.\end{cases}\label{eq:opt_prob_P1_defn}
\end{equation}
Now we use the following lemma to simplify $\mathcal{P}_{1}$ without
losing gDoF.
\begin{lem}
The gDoF achieved in $\mathcal{P}_{1}$ can be achieved by a point
mass distribution, i.e.,
\[
\text{gDoF}\brac{\mathcal{P}_{1}}=\text{gDoF}\brac{\mathcal{P}_{7}},
\]
where $\mathcal{P}_{7}$ is the following: {\normalfont
\begin{equation}
\mathcal{P}_{7}:\begin{cases}
\underset{}{\text{maximize}}\ f\brac{\abs a^{2},\abs b^{2},\abs c^{2}} & \text{ with}\\
\abs a^{2}\leq T,\abs b^{2}\leq T,\abs c^{2}\leq T.
\end{cases}
\end{equation}
} \label{lem:discretization}
\end{lem}
\begin{IEEEproof}
[Proof idea]The proof proceeds in several steps:

Step 1: Show that there exists a discretization (over an infinite
set) for any distribution of $\brac{\abs a^{2},\abs b^{2},\abs c^{2}}$
that does not incur a loss in gDoF.

Step 2: Show that the discretization can be limited to a finite set
without incurring a loss in gDoF.

Step 3: View the problem as a linear program with two constraints,
and show that there is an optimal distribution with just two mass
points.

Step 4: Show that the two mass points can be collapsed to a single
point using arguments of symmetry.

The details of the proof are given in Appendix \ref{app:proof_discretization}.
\end{IEEEproof}
Changing the variables from $\brac{\abs a^{2},\abs b^{2},\abs c^{2}}$
to $\brac{\gamma_{a},\gamma_{b},\gamma_{c}}$ with the substitution
$\abs a^{2}=\snr^{-\gamma_{a}},$ $\abs b^{2}=\snr^{-\gamma_{b}},$
$\abs c^{2}=\snr^{-\gamma_{c}}$, it is clear that
\[
\text{gDoF}\brac{\mathcal{P}_{1}}=\text{gDoF}\brac{\mathcal{P}_{7}}=\brac{\mathcal{P}_{8}},
\]
where $\mathcal{P}_{8}$ is the following:
\begin{equation}
\mathcal{P}_{8}:\begin{cases}
\underset{}{\text{maximize}}\ f_{\gamma}\brac{\gamma_{a},\gamma_{b},\gamma_{c}} & \text{}\\
\gamma_{a}\geq0,\gamma_{b}\geq0,\gamma_{c}\geq0,
\end{cases}\label{eq:2x2_outer_gdof_opt_problem}
\end{equation}
with
\begin{align}
 & f_{\gamma}\brac{\gamma_{a},\gamma_{b},\gamma_{c}}\nonumber \\
 & \begin{aligned}=\max\left(\vphantom{a^{a^{a^{a^{a}}}}}\right. & \max\brac{-\gamma_{a}+\gamma_{11},-\gamma_{b}+\gamma_{12},0}+\max\brac{-\gamma_{a}+\gamma_{21},-\gamma_{b}+\gamma_{22},0},\\
 & \left.\vphantom{a^{a^{a^{a^{a}}}}}\max\brac{-\gamma_{c}+\gamma_{12},0}+\max\brac{-\gamma_{c}+\gamma_{22},0}\right)
\end{aligned}
\nonumber \\
 & \begin{alignedat}{1}\quad+\brac{T-1}\max\left(\vphantom{a^{a^{a^{a^{a}}}}}\right. & \max\brac{-\gamma_{a}+\gamma_{11},0}+\max\brac{-\gamma_{c}+\gamma_{22},0},\\
 & \left.\vphantom{a^{a^{a^{a^{a}}}}}\gamma_{b}+\max\brac{\gamma_{12},\gamma_{22}},\max\brac{-\gamma_{a}+\gamma_{21},0}+\max\brac{-\gamma_{c}+\gamma_{12},0}\right)
\end{alignedat}
\nonumber \\
 & \quad-\max\brac{-\gamma_{a}+\gamma_{11},-\gamma_{b}+\gamma_{12},-\gamma_{c}+\gamma_{12},-\gamma_{a}-\gamma_{c}+\gamma_{11}+\gamma_{12},0}\nonumber \\
 & \quad-\max\brac{-\gamma_{a}+\gamma_{21},-\gamma_{b}+\gamma_{22},-\gamma_{c}+\gamma_{22},-\gamma_{a}-\gamma_{c}+\gamma_{21}+\gamma_{22},0}.\label{eq:2x2_piecewiselinear_outerbound}
\end{align}
For a $2\times2$ MIMO channel with two different $\snr$ exponents,
one in the direct links and another in the crosslinks, we have $\gamma_{11}=\gamma_{22}=\gamma_{D},\ \gamma_{CL}=\gamma_{12}=\gamma_{21}$.
Also, without loss of generality, it can be assumed that $\gamma_{D}>\gamma_{CL}$.
By inspection of the optimization problem, it is clear that we can
also restrict $\gamma_{a}\leq\gamma_{D},\gamma_{b}\leq\gamma_{D},\gamma_{c}\leq\gamma_{D}$
without affecting the solution. With these additional constraints,
we can simplify $\mathcal{P}_{8}$ to $\mathcal{P}_{9}$ with $\mathcal{P}_{9}$
defined as the following:
\begin{equation}
\mathcal{P}_{9}:\begin{cases}
\underset{}{\text{maximize}}\ \max\brac{-2\gamma_{a}+\gamma_{CL},-\gamma_{a}+\gamma_{D}-\gamma_{b},-2\gamma_{b}+\gamma_{CL},-2\gamma_{c}+\gamma_{CL},-\gamma_{c}}\\
\qquad\qquad\qquad\qquad+\brac{T-1}\max\brac{-\gamma_{a}+\gamma_{D}-\gamma_{c},-\gamma_{b}}+T\gamma_{D}-t_{1}-t_{2} & \text{}\\
t_{1}=\max\brac{-\gamma_{a}+\gamma_{D},-\gamma_{b}+\gamma_{CL},-\gamma_{a}-\gamma_{c}+\gamma_{D}+\gamma_{CL}}\\
t_{2}=\max\brac{-\gamma_{b}+\gamma_{D},-\gamma_{c}+\gamma_{D},-\gamma_{a}-\gamma_{c}+\gamma_{D}+\gamma_{CL}}\\
0\leq\gamma_{a}\leq\gamma_{D},0\leq\gamma_{b}\leq\gamma_{D},0\leq\gamma_{c}\leq\gamma_{D}.
\end{cases}
\end{equation}
Using standard linear programming arguments, $\mathcal{P}_{9}$ has
a solution for $\brac{\gamma_{a},\gamma_{b},\gamma_{c},t_{1},t_{2}}$
in one of the corner points of the following region:
\[
\mathcal{R}:\cbrac{\begin{alignedat}{1}0\leq\gamma_{a}\leq\gamma_{D}; & \ 0\leq\gamma_{b}\leq\gamma_{D}\\
0\leq\gamma_{c}\leq\gamma_{D}; & \ \\
t_{1}\geq-\gamma_{a}+\gamma_{D}; & \ t_{1}\geq-\gamma_{b}+\gamma_{CL}\\
t_{1}\geq-\gamma_{a}-\gamma_{c}+\gamma_{D}+\gamma_{CL}; & \ t_{2}\geq-\gamma_{b}+\gamma_{D}\\
t_{2}\geq-\gamma_{a}-\gamma_{c}+\gamma_{D}+\gamma_{CL}; & \ t_{2}\geq-\gamma_{c}+\gamma_{D}
\end{alignedat}
}.
\]
This can be seen by considering case by case for $\mathcal{P}_{9}$,
depending on which term inside the $\max\brac{\cdot}$'s could come
out in the objective function, and noting that $\underset{\gamma_{a},\gamma_{b},\gamma_{c},t_{1},t_{2}}{\text{maximize}}\ \max\brac{f_{1},f_{2}}$
is same as $\max\brac{\underset{\gamma_{a},\gamma_{b},\gamma_{c},t_{1},t_{2}}{\text{maximize}}\brac{f_{1}},\underset{\gamma_{a},\gamma_{b},\gamma_{c},t_{1},t_{2}}{\text{maximize}}\brac{f_{2}}}$
for linear $f_{1},f_{2}$.

Suppose $-2\gamma_{a}+\gamma_{CL}=\max\brac{-2\gamma_{a}+\gamma_{CL},-\gamma_{a}+\gamma_{D}-\gamma_{b},-2\gamma_{b}+\gamma_{CL},-2\gamma_{c}+\gamma_{CL},-\gamma_{c}}$
and $-\gamma_{a}+\gamma_{D}-\gamma_{c}=\max\brac{-\gamma_{a}+\gamma_{D}-\gamma_{c},-\gamma_{b}}$,
then $\mathcal{P}_{9}$ has a solution in one of the corner points
of $\mathcal{R}$. This is true for all possible cases of the values
of the two $\max\brac{}$'s. Hence $\mathcal{P}_{9}$ itself has a
solution in one of the corner points of $\mathcal{R}$.

We code in Mathematica to find all the corner points of $\mathcal{R}$
and find the maximum across the corner points. Finding all the corner
points and the subsequent calculations are mechanical, and we believe,
does not add value to the paper and would only further lengthen the
paper. So we have deferred it to the software. The result is still
mathematically rigorous, and we suggest the use of Mathematica to
verify some formulas rather than perform long calculations. We obtain
the solution in Table \ref{tab:Sol_outerbd_2x2_sym_mimo}. Our Mathematica
code is available online at \url{https://arxiv.org/src/1705.07355v5/anc/Sym_mimo_outerbound.nb}.
This code uses $\gamma_{D}=1,\gamma_{CL}=1-\epsilon$ and we can obtain
the general solution with a simple scaling.

\section{Conclusions\label{sec:Conclusion}}

We considered the noncoherent MIMO channel with link strengths scaled
with different exponents of $\snr$. Under this model, we derived
a structure for the capacity achieving input distribution. We showed
that for $T=1$, the gDoF is zero for a MIMO channel of any size.
Also for SIMO channels and MISO channels, we proved that selecting
the best antenna can achieve the gDoF. We derived the gDoF for the
$2\times2$  MIMO channel with two different exponents in the direct
and cross links and showed that both the antennas are always needed
to achieve the gDoF. Also, training-based schemes were shown to be
suboptimal for this $2\times2$ MIMO channel with two different exponents.
We extended this observation to an $M\times M$  MIMO channel with
two different exponents in the direct and cross links; we demonstrated
a strategy that could achieve larger gDoF than training-based schemes.
A possible direction for future work would be to try to derive the
gDoF of $M\times M$ MIMO channels with two different exponents in
the direct and cross links. A subsequent step would be to look into
the case of MIMO channels with arbitrary size and arbitrary SNR exponents.
The outer bounds for larger MIMO channels seem to be a challenge at
the moment, our outer bounds for the $2\times2$ MIMO channel illustrate
some of the difficulties: we used a Gram-Schmidt process for the LQ
decomposition of matrices and developed new lemmas to bound the terms
in the mutual information expression. The same methods do not seem
to be directly applicable to larger MIMO systems. Another line of
work is to study the gDoF for noncoherent relay channels.

\appendices{}

\section{Proof of Lemma \ref{lem:isotropic_entropy_to_radial}\label{app:proof_lemma_isotropic_entropy_to_radial}}

Here we derive the formula for calculating $h\brac{\sbrac{\xi_{1},\xi_{2},\ldots,\xi_{n}}Q}$
with $\sbrac{\xi_{1},\xi_{2},\ldots,\xi_{n}}$ being an arbitrary
complex random vector and $Q$ being an $n\times n$ isotropically
distributed random unitary matrix independent of $\xi_{i}$. We do
this by noting that in radial coordinates, the distribution of $\sbrac{\xi_{1},\xi_{2},\ldots,\xi_{n}}Q$
is dependent only on the radius. Let
\[
V=\sbrac{\xi_{1},\xi_{2},\ldots,\xi_{n}}Q.
\]
Now for any fixed $n\times n$ unitary matrix $Q'$, the vectors $V$
and $VQ'$ have the same distribution due to the property of isotropic
distribution. Hence for any $v_{1},v_{2}\in\mathbb{C}^{n}$, if $\twonorm{v_{1}}=\twonorm{v_{2}}$,
then
\begin{equation}
p_{v}\brac{v_{1}}=p_{v}\brac{v_{2}},
\end{equation}
since there exists a unitary matrix $Q''$ such that $v_{1}Q''=v_{2}$.
One such $Q''$ can be obtained using Householder transformation.
Now the probability distribution can be viewed in $\mathbb{R}^{2n}$,
and we use the $2n$ dimensional vector $U$. Let
\begin{equation}
\varUpsilon=\sum\abs{\xi_{i}}^{2}.
\end{equation}
Let $\brac{r,\overline{\theta}}$ be the radial coordinates, $\brac{t,\overline{\theta}}$
be similar coordinates but with $t=r^{2}$. Let $p_{u,t}\brac{t,\overline{\theta}}=p_{u}\brac{u\brac{t,\overline{\theta}}}$
be obtained from $p_{u}\brac u$ by expressing $u$ in $\brac{t,\overline{\theta}}$
coordinates. Similarly $p_{u,r}\brac{r,\overline{\theta}}=p_{u}\brac{u\brac{r,\overline{\theta}}}$.

The $2n-1$ dimensional surface area (embedded in a $2n$ dimensional
Euclidean) is $\brac{\frac{2\pi^{n}}{\Gamma\brac n}}r^{2n-1}$. Hence
\[
\brac{\frac{2\pi^{n}}{\Gamma\brac n}}p_{u,r}\brac{r,\overline{\theta}}r^{2n-1}dr
\]
 is the probability that $\abs U\in\sbrac{r,r+dr}$. Hence $\brac{\frac{\pi^{n}}{\Gamma\brac n}}p_{u,t}\brac{t,\overline{\theta}}t^{n-1}dt$
is the probability that $\varUpsilon=\twonorm U^{2}\in\sbrac{t,t+dt}$.
Hence
\begin{align}
\brac{\frac{\pi^{n}}{\Gamma\brac n}}p_{u,t}\brac{t,\overline{\theta}}t^{n-1} & =p_{\varUpsilon}\brac t\\
p_{u,t}\brac{t,\overline{\theta}} & =p_{\varUpsilon}\brac t\frac{1}{t^{n-1}\brac{\frac{\pi^{n}}{\Gamma\brac n}}}.\label{eq:pdf_sumsquared}
\end{align}
Now
\begin{align}
h\brac U & =-\int p_{u}\brac u\lgbrac{p_{u}\brac u}du\\
 & \overset{\brac i}{=}-\int p_{u}\brac{u\brac{r,\overline{\theta}}}\lgbrac{p_{u}\brac{u\brac{r,\overline{\theta}}}}\brac{\frac{2\pi^{n}}{\Gamma\brac n}}r^{2n-1}dr\\
 & \overset{\brac{ii}}{=}-\int p_{u}\brac{u\brac{t,\overline{\theta}}}\lgbrac{p_{u}\brac{u\brac{t,\overline{\theta}}}}\brac{\frac{\pi^{n}}{\Gamma\brac n}}t^{n-1}dt\\
 & \overset{\brac{iii}}{=}-\int p_{\varUpsilon}\brac t\lgbrac{p_{\varUpsilon}\brac t\frac{1}{t^{n-1}\brac{\frac{\pi^{n}}{\Gamma\brac n}}}}dt\\
 & =-\int p_{\varUpsilon}\brac t\lgbrac{p_{\varUpsilon}\brac t}dt+\lgbrac{\frac{\pi^{n}}{\Gamma\brac n}}+\brac{n-1}\int p_{\varUpsilon}\brac t\lgbrac tdt\\
 & =h\brac{\varUpsilon}+\brac{n-1}\expect{\lgbrac{\varUpsilon}}+\lgbrac{\frac{\pi^{n}}{\Gamma\brac n}}\\
 & =h\brac{\sum\abs{\xi_{i}}^{2}}+\brac{n-1}\expect{\lgbrac{\sum\abs{\xi_{i}}^{2}}}+\lgbrac{\frac{\pi^{n}}{\Gamma\brac n}},
\end{align}
where $\brac i$ is by change of variables to $\brac{r,\overline{\theta}}$,
then integrating over $\overline{\theta}$ and noting that $p_{u}\brac{u\brac{r,\overline{\theta}}}$
is independent of $\overline{\theta}$. The step $\brac{ii}$ is by
change of variables to $\brac{t,\overline{\theta}}$, $\brac{iii}$
is using (\ref{eq:pdf_sumsquared}).

\section{Proof of Lemma \ref{lem:discretization}\label{app:proof_discretization}}

Here we consider the optimization problem $\mathcal{P}_{1}$ from
(\ref{eq:opt_prob_P1_defn}) on page \pageref{eq:opt_prob_P1_defn}
and show that its objective function $\expect{f\brac{\abs a^{2},\abs b^{2},\abs c^{2}}}$
can be optimized for gDoF by a point mass distribution. We have the
form for $f\brac{\abs a^{2},\abs b^{2},\abs c^{2}}$ as
\begin{align}
 & f\brac{\abs a^{2},\abs b^{2},\abs c^{2}}\nonumber \\
 & =\lgbrac{\brac{\abs a^{2}\rho_{11}^{2}+\abs b^{2}\rho_{12}^{2}+1}\brac{\abs a^{2}\rho_{21}^{2}+\abs b^{2}\rho_{22}^{2}+1}+\brac{\abs c^{2}\rho_{12}^{2}+1}\brac{\abs c^{2}\rho_{22}^{2}+1}}\nonumber \\
 & \qquad+\brac{T-1}\lgbrac{\brac{\abs a^{2}\rho_{11}^{2}+1}\brac{\abs c^{2}\rho_{22}^{2}+1}+\abs b^{2}\brac{\rho_{12}^{2}+\rho_{22}^{2}}+\brac{\abs a^{2}\rho_{21}^{2}+1}\brac{\abs c^{2}\rho_{12}^{2}+1}}\nonumber \\
 & \qquad-\lgbrac{\brac{1+\abs a^{2}\rho_{11}^{2}}\brac{1+\abs c^{2}\rho_{12}^{2}}+\abs b^{2}\rho_{12}^{2}}\nonumber \\
 & \qquad-\lgbrac{\brac{1+\abs a^{2}\rho_{21}^{2}}\brac{1+\abs c^{2}\rho_{22}^{2}}+\abs b^{2}\rho_{22}^{2}}.\label{eq:objective_fn_local}
\end{align}
Now
\begin{align}
 & \pderiv{}{\abs a^{2}}f\brac{\abs a^{2},\abs b^{2},\abs c^{2}}\nonumber \\
 & =\frac{\rho_{11}^{2}\brac{\abs a^{2}\rho_{21}^{2}+\abs b^{2}\rho_{22}^{2}+1}+\brac{\abs a^{2}\rho_{11}^{2}+\abs b^{2}\rho_{12}^{2}+1}\rho_{21}^{2}}{\brac{\abs a^{2}\rho_{11}^{2}+\abs b^{2}\rho_{12}^{2}+1}\brac{\abs a^{2}\rho_{21}^{2}+\abs b^{2}\rho_{22}^{2}+1}+\brac{\abs c^{2}\rho_{12}^{2}+1}\brac{\abs c^{2}\rho_{22}^{2}+1}}\nonumber \\
 & \qquad+\brac{T-1}\frac{\rho_{11}^{2}\brac{\abs c^{2}\rho_{22}^{2}+1}+\brac{\abs c^{2}\rho_{12}^{2}+1}\rho_{21}^{2}}{\brac{\abs a^{2}\rho_{11}^{2}+1}\brac{\abs c^{2}\rho_{22}^{2}+1}+\abs b^{2}\brac{\rho_{12}^{2}+\rho_{22}^{2}}+\brac{\abs a^{2}\rho_{21}^{2}+1}\brac{\abs c^{2}\rho_{12}^{2}+1}}\nonumber \\
 & \qquad-\frac{\brac{\rho_{11}^{2}}\brac{1+\abs c^{2}\rho_{12}^{2}}}{\brac{1+\abs a^{2}\rho_{11}^{2}}\brac{1+\abs c^{2}\rho_{12}^{2}}+\abs b^{2}\rho_{12}^{2}}\nonumber \\
 & \qquad-\frac{\brac{\rho_{21}^{2}}\brac{1+\abs c^{2}\rho_{22}^{2}}}{\brac{1+\abs a^{2}\rho_{21}^{2}}\brac{1+\abs c^{2}\rho_{22}^{2}}+\abs b^{2}\rho_{22}^{2}}.
\end{align}
Hence
\begin{align}
\abs{\pderiv{}{\abs a^{2}}f\brac{\abs a^{2},\abs b^{2},\abs c^{2}}} & \leq\rho_{11}^{2}+\rho_{21}^{2}+\brac{T-1}\brac{\rho_{11}^{2}+\rho_{21}^{2}}+\brac{\rho_{11}^{2}}+\brac{\rho_{21}^{2}}\\
 & \leq2\brac{T+1}\max_{i,j}\rho_{ij}^{2}.
\end{align}
Similarly
\begin{align}
\abs{\pderiv{}{\abs b^{2}}f\brac{\abs a^{2},\abs b^{2},\abs c^{2}}} & \leq2\brac{T+1}\max_{i,j}\rho_{ij}^{2},\\
\abs{\pderiv{}{\abs c^{2}}f\brac{\abs a^{2},\abs b^{2},\abs c^{2}}} & \leq2\brac{T+1}\max_{i,j}\rho_{ij}^{2}
\end{align}
holds. Let $\rho_{*}^{2}=\max_{i,j}\rho_{ij}^{2}$. Now with $\Delta=1/\brac{2\brac{T+1}\rho_{*}^{2}}$,
if $\left\Vert \brac{\abs a^{2},\abs b^{2},\abs c^{2}}-\brac{\abs{a'}^{2},\abs{b'}^{2},\abs{c'}^{2}}\vphantom{a^{a^{a^{a}}}}\right\Vert \leq\sqrt{3}\Delta$,
then
\begin{align}
 & \abs{f\brac{\abs a^{2},\abs b^{2},\abs c^{2}}-f\brac{\abs{a'}^{2},\abs{b'}^{2},\abs{c'}^{2}}}\\
 & \leq\left\Vert \sbrac{2\brac{T+1}\rho_{*}^{2},2\brac{T+1}\rho_{*}^{2},2\brac{T+1}\rho_{*}^{2}}\vphantom{a^{^{a^{a^{a}}}}}\right\Vert \sqrt{3}\Delta\\
 & \leq3.
\end{align}
Hence by considering a discrete version of the problem as
\begin{equation}
\mathcal{P}_{2}:\begin{cases}
\underset{\expect{\abs a^{2}+\abs b^{2}+\abs c^{2}}\leq T}{\text{maximize}}\expect{f\brac{\abs a^{2},\abs b^{2},\abs c^{2}}} & \text{}\\
\text{Support}\brac{\abs a^{2},\abs b^{2},\abs c^{2}}=\cbrac{0,\Delta,2\Delta,\ldots,\infty}^{3}
\end{cases}
\end{equation}
the optimum value achieved is within $3$ of the optimum value of
$\mathcal{P}_{1}$. Hence for an outer bound on gDoF, it is sufficient
to solve $\mathcal{P}_{2}$.
\begin{equation}
\text{gDoF}\brac{\mathcal{P}_{1}}=\text{gDoF}\brac{\mathcal{P}_{2}}.
\end{equation}
We will now show that it is sufficient to restrict $\text{Support}\brac{\abs a^{2},\abs b^{2},\abs c^{2}}=\cbrac{0,\Delta,2\Delta,\ldots,\left\lfloor \rho_{*}^{4}\right\rfloor \Delta}^{3}$
for an outer bound on gDoF.

Let the optimum value of $\mathcal{P}_{2}$ be achieved by a probability
distribution $\cbrac{p_{i}^{*}}$ at the points $\cbrac{\brac{l_{1i}^{*}\Delta,l_{2i}^{*}\Delta,l_{3i}^{*}\Delta}}$
with $l_{ji}^{*}\in\mathbb{Z}$. Let
\begin{equation}
S_{1}=\cbrac{i:\max\brac{l_{1i}^{*},l_{2i}^{*},l_{3i}^{*}}\leq\rho_{*}^{4}},
\end{equation}
\begin{equation}
S_{2}=\cbrac{i:\max\brac{l_{1i}^{*},l_{2i}^{*},l_{3i}^{*}}>\rho_{*}^{4}}
\end{equation}
and let $\max\brac{l_{1i}^{*},l_{2i}^{*},l_{3i}^{*}}=l_{Mi}^{*}$
for labeling. The optimum value $\brac{\mathcal{P}_{2}}$ is given
by
\begin{align}
\brac{\mathcal{P}_{2}} & =\sum_{i\in S_{1}}p_{i}^{*}f\brac{l_{1i}^{*}\Delta,l_{2i}^{*}\Delta,l_{3i}^{*}\Delta}+\sum_{i\in S_{2}}p_{i}^{*}f\brac{l_{1i}^{*}\Delta,l_{2i}^{*}\Delta,l_{3i}^{*}\Delta}.
\end{align}
We will now show that $\sum_{i\in S_{2}}p_{i}^{*}f\brac{l_{1i}^{*}\Delta,l_{2i}^{*}\Delta,l_{3i}^{*}\Delta}$
does not contribute to the gDoF. The points in $S_{2}$ have large
power and hence they have low probability due to power constraints;
this ends up limiting the contribution to gDoF. We prove this precisely
in the following steps. Using the structure of $f\brac{\abs a^{2},\abs b^{2},\abs c^{2}}$
and $\Delta=1/\brac{2\brac{T+1}\rho_{*}^{2}}$, we can bound
\begin{align}
\abs{f\brac{l_{1i}^{*}\Delta,l_{2i}^{*}\Delta,l_{3i}^{*}\Delta}} & \leq\lgbrac{\brac{2l_{Mi}^{*}+1}\brac{2l_{Mi}^{*}+1}+\brac{l_{Mi}^{*}+1}\brac{l_{Mi}^{*}+1}}\nonumber \\
 & \qquad+\brac{T-1}\lgbrac{\brac{l_{Mi}^{*}+1}\brac{l_{Mi}^{*}+1}+2l_{Mi}^{*}+\brac{l_{Mi}^{*}+1}\brac{l_{Mi}^{*}+1}}\nonumber \\
 & \qquad+2\lgbrac{\brac{1+l_{Mi}^{*}}\brac{1+l_{Mi}^{*}}+l_{Mi}^{*}}\\
 & \leq\brac{T+2}\lgbrac{\brac{2l_{Mi}^{*}+1}\brac{2l_{Mi}^{*}+1}3}\\
 & =2\brac{T+2}\lgbrac{2l_{Mi}^{*}+1}+\brac{T+2}\lgbrac 3.
\end{align}
Hence
\begin{align}
 & \abs{\sum_{i\in S_{2}}p_{i}^{*}f\brac{l_{1i}^{*}\Delta,l_{2i}^{*}\Delta,l_{3i}^{*}\Delta}}\nonumber \\
 & \leq\sum_{i\in S_{2}}p_{i}^{*}2\brac{T+2}\lgbrac{2l_{Mi}^{*}+1}+\brac{T+2}\lgbrac 3\\
 & \overset{\brac i}{\leq}2\brac{T+2}\brac{\sum_{i\in S_{2}}p_{i}^{*}}\lgbrac{2\frac{\sum_{i\in S_{2}}p_{i}^{*}l_{Mi}^{*}}{\sum_{j\in S_{2}}p_{j}^{*}}+1}+\brac{T+2}\lgbrac 3\\
 & \overset{\brac{ii}}{\leq}2\brac{T+2}\brac{\sum_{i\in S_{2}}p_{i}^{*}}\lgbrac{2\frac{T}{\Delta\sum_{j\in S_{2}}p_{j}^{*}}+1}+\brac{T+2}\lgbrac 3\\
 & =2\brac{T+2}\brac{\sum_{i\in S_{2}}p_{i}^{*}}\lgbrac{2\frac{T}{\Delta}+\sum_{j\in S_{2}}p_{j}^{*}}\nonumber \\
 & \qquad-2\brac{T+2}\brac{\sum_{i\in S_{2}}p_{i}^{*}}\lgbrac{\sum_{j\in S_{2}}p_{j}^{*}}+\brac{T+2}\lgbrac 3\\
 & \overset{\brac{iii}}{\leq}2\brac{T+2}\brac{\sum_{i\in S_{2}}p_{i}^{*}}\lgbrac{2\frac{T}{\Delta}+1}+2\brac{T+2}\frac{\lgbrac e}{e}+\brac{T+2}\lgbrac 3\\
 & \overset{\brac{iv}}{\leq}2\brac{T+2}\brac{\frac{T}{\rho_{*}^{4}\Delta}}\lgbrac{2\frac{T}{\Delta}+1}+2\brac{T+2}\frac{\lgbrac e}{e}+\brac{T+2}\lgbrac 3\\
 & \overset{\brac v}{=}2\brac{T+2}\brac{\frac{2T\brac{T+1}}{\rho_{*}^{2}}}\lgbrac{4T\brac{T+1}\rho_{*}^{2}+1}\nonumber \\
 & \qquad+2\brac{T+2}\frac{\lgbrac e}{e}+\brac{T+2}\lgbrac 3\\
 & \overset{\brac{vi}}{\leq}2\brac{T+2}\brac{2T\brac{T+1}}\brac{4T\brac{T+1}+1}\frac{\lgbrac e}{e}\\
 & \qquad+2\brac{T+2}\frac{\lgbrac e}{e}+\brac{T+2}\lgbrac 3\\
 & =r_{1}\brac T\text{ independent of }\snr,
\end{align}
where $\brac i$ is due to Jensen's inequality, $\brac{ii}$ is due
to the power constraint $\sum_{i\in S_{2}}p_{i}^{*}l_{Mi}^{*}\Delta\leq T\Rightarrow\sum_{i\in S_{2}}p_{i}^{*}l_{Mi}^{*}\leq\frac{T}{\Delta}$,
$\brac{iii}$ is due to the fact $0\leq\brac{\sum_{i\in S_{2}}p_{i}^{*}}\leq1$
and $-x\lgbrac x\leq\frac{\lgbrac e}{e}$ for $x\in[0,1]$, $\brac{iv}$
is due to the fact $\sum_{i\in S_{2}}p_{i}^{*}l_{Mi}^{*}\Delta\leq T$
(power constraint) and $\rho_{*}^{4}\Delta<l_{Mi}^{*}\Delta$ in $S_{2}$,
hence $\sum_{i\in S_{2}}p_{i}^{*}\rho_{*}^{4}\Delta\leq T$ and $\sum_{i\in S_{2}}p_{i}^{*}\leq\frac{T}{\rho_{*}^{4}\Delta}$,
$\brac v$ is using $\Delta=\frac{1}{2\brac{T+1}\rho_{*}^{2}}$ and
$\brac{vi}$ is due to the fact $\frac{1}{x}\lgbrac x\leq\frac{\lgbrac e}{e}$
for $x\in[1,+\infty)$ and assuming $\rho_{*}^{2}>1.$ (If $\rho_{*}^{2}\leq1$,
then the system has zero gDoF, so we consider only $\rho_{*}^{2}>1$).
Hence it follows that
\[
\abs{\brac{\mathcal{P}_{2}}-\sum_{i\in S_{1}}p_{i}^{*}f\brac{l_{1i}^{*}\Delta,l_{2i}^{*}\Delta,l_{3i}^{*}\Delta}}=\abs{\sum_{i\in S_{2}}p_{i}^{*}f\brac{l_{1i}^{*}\Delta,l_{2i}^{*}\Delta,l_{3i}^{*}\Delta}}\leq r_{1}\brac T.
\]
Hence it follows that
\begin{equation}
\mathcal{P}_{3}:\begin{cases}
\underset{\expect{\abs a^{2}+\abs b^{2}+\abs c^{2}}\leq T}{\text{maximize}}\expect{f\brac{\abs a^{2},\abs b^{2},\abs c^{2}}}\\
\text{Support}\brac{\abs a^{2},\abs b^{2},\abs c^{2}}=S_{1}
\end{cases}
\end{equation}
achieves the same gDoF as $\mathcal{P}_{2}$, because any non-zero
probability outside $S_{1}$ in $\mathcal{P}_{2}$ can be assigned
to $\brac{0,0,0}$ in $\mathcal{P}_{3}$ by changing the value of
the objective function by a constant independent of $\snr$. Hence
\begin{equation}
\text{gDoF}\brac{\mathcal{P}_{1}}=\text{gDoF}\brac{\mathcal{P}_{2}}=\text{gDoF}\brac{\mathcal{P}_{3}}.
\end{equation}
Now $\mathcal{P}_{3}$ is a linear program with a finite number of
variables and constraints. (Also $\mathcal{P}_{3}$ has a finite optimum
value because of Jensen's inequality.) The variables are $\cbrac{p_{i}}_{i\in S_{1}}$
and the maximum number of nontrivial active constraints on $\cbrac{p_{i}}_{i\in S_{1}}$
is two, derived from
\begin{equation}
\expect{\abs a^{2}+\abs b^{2}+\abs c^{2}}=T,\ \sum p_{i}=1.
\end{equation}
Trivial constraints are $p_{i}\geq0,\ i\in S_{1}$. Hence by the theory
of linear optimization, there exists an optimal $\cbrac{p_{i}^{*}}_{i\in S_{1}}$
with at most two nonzero values. Hence it follows that
\begin{equation}
\mathcal{P}_{4}:\begin{cases}
\underset{}{\text{maximize}}\sum_{i=1}^{2}p_{i}f_{1}\brac{\abs{a_{i}}^{2},\abs{b_{i}}^{2},\abs{c_{i}}^{2}} & \text{}\\
\sum_{i=1}^{2}p_{i}\brac{\abs{a_{i}}^{2}+\abs{b_{i}}^{2}+\abs{c_{i}}^{2}}\leq T,\\
\sum p_{i}=1,\\
\abs{a_{i}}^{2},\abs{b_{i}}^{2},\abs{c_{i}}^{2}\geq0
\end{cases}
\end{equation}
has $\brac{\mathcal{P}_{4}}\geq\brac{\mathcal{P}_{3}}$. Note that
we have allowed $\brac{\abs{a_{i}}^{2},\abs{b_{i}}^{2},\abs{c_{i}}^{2}}_{i=1}^{2}$
to be real positive variables to be optimized. However, it is also
clear that $\brac{\mathcal{P}_{4}}\leq\brac{\mathcal{P}_{1}}$. Hence
\begin{equation}
\text{gDoF}\brac{\mathcal{P}_{1}}=\text{gDoF}\brac{\mathcal{P}_{2}}=\text{gDoF}\brac{\mathcal{P}_{3}}=\text{gDoF}\brac{\mathcal{P}_{4}}.
\end{equation}
Now consider
\begin{equation}
\mathcal{P}_{5}:\begin{cases}
\underset{}{\text{maximize}}\sum_{i=1}^{2}p_{i}f_{1}\brac{\abs{a_{i}}^{2},\abs{b_{i}}^{2},\abs{c_{i}}^{2}} & \text{}\\
p_{i}\abs{a_{i}}^{2}\leq T,p_{i}\abs{b_{i}}^{2}\leq T,p_{i}\abs{c_{i}}^{2}\leq T,\\
\sum p_{i}=1,\\
\abs{a_{i}}^{2},\abs{b_{i}}^{2},\abs{c_{i}}^{2}\geq0.
\end{cases}
\end{equation}
It can be easily shown that $\text{gDoF}\brac{\mathcal{P}_{4}}=\text{gDoF}\brac{\mathcal{P}_{5}}$,
we omit the proof.
\begin{lem}
Adding the constraints $\abs{a_{i}}^{2}\leq T,\abs{b_{i}}^{2}\leq T,\abs{c_{i}}^{2}\leq T$
does not change the gDoF of $\mathcal{P}_{5}$.
\end{lem}
\begin{IEEEproof}
We have
\begin{align}
 & f\brac{\abs a^{2},\abs b^{2},\abs c^{2}}\nonumber \\
 & =\lgbrac{\brac{\abs a^{2}\rho_{11}^{2}+\abs b^{2}\rho_{12}^{2}+1}\brac{\abs a^{2}\rho_{21}^{2}+\abs b^{2}\rho_{22}^{2}+1}+\brac{\abs c^{2}\rho_{12}^{2}+1}\brac{\abs c^{2}\rho_{22}^{2}+1}}\nonumber \\
 & \qquad+\brac{T-1}\lgbrac{\brac{\abs a^{2}\rho_{11}^{2}+1}\brac{\abs c^{2}\rho_{22}^{2}+1}+\abs b^{2}\brac{\rho_{12}^{2}+\rho_{22}^{2}}+\brac{\abs a^{2}\rho_{21}^{2}+1}\brac{\abs c^{2}\rho_{12}^{2}+1}}\nonumber \\
 & \qquad-\lgbrac{\brac{1+\abs a^{2}\rho_{11}^{2}}\brac{1+\abs c^{2}\rho_{12}^{2}}+\abs b^{2}\rho_{12}^{2}}\nonumber \\
 & \qquad-\lgbrac{\brac{1+\abs a^{2}\rho_{21}^{2}}\brac{1+\abs c^{2}\rho_{22}^{2}}+\abs b^{2}\rho_{22}^{2}}.
\end{align}
Suppose $\abs{a_{i}}^{2}>T$ and consider
\[
t_{1}=p_{i}\lgbrac{\brac{\abs{a_{i}}^{2}\rho_{11}^{2}+\abs{b_{i}}^{2}\rho_{12}^{2}+1}\brac{\abs{a_{i}}^{2}\rho_{21}^{2}+\abs{b_{i}}^{2}\rho_{22}^{2}+1}+\brac{\abs{c_{i}}^{2}\rho_{12}^{2}+1}\brac{\abs{c_{i}}^{2}\rho_{22}^{2}+1}}.
\]
We will show that setting $\abs{a_{i}}^{2}=T$ would change the value
of $t_{1}$ only by a constant independent of $\snr$. The other terms
have a similar structure and can be handled in a similar way. If $\brac{\abs{c_{i}}^{2}\rho_{12}^{2}+1}\brac{\abs{c_{i}}^{2}\rho_{22}^{2}+1}>\brac{\abs{a_{i}}^{2}\rho_{11}^{2}+\abs{b_{i}}^{2}\rho_{12}^{2}+1}\brac{\abs{a_{i}}^{2}\rho_{21}^{2}+\abs{b_{i}}^{2}\rho_{22}^{2}+1}$,
then the claim is trivially true; we can replace $\abs{a_{i}}^{2}>T$
with $\abs{a_{i}}^{2}=T$ while changing the value of $t_{1}$ by
only a constant. Otherwise
\begin{align}
t_{1} & \eqdof p_{i}\lgbrac{\brac{\abs{a_{i}}^{2}\rho_{11}^{2}+\abs{b_{i}}^{2}\rho_{12}^{2}+1}\brac{\abs{a_{i}}^{2}\rho_{21}^{2}+\abs{b_{i}}^{2}\rho_{22}^{2}+1}}\\
 & =\underset{t_{11}}{\underbrace{p_{i}\lgbrac{\abs{a_{i}}^{2}\rho_{11}^{2}+\abs{b_{i}}^{2}\rho_{12}^{2}+1}}}+\underset{t_{12}}{\underbrace{p_{i}\lgbrac{\abs{a_{i}}^{2}\rho_{21}^{2}+\abs{b_{i}}^{2}\rho_{22}^{2}+1}}}.
\end{align}
Now consider $t_{11}=p_{i}\lgbrac{\abs{a_{i}}^{2}\rho_{11}^{2}+\abs{b_{i}}^{2}\rho_{12}^{2}+1}$.
If $\abs{a_{i}}^{2}\rho_{11}^{2}<\abs{b_{i}}^{2}\rho_{12}^{2}+1$,
then we can replace $\abs{a_{i}}^{2}>T$ with $\abs{a_{i}}^{2}=T$
without losing gDoF.

If $\abs{a_{i}}^{2}\rho_{11}^{2}>\abs{b_{i}}^{2}\rho_{12}^{2}+1$,
then $t_{11}\eqdof p_{i}\lgbrac{\abs{a_{i}}^{2}\rho_{11}^{2}+1}$,
where the approximation is tight within a constant (constant less
than 1). Now if we replace $\abs{a_{i}}^{2}>T$ with $\abs{a_{i}}^{2}=T$,
the difference arising is bounded independent of $\snr$, as seen
below:
\begin{align}
p_{i}\lgbrac{\abs{a_{i}}^{2}\rho_{11}^{2}+1} & \overset{\brac i}{\leq}p_{i}\lgbrac{\frac{T}{p_{i}}\rho_{11}^{2}+1}\\
 & =p_{i}\lgbrac{T\rho_{11}^{2}+p_{i}}-p_{i}\lgbrac{p_{i}}\\
 & \leq p_{i}\lgbrac{T\rho_{11}^{2}+1}-p_{i}\lgbrac{p_{i}},
\end{align}
where $\brac i$ is because $p_{i}\abs{a_{i}}^{2}\leq T$ due to the
power constraint. We also have $\abs{a_{i}}^{2}>T$, hence it follows
that
\begin{align}
\abs{p_{i}\lgbrac{\abs{a_{i}}^{2}\rho_{11}^{2}+1}-p_{i}\lgbrac{T\rho_{11}^{2}+1}} & \leq\abs{p_{i}\lgbrac{p_{i}}}\leq\frac{\lgbrac e}{e}.
\end{align}
Following the same logic for the other terms, it can be shown that
adding the constraints $\abs{a_{i}}^{2}\leq T,\ \abs{b_{i}}^{2}\leq T,\ \abs{c_{i}}^{2}\leq T$
does not change the gDoF of $\mathcal{P}_{5}$.
\end{IEEEproof}
With the additional constraints $\abs{a_{i}}^{2}\leq T,\ \abs{b_{i}}^{2}\leq T,\ \abs{c_{i}}^{2}\leq T$,
the existing constraints $p_{i}\abs{a_{i}}^{2}\leq T,\ p_{i}\abs{b_{i}}^{2}\leq T,\ p_{i}\abs{c_{i}}^{2}\leq T$
become redundant. Hence we have $\text{gDoF}\brac{\mathcal{P}_{5}}=\text{gDoF}\brac{\mathcal{P}_{6}}$
for $\mathcal{P}_{6}$ defined as
\begin{equation}
\mathcal{P}_{6}:\begin{cases}
\underset{}{\text{maximize}}\ \sum_{i=1}^{2}p_{i}f\brac{\abs{a_{i}}^{2},\abs{b_{i}}^{2},\abs{c_{i}}^{2}} & \text{}\\
\abs{a_{i}}^{2}\leq T,\ \abs{b_{i}}^{2}\leq T,\ \abs{c_{i}}^{2}\leq T,\\
\sum p_{i}=1.
\end{cases}
\end{equation}
It is clear from the structure of $\mathcal{P}_{6}$ that the solution
has $\brac{\abs{a_{1}}^{2},\abs{b_{1}}^{2},\abs{c_{1}}^{2}}=\brac{\abs{a_{2}}^{2},\abs{b_{2}}^{2},\abs{c_{2}}^{2}}$.
Hence it suffices to solve $\mathcal{P}_{7}$ defined as
\begin{equation}
\mathcal{P}_{7}:\begin{cases}
\underset{}{\text{maximize}}\ f\brac{\abs a^{2},\abs b^{2},\abs c^{2}} & \text{}\\
\abs a^{2}\leq T,\ \abs b^{2}\leq T,\ \abs c^{2}\leq T.
\end{cases}
\end{equation}

\bibliographystyle{IEEEtran}
\bibliography{C:/Users/Joyson/Desktop/Project/Bibtex/bibJournalList,C:/Users/Joyson/Desktop/Project/Bibtex/references}

\begin{thebibliography}{10}
\providecommand{\url}[1]{#1}
\csname url@samestyle\endcsname
\providecommand{\newblock}{\relax}
\providecommand{\bibinfo}[2]{#2}
\providecommand{\BIBentrySTDinterwordspacing}{\spaceskip=0pt\relax}
\providecommand{\BIBentryALTinterwordstretchfactor}{4}
\providecommand{\BIBentryALTinterwordspacing}{\spaceskip=\fontdimen2\font plus
\BIBentryALTinterwordstretchfactor\fontdimen3\font minus
  \fontdimen4\font\relax}
\providecommand{\BIBforeignlanguage}[2]{{%
\expandafter\ifx\csname l@#1\endcsname\relax
\typeout{** WARNING: IEEEtran.bst: No hyphenation pattern has been}%
\typeout{** loaded for the language `#1'. Using the pattern for}%
\typeout{** the default language instead.}%
\else
\language=\csname l@#1\endcsname
\fi
#2}}
\providecommand{\BIBdecl}{\relax}
\BIBdecl

\bibitem{Joyson_2x2MIMO_isit}
J.~Sebastian, A.~Sengupta, and S.~N. Diggavi, ``On capacity of noncoherent
  {MIMO} with asymmetric link strengths,'' in \emph{IEEE International
  Symposium on Information Theory}, June 2017, pp. 541--545.

\bibitem{marzetta1999capacity}
T.~L. Marzetta and B.~M. Hochwald, ``Capacity of a mobile multiple-antenna
  communication link in {Rayleigh} flat fading,'' \emph{IEEE Transactions on
  Information Theory}, vol.~45, no.~1, pp. 139--157, 1999.

\bibitem{Zheng_Tse_Grassmann_MIMO}
L.~Zheng and D.~N.~C. Tse, ``{Communication on the Grassmann manifold: a
  geometric approach to the noncoherent multiple-antenna channel},'' \emph{IEEE
  Transactions on Information Theory}, vol.~48, no.~2, pp. 359--383, Feb 2002.

\bibitem{Wang_2013_noncoh_large_MIMO}
W.~Yang, G.~Durisi, and E.~Riegler, ``On the capacity of large-{MIMO}
  block-fading channels,'' \emph{IEEE Journal on Selected Areas in
  Communications}, vol.~31, no.~2, pp. 117--132, February 2013.

\bibitem{Taricco_Elia_97}
G.~Taricco and M.~Elia, ``Capacity of fading channel with no side
  information,'' \emph{Electronics Letters}, vol.~33, no.~16, pp. 1368--1370,
  Jul 1997.

\bibitem{Abou_Faycal_noncoherent}
I.~C. Abou-Faycal, M.~D. Trott, and S.~Shamai, ``{The capacity of discrete-time
  memoryless Rayleigh-fading channels},'' \emph{IEEE Transactions on
  Information Theory}, vol.~47, no.~4, pp. 1290--1301, May 2001.

\bibitem{lapidoth2003capacity}
A.~Lapidoth and S.~M. Moser, ``Capacity bounds via duality with applications to
  multiple-antenna systems on flat-fading channels,'' \emph{IEEE Transactions
  on Information Theory}, vol.~49, no.~10, pp. 2426--2467, 2003.

\bibitem{etkin_tse_no_fb_IC}
R.~H. Etkin, D.~N.~C. Tse, and H.~Wang, ``Gaussian interference channel
  capacity to within one bit,'' \emph{IEEE Transactions on Information Theory},
  vol.~54, no.~12, pp. 5534--5562, 2008.

\bibitem{gDoF_K_user_IC}
S.~A. {Jafar} and S.~{Vishwanath}, ``Generalized degrees of freedom of the
  symmetric gaussian $k$-user interference channel,'' \emph{IEEE Transactions
  on Information Theory}, vol.~56, no.~7, pp. 3297--3303, July 2010.

\bibitem{gDoF_MIMO_IC}
S.~{Karmakar} and M.~K. {Varanasi}, ``The generalized degrees of freedom region
  of the mimo interference channel and its achievability,'' \emph{IEEE
  Transactions on Information Theory}, vol.~58, no.~12, pp. 7188--7203, Dec
  2012.

\bibitem{Morgenshtern_SIMO_correl_2013}
V.~I. Morgenshtern, E.~Riegler, W.~Yang, G.~Durisi, S.~Lin, B.~Sturmfels, and
  H.~Bolcskei, ``Capacity pre-log of noncoherent {SIMO} channels via
  {Hironaka's} theorem,'' \emph{IEEE Transactions on Information Theory},
  vol.~59, no.~7, pp. 4213--4229, July 2013.

\bibitem{Koliander_MIMO_correl_2014}
G.~Koliander, E.~Riegler, G.~Durisi, and F.~Hlawatsch, ``Degrees of freedom of
  generic block-fading {MIMO} channels without apriori channel state
  information,'' \emph{IEEE Transactions on Information Theory}, vol.~60,
  no.~12, pp. 7760--7781, Dec 2014.

\bibitem{Lapidoth_network}
A.~Lapidoth, ``On the high-{SNR} capacity of noncoherent networks,'' \emph{IEEE
  Transactions on Information Theory}, vol.~51, no.~9, pp. 3025--3036, Sept
  2005.

\bibitem{Koch2013}
T.~Koch and G.~Kramer, ``On noncoherent fading relay channels at high
  signal-to-noise ratio,'' \emph{IEEE Transactions on Information Theory},
  vol.~59, no.~4, pp. 2221--2241, April 2013.

\bibitem{Gohary_non_coherent_2014}
R.~H. Gohary and H.~Yanikomeroglu, ``Grassmannian signalling achieves tight
  bounds on the ergodic high-{SNR} capacity of the noncoherent {MIMO}
  full-duplex relay channel,'' \emph{IEEE Transactions on Information Theory},
  vol.~60, no.~5, pp. 2480--2494, May 2014.

\bibitem{bhushan2014network_densification}
N.~Bhushan, J.~Li, D.~Malladi, R.~Gilmore, D.~Brenner, A.~Damnjanovic,
  R.~Sukhavasi, C.~Patel, and S.~Geirhofer, ``Network densification: the
  dominant theme for wireless evolution into {5G},'' \emph{IEEE Communications
  Magazine}, vol.~52, no.~2, pp. 82--89, 2014.

\bibitem{wu2015cloud}
J.~Wu, Z.~Zhang, Y.~Hong, and Y.~Wen, ``{Cloud radio access network (C-RAN): a
  primer},'' \emph{IEEE Network}, vol.~29, no.~1, pp. 35--41, 2015.

\bibitem{Irmer_comp_2011}
R.~Irmer, H.~Droste, P.~Marsch, M.~Grieger, G.~Fettweis, S.~Brueck, H.~Mayer,
  L.~Thiele, and V.~Jungnickel, ``Coordinated multipoint: Concepts,
  performance, and field trial results,'' \emph{IEEE Communications Magazine},
  vol.~49, no.~2, pp. 102--111, February 2011.

\bibitem{karakus_d2d_sidechannel_2017}
C.~Karakus and S.~N. Diggavi, ``Enhancing multiuser {MIMO} through
  opportunistic {D2D} cooperation,'' \emph{IEEE Transactions on Wireless
  Communications}, vol.~16, no.~9, pp. 5616--5629, Sept 2017.

\bibitem{Martina_gdof_MIMO}
M.~Cardone, D.~Tuninetti, R.~Knopp, and U.~Salim, ``Gaussian half-duplex relay
  networks: Improved constant gap and connections with the assignment
  problem,'' \emph{IEEE Transactions on Information Theory}, vol.~60, no.~6,
  pp. 3559--3575, Jun 2014.

\bibitem{Joyson_fading}
J.~Sebastian, C.~Karakus, S.~N. Diggavi, and I.~H. Wang, ``Rate splitting is
  approximately optimal for fading {Gaussian} interference channels,'' in
  \emph{Annual Allerton Conference on Communication, Control, and Computing},
  Sept 2015, pp. 315--321.

\bibitem{abramowitz1964handbook}
M.~Abramowitz and I.~A. Stegun, \emph{Handbook of mathematical functions: with
  formulas, graphs, and mathematical tables}.\hskip 1em plus 0.5em minus
  0.4em\relax Courier Corporation, 1964, no.~55.

\bibitem{cover2012elements}
\BIBentryALTinterwordspacing
T.~Cover and J.~Thomas, \emph{Elements of Information Theory}.\hskip 1em plus
  0.5em minus 0.4em\relax Wiley, 2012. [Online]. Available:
  \url{https://books.google.com/books?id=VWq5GG6ycxMC}
\BIBentrySTDinterwordspacing

\end{thebibliography}

\clearpage

\section{Proof of Theorem \ref{thm:decompose_graph}: Decomposing into disjoint
parts of MIMO channels\label{app:decompose_graph}}

Here we prove that for a MIMO system whose channel can be decomposed
into disjoint parts, the capacity can be achieved by allocating power
to the disjoint parts separately. Let the channel matrix $G$ of the
system be block diagonal as $G=\text{diag}\brac{G_{1},\ldots,G_{K}}$,
where $G_{i}$ are the diagonal blocks corresponding to the disjoint
parts of the channel, then the capacity $C\brac{P,\text{diag}\brac{G_{1},\ldots,G_{K}}}$
of the channel for a power $P$ can be achieved by splitting power
across the blocks, \emph{i.e.,} $C\brac{P,\text{diag}\brac{G_{1},\ldots,G_{K}}}=\max_{P_{1}+\cdots+P_{K}\leq P}\brac{C\brac{P_{1},G_{1}}+\cdots+C\brac{P_{K},G_{K}}}$.
We just need to show that for $G=\text{diag}\brac{G_{1},G_{2}}$,
the capacity of the channel can be achieved by a power splitting across
the two blocks of channels $G_{1},G_{2}$ \emph{i.e.,}
\begin{equation}
C\brac{P,\text{diag}\brac{G_{1},G_{2}}}=\max_{P_{1}+P_{2}\leq P}\brac{C\brac{P_{1},G_{1}}+C\brac{P_{2},G_{2}}}
\end{equation}
and the general result for multiple disjoint parts in the MIMO channel
will follow due to induction. We have

\begin{align}
h\brac Y & \overset{\brac{ii}}{\leq}h\brac{Y_{G1}}+h\brac{Y_{G2}}\\
h\brac{\rline YX} & =h\brac{\rline{Y_{G1}Y_{G2}}X_{G1}X_{G2}}\\
 & =h\brac{\rline{Y_{G1}}X_{G1}X_{G2}}+h\brac{\rline{Y_{G2}}Y_{G1}X_{G1}X_{G2}}\\
 & \overset{\brac{ii}}{=}h\brac{\rline{Y_{G1}}X_{G2}}+h\brac{\rline{Y_{G2}}X_{G2}},
\end{align}
where $\brac i$ is because conditioning reduces entropy and $\brac{ii}$
is because $X_{G2}-X_{G1}-Y_{G1}$ and $\brac{X_{G1},Y_{G1}}-X_{G2}-Y_{G2}$
are Markov chains.

Hence
\begin{align}
I\brac{X;Y} & \leq I\brac{X_{G1};Y_{G1}}+I\brac{X_{G2};Y_{G2}}
\end{align}
subject to $\expect{\twonorm{X_{G1}}^{2}+\twonorm{X_{G2}}^{2}}\leq P$.
The RHS can be achieved by treating the two blocks of channels $G_{1},G_{2}$
separately with a power allocation, hence
\begin{equation}
C\brac{P,\text{diag}\brac{G_{1},G_{2}}}=\max_{P_{1}+P_{2}\leq P}\brac{C\brac{P_{1},G_{1}}+C\brac{P_{2},G_{2}}}.
\end{equation}

\section{Inner bound for the $2\times2$  MIMO channel\label{app:Inner-bound-for2x2mimo}}

Here we prove the achievability result from Theorem \ref{thm:2x2-sym-mimo}
for the $2\times2$ MIMO channel with exponents $\gamma_{D}$ in the
direct links and $\gamma_{CL}$ in the crosslinks. We use the input
distribution
\[
X=\left[\begin{array}{cccccc}
a & 0 & 0 & . & . & 0\\
\eta & c & 0 & . & . & 0
\end{array}\right]Q
\]
 with constants $a,c$ and $\eta\sim\mathcal{CN}\brac{0,\abs b^{2}}$
with
\[
\abs a^{2}=\snr^{\gamma_{a}},\abs b^{2}=\snr^{\gamma_{b}},\abs c^{2}=\snr^{\gamma_{c}},\gamma_{a}\leq0,\gamma_{b}\leq0,\gamma_{c}\leq0.
\]
 With this choice, we proceed to lower bound $I\brac{X;Y}$.
\begin{align}
I\brac{X;Y} & =h\brac Y-h\brac{\rline YX}\\
h\brac Y & =h\brac{GX+W}\\
 & \geq h\brac{GX}\\
 & =h\brac{\left[\begin{array}{cc}
g_{11} & g_{12}\\
g_{21} & g_{22}
\end{array}\right]\left[\begin{array}{cccccc}
a & 0 & 0 & . & . & 0\\
\eta & c & 0 & . & . & 0
\end{array}\right]Q}.
\end{align}
Now

\begin{align}
 & \left[\begin{array}{cc}
g_{11} & g_{12}\\
g_{21} & g_{22}
\end{array}\right]\left[\begin{array}{cccccc}
a & 0 & 0 & . & . & 0\\
\eta & c & 0 & . & . & 0
\end{array}\right]\nonumber \\
 & =\left[\begin{array}{cccccc}
ag_{11}+\eta g_{12} & cg_{12} & 0 & . & . & 0\\
ag_{21}+\eta g_{22} & cg_{22} & 0 & . & . & 0
\end{array}\right]\\
 & =\left[\begin{array}{cccccc}
\sqrt{\abs{ag_{11}+\eta g_{12}}^{2}+\abs{cg_{12}}^{2}} & 0 & 0 & . & . & 0\\
\frac{\brac{ag_{21}+\eta g_{22}}\brac{ag_{11}+\eta g_{12}}^{*}+\abs c^{2}g_{22}g_{12}^{*}}{\sqrt{\abs{ag_{11}+\eta g_{12}}^{2}+\abs{cg_{12}}^{2}}} & \frac{\brac{ag_{21}+\eta g_{22}}cg_{12}-cg_{22}\brac{ag_{11}+\eta g_{12}}}{\sqrt{\abs{ag_{11}+\eta g_{12}}^{2}+\abs{cg_{12}}^{2}}} & 0 & . & . & 0
\end{array}\right]\Phi,
\end{align}
where in the last step, we performed an LQ transformation and $\Phi$
is unitary. Hence due to the property of isotropic unitary matrices
and steps similar to (\ref{eq:2x2LQsimplification_begin}) to (\ref{eq:2x2LQsimplification_end})
in Section \ref{subsec:2x2MIMO}, we get:
\begin{align}
 & h\brac{GX}\nonumber \\
 & \overset{\brac i}{=}h\brac{\sqrt{\abs{ag_{11}+\eta g_{12}}^{2}+\abs{cg_{12}}^{2}}\overline{q_{1}}^{\brac T}}\nonumber \\
 & \qquad+h\brac{\rline{\sbrac{\frac{\brac{ag_{21}+\eta g_{22}}\brac{ag_{11}+\eta g_{12}}^{*}+\abs c^{2}g_{22}g_{12}^{*}}{\sqrt{\abs{ag_{11}+\eta g_{12}}^{2}+\abs{cg_{12}}^{2}}},\frac{ac\brac{g_{12}g_{21}-g_{11}g_{22}}}{\sqrt{\abs{ag_{11}+\eta g_{12}}^{2}+\abs{cg_{12}}^{2}}}\overline{q}_{2}^{\brac{T-1}}}}\xi_{11}}\nonumber \\
 & \overset{\brac{ii}}{=}h\brac{\sqrt{\abs{ag_{11}+\eta g_{12}}^{2}+\abs{cg_{12}}^{2}}\overline{q_{1}}^{\brac T}}-T\expect{\lgbrac{\abs{ag_{11}+\eta g_{12}}^{2}+\abs{cg_{12}}^{2}}}\nonumber \\
 & \qquad+\underset{\alpha}{\underbrace{h\brac{\rline{\sbrac{\brac{ag_{21}+\eta g_{22}}\brac{ag_{11}+\eta g_{12}}^{*}+\abs c^{2}g_{22}g_{12}^{*},\ ac\brac{g_{12}g_{21}-g_{11}g_{22}}\overline{q}_{2}^{\brac{T-1}}}\vphantom{a^{a^{a^{a^{a^{a^{a}}}}}}}}\xi_{11}}}},\label{eq:innerbound_2x2_h(gaq)}
\end{align}
where $\overline{q_{1}}^{\brac i}$ denotes an $i$ dimensional isotropically
distributed random unit vector and $\xi_{11}=\sqrt{\abs{ag_{11}+\eta g_{12}}^{2}+\abs{cg_{12}}^{2}}$.
The step $\brac i$ involved the simplification $\brac{ag_{21}+\eta g_{22}}cg_{12}-cg_{22}\brac{ag_{11}+\eta g_{12}}=ac\brac{g_{12}g_{21}-g_{11}g_{22}}$
and the step $\brac i$ involved moving $\sqrt{\abs{ag_{11}+\eta g_{12}}^{2}+\abs{cg_{12}}^{2}}$
from the denominator. Now{\small{}
\begin{align}
 & \alpha\nonumber \\
 & =h\brac{\rline{ac\brac{g_{12}g_{21}-g_{11}g_{22}}\overline{q}_{2}^{\brac{T-1}}\vphantom{a^{a^{a^{a}}}}}\xi_{11}}\nonumber \\
 & \qquad+h\brac{\rline{\brac{ag_{21}+\eta g_{22}}\brac{ag_{11}+\eta g_{12}}^{*}+\abs c^{2}g_{22}g_{12}^{*}\vphantom{a^{a^{a^{a}}}}}ac\brac{g_{12}g_{21}-g_{11}g_{22}}\overline{q}_{2}^{\brac{T-1}},\xi_{11}}\nonumber \\
 & \overset{\brac i}{\geq}+h\brac{\rline{\abs{ac\brac{g_{12}g_{21}-g_{11}g_{22}}}^{2}\vphantom{a^{a^{a^{a}}}}}\xi_{11}}+\brac{T-2}\expect{\lgbrac{\abs{ac\brac{g_{12}g_{21}-g_{11}g_{22}}}^{2}}}+\lgbrac{\frac{\pi^{T-1}}{\Gamma\brac{T-1}}}\nonumber \\
 & \qquad+h\brac{\rline{\brac{ag_{21}+\eta g_{22}}\brac{ag_{11}+\eta g_{12}}^{*}+\abs c^{2}g_{22}g_{12}^{*}\vphantom{a^{a^{a^{a}}}}}ac\brac{g_{12}g_{21}-g_{11}g_{22}},\overline{q}_{2}^{\brac{T-1}},\xi_{11}}\nonumber \\
 & \overset{\brac{ii}}{=}h\brac{\rline{ac\brac{g_{12}g_{21}-g_{11}g_{22}}\vphantom{a^{a^{a^{a}}}}}\xi_{11}}-\lgbrac{\pi}+\brac{T-2}\expect{\lgbrac{\abs{ac\brac{g_{12}g_{21}-g_{11}g_{22}}}^{2}}}\nonumber \\
 & \qquad+\lgbrac{\frac{\pi^{T-1}}{\Gamma\brac{T-1}}}+h\brac{\rline{\brac{ag_{21}+\eta g_{22}}\brac{ag_{11}+\eta g_{12}}^{*}+\abs c^{2}g_{22}g_{12}^{*}\vphantom{a^{a^{a^{a}}}}}ac\brac{g_{12}g_{21}-g_{11}g_{22}},\xi_{11}}\nonumber \\
 & \overset{\brac{iii}}{=}\brac{T-2}\expect{\lgbrac{\abs{acg_{12}g_{21}-acg_{11}g_{22}}^{2}}}+\lgbrac{\frac{\pi^{T-2}}{\Gamma\brac{T-1}}}+2\expect{\lgbrac{\abs{ag_{11}+\eta g_{12}}^{2}+\abs{cg_{12}}^{2}}}\nonumber \\
 & \qquad h\brac{\rline{\sbrac{\frac{\brac{ag_{21}+\eta g_{22}}\brac{ag_{11}+\eta g_{12}}^{*}+\abs c^{2}g_{22}g_{12}^{*}}{\sqrt{\abs{ag_{11}+\eta g_{12}}^{2}+\abs{cg_{12}}^{2}}},\frac{ac\brac{g_{12}g_{21}-g_{11}g_{22}}}{\sqrt{\abs{ag_{11}+\eta g_{12}}^{2}+\abs{cg_{12}}^{2}}}}}\xi_{11}}\nonumber \\
 & \overset{\brac{iv}}{\geq}\brac{T-2}\expect{\lgbrac{\abs{acg_{12}g_{21}-acg_{11}g_{22}}^{2}}}+\lgbrac{\frac{\pi^{T-2}}{\Gamma\brac{T-1}}}+2\expect{\lgbrac{\abs{ag_{11}+\eta g_{12}}^{2}+\abs{cg_{12}}^{2}}}\nonumber \\
 & \qquad h\brac{\rline{\sbrac{\frac{\brac{ag_{21}+\eta g_{22}}\brac{ag_{11}+\eta g_{12}}^{*}+\abs c^{2}g_{22}g_{12}^{*}}{\sqrt{\abs{ag_{11}+\eta g_{12}}^{2}+\abs{cg_{12}}^{2}}},\frac{\brac{ag_{21}+\eta g_{22}}cg_{12}-cg_{22}\brac{ag_{11}+\eta g_{12}}}{\sqrt{\abs{ag_{11}+\eta g_{12}}^{2}+\abs{cg_{12}}^{2}}}}}ag_{11}+\eta g_{12},g_{12}}\nonumber \\
 & \overset{\brac v}{=}\brac{T-2}\expect{\lgbrac{\abs{acg_{12}g_{21}-acg_{11}g_{22}}^{2}}}+\lgbrac{\frac{\pi^{T-2}}{\Gamma\brac{T-1}}}+2\expect{\lgbrac{\abs{ag_{11}+\eta g_{12}}^{2}+\abs{cg_{12}}^{2}}}\nonumber \\
 & \qquad+h\brac{\rline{\left[\begin{array}{cc}
ag_{21}+\eta g_{22} & cg_{22}\end{array}\right]\vphantom{a^{a^{a^{a}}}}}ag_{11}+\eta g_{12},g_{12}},\label{eq:innerbound_2x2_alpha}
\end{align}
}where $\brac i$ is using Lemma \ref{lem:isotropic_entropy_to_radial}
on page \pageref{lem:isotropic_entropy_to_radial} for the first term
and the fact that conditioning reduces entropy for the second term,
$\brac{ii}$ is using Lemma \ref{lem:isotropic_entropy_to_radial}
on $h\brac{\rline{ac\brac{g_{12}g_{21}-g_{11}g_{22}}}\xi_{11}}$.
Note that with $\theta\sim\text{Unif}\sbrac{0,2\pi}$ independent
of other random variables, $\rline{ac\brac{g_{12}g_{21}-g_{11}g_{22}}e^{i\theta}}\xi_{11}$
and $\rline{ac\brac{g_{12}g_{21}-g_{11}g_{22}}}\xi_{11}$ have the
same distribution; $e^{i\theta}$ is the unitary distribution in one
dimension; hence Lemma \ref{lem:isotropic_entropy_to_radial} can
be applied. The step $\brac{iii}$ is using $\xi_{11}=\sqrt{\abs{ag_{11}+\eta g_{12}}^{2}+\abs{cg_{12}}^{2}}$
and rearranging the terms, $\brac{iv}$ is because conditioning reduces
entropy and $\brac v$ is by a unitary transformation on the last
term. Hence by substituting (\ref{eq:innerbound_2x2_alpha}) in (\ref{eq:innerbound_2x2_h(gaq)}),
we have
\begin{align}
 & h\brac{GX}\nonumber \\
 & \geq h\brac{\sqrt{\abs{ag_{11}+\eta g_{12}}^{2}+\abs{cg_{12}}^{2}}\overline{q_{1}}^{\brac T}}-\brac{T-2}\expect{\lgbrac{\abs{ag_{11}+\eta g_{12}}^{2}+\abs{cg_{12}}^{2}}}\nonumber \\
 & \qquad+\lgbrac{\frac{\pi^{T-2}}{\Gamma\brac{T-1}}}+\brac{T-2}\expect{\lgbrac{\abs{acg_{12}g_{21}-acg_{11}g_{22}}^{2}}}\nonumber \\
 & \qquad+h\brac{\rline{\left[\begin{array}{cc}
ag_{21}+\eta g_{22} & cg_{22}\end{array}\right]\vphantom{a^{a^{a^{a}}}}}ag_{11}+\eta g_{12},g_{12}}\\
 & \overset{\brac i}{=}h\brac{\abs{ag_{11}+\eta g_{12}}^{2}+\abs{cg_{12}}^{2}}+\expect{\lgbrac{\abs{ag_{11}+\eta g_{12}}^{2}+\abs{cg_{12}}^{2}}}+\lgbrac{\frac{\pi^{T-2}}{\Gamma\brac{T-1}}}\nonumber \\
 & \qquad+\lgbrac{\frac{\pi^{T}}{\Gamma\brac T}}+\brac{T-2}\expect{\lgbrac{\abs{acg_{12}g_{21}-acg_{11}g_{22}}^{2}}}\nonumber \\
 & \qquad+h\brac{\rline{\left[\begin{array}{cc}
ag_{21}+\eta g_{22} & cg_{22}\end{array}\right]\vphantom{a^{a^{a^{a}}}}}ag_{11}+\eta g_{12},g_{12}},\label{eq:innerbound_2x2_H(GAQ)_step2}
\end{align}
where $\brac i$ is using Lemma \ref{lem:isotropic_entropy_to_radial}
on $h\brac{\sqrt{\abs{ag_{11}+\eta g_{12}}^{2}+\abs{cg_{12}}^{2}}\overline{q_{1}}^{\brac T}}$.
Also
\begin{align}
h\brac{\rline{\left[\begin{array}{cc}
ag_{21}+\eta g_{22} & cg_{22}\end{array}\right]\vphantom{a^{a^{a^{a}}}}}ag_{11}+\eta g_{12},g_{12}} & =h\brac{cg_{22}}+h\brac{\rline{ag_{21}+\eta g_{22}}ag_{11}+\eta g_{12},g_{12},g_{22}}.\label{eq:innerbound_2x2_h(gaq)_part1}
\end{align}
Using $\eta\sim\mathcal{CN}\brac{0,\abs b^{2}}$, we have
\begin{align}
 & h\brac{\rline{ag_{21}+\eta g_{22}\vphantom{a^{a^{a^{a}}}}}ag_{11}+\eta g_{12},g_{12},g_{22}}\nonumber \\
 & =h\brac{\rline{ag_{21}+\eta g_{22},ag_{11}+\eta g_{12}\vphantom{a^{a^{a^{a}}}}}g_{12},g_{22}}-h\brac{\rline{ag_{11}+\eta g_{12}\vphantom{a^{a^{a^{a}}}}}g_{12},g_{22}}\\
 & =\expect{\lgbrac{\abs{\begin{array}{cc}
\abs a^{2}\rho_{21}^{2}+\abs b^{2}\abs{g_{22}}^{2} & \abs b^{2}g_{22}g_{12}^{*}\\
\abs b^{2}g_{22}^{*}g_{12} & \abs a^{2}\rho_{11}^{2}+\abs b^{2}\abs{g_{12}}^{2}
\end{array}}}}\nonumber \\
 & \qquad-\expect{\lgbrac{\abs a^{2}\rho_{11}^{2}+\abs b^{2}\abs{g_{12}}^{2}}}+\lgbrac{\pi e}\label{eq:innerbound_2x2_simplification_simulation}\\
 & \overset{\brac i}{\geq}\lgbrac{\abs a^{4}\snr^{\gamma_{11}+\gamma_{21}}+\abs a^{2}\abs b^{2}\snr^{\gamma_{12}+\gamma_{21}}+\abs a^{2}\abs b^{2}\snr^{\gamma_{11}+\gamma_{22}}}\nonumber \\
 & \qquad-\lgbrac{\abs a^{2}\snr^{\gamma_{11}}+\abs b^{2}\snr^{\gamma_{12}}}+\lgbrac{\pi e}-2\gamma_{E}\lgbrac e,\label{eq:innerbound_2x2_h(gaq)_part2}
\end{align}
 where $\brac i$ is using Lemma \ref{fact:Jensens_gap} from page
\pageref{fact:Jensens_gap} on $\abs{g_{22}}^{2}$ and $\abs{g_{12}}^{2}$.
Now substituting (\ref{eq:innerbound_2x2_h(gaq)_part2}) and (\ref{eq:innerbound_2x2_h(gaq)_part1})
in (\ref{eq:innerbound_2x2_H(GAQ)_step2}) we get
\begin{align}
h\brac{GX} & \geq h\brac{\abs{ag_{11}+\eta g_{12}}^{2}+\abs{cg_{12}}^{2}}+\expect{\lgbrac{\abs{ag_{11}+\eta g_{12}}^{2}+\abs{cg_{12}}^{2}}}+\lgbrac{\frac{\pi^{T-2}}{\Gamma\brac{T-1}}}\nonumber \\
 & \qquad+\lgbrac{\frac{\pi^{T}}{\Gamma\brac T}}+\brac{T-2}\expect{\lgbrac{\abs{acg_{12}g_{21}-acg_{11}g_{22}}^{2}}}+h\brac{cg_{22}}\nonumber \\
 & \qquad+\lgbrac{\abs a^{4}\snr^{\gamma_{11}+\gamma_{21}}+\abs a^{2}\abs b^{2}\snr^{\gamma_{12}+\gamma_{21}}+\abs a^{2}\abs b^{2}\snr^{\gamma_{11}+\gamma_{22}}}\nonumber \\
 & \qquad-\lgbrac{\abs a^{2}\snr^{\gamma_{11}}+\abs b^{2}\snr^{\gamma_{12}}}+\lgbrac{\pi e}-2\gamma_{E}\lgbrac e.\label{eq:hGX_inner}
\end{align}
 Now we use our choice $\eta\sim\mathcal{CN}\brac{0,\abs b^{2}}$,
$\abs a^{2}=\snr^{-\gamma_{a}},\abs b^{2}=\snr^{-\gamma_{b}},\abs c^{2}=\snr^{-\gamma_{c}},\gamma_{a}\geq0,\gamma_{b}\geq0,\gamma_{c}\geq0$.
We then have
\begin{align}
 & h\brac{\abs{ag_{11}+\eta g_{12}}^{2}+\abs{cg_{12}}^{2}}\nonumber \\
 & \overset{\brac i}{\geq}\max\brac{h\brac{\abs{ag_{11}+\eta g_{12}}^{2}},h\brac{\abs{cg_{12}}^{2}}}\\
 & \geq\max\brac{h\brac{\rline{\abs{ag_{11}+\eta g_{12}}^{2}\vphantom{a^{a^{a^{a}}}}}g_{12}},h\brac{\abs{cg_{12}}^{2}}}\\
 & \overset{\brac{ii}}{\eqdof}\max\brac{\expect{\lgbrac{\snr^{-\gamma_{a}+\gamma_{11}}+\snr{}^{-\gamma_{b}}\abs{g_{12}}^{2}}},\lgbrac{\snr^{-\gamma_{c}+\gamma_{12}}}}\\
 & \eqdof\max\brac{\lgbrac{\snr^{-\gamma_{a}+\gamma_{11}}+\snr^{-\gamma_{b}+\gamma_{12}}},\lgbrac{\snr^{-\gamma_{c}+\gamma_{12}}}},
\end{align}
where $\brac i$ is using the fact that conditioning reduces entropy,
$\brac{ii}$ is using the property of exponential distributions and
$\brac{iii}$ is using Lemma \ref{fact:Jensens_gap}. Hence
\begin{align*}
\underset{\snr\rightarrow\infty}{\text{lim}}h\brac{\abs{ag_{11}+\eta g_{12}}^{2}+\abs{cg_{12}}^{2}}/\lgbrac{\snr} & \geq\max\brac{-\gamma_{a}+\gamma_{11},-\gamma_{b}+\gamma_{12},-\gamma_{c}+\gamma_{12}}.
\end{align*}
Now
\begin{align}
 & \expect{\lgbrac{\abs{ag_{11}+\eta g_{12}}^{2}+\abs{cg_{12}}^{2}}}\nonumber \\
 & \geq\max\brac{\expect{\lgbrac{\abs{ag_{11}+\eta g_{12}}^{2}}},\expect{\lgbrac{\abs{cg_{12}}^{2}}}}\\
 & \overset{\brac i}{\eqdof}\max\brac{\lgbrac{\snr^{-\gamma_{a}+\gamma_{11}}+\snr^{\gamma_{b}+\gamma_{12}}},\lgbrac{\snr^{-\gamma_{c}+\gamma_{12}}}},
\end{align}
where $\brac i$ is using Lemma \ref{fact:Jensens_gap}. Also
\begin{align}
\expect{\lgbrac{\abs{acg_{12}g_{21}-acg_{11}g_{22}}^{2}}} & \eqdof\lgbrac{\snr^{-\gamma_{a}-\gamma_{c}+\gamma_{12}+\gamma_{21}}+\snr^{-\gamma_{a}-\gamma_{c}+\gamma_{11}+\gamma_{22}}}
\end{align}
using Lemma \ref{fact:Jensens_gap} repeatedly. Similarly evaluating
other terms in \ref{eq:hGX_inner}, we get
\begin{align}
 & \underset{\snr\rightarrow\infty}{\text{lim}}\frac{h\brac{GX}}{\lgbrac{\snr}}\nonumber \\
 & \geq2\max\brac{-\gamma_{a}+\gamma_{11},-\gamma_{b}+\gamma_{12},-\gamma_{c}+\gamma_{12}}\nonumber \\
 & \quad+\brac{T-2}\brac{-\gamma_{a}-\gamma_{c}+\max\brac{\gamma_{12}+\gamma_{21},\gamma_{11}+\gamma_{22}}}-\gamma_{c}+\gamma_{22}\nonumber \\
 & \quad+\max\brac{-2\gamma_{a}+\gamma_{11}+\gamma_{21},-\gamma_{a}-\gamma_{b}+\gamma_{12}+\gamma_{21},-\gamma_{a}-\gamma_{b}+\gamma_{11}+\gamma_{22}}\nonumber \\
 & \quad-\max\brac{-\gamma_{a}+\gamma_{11},-\gamma_{b}+\gamma_{12}}.\label{eq:dof_2x2ach1}
\end{align}
Also, using (\ref{eq:h(Y|X)}), (\ref{eq:h(Y(n)|X)}) we have
\begin{align}
 & h\brac{Y|X}\nonumber \\
 & =\expect{\lgbrac{\abs a^{2}\rho_{11}^{2}+\abs{\eta}^{2}\rho_{12}^{2}+\abs c^{2}\rho_{12}^{2}+\abs a^{2}\abs c^{2}\rho_{11}^{2}\rho_{12}^{2}+1}}\nonumber \\
 & \quad+\expect{\lgbrac{\abs a^{2}\rho_{21}^{2}+\abs{\eta}^{2}\rho_{22}^{2}+\abs c^{2}\rho_{22}^{2}+\abs a^{2}\abs c^{2}\rho_{21}^{2}\rho_{22}^{2}+1}}\nonumber \\
 & \quad+2T\lgbrac{\pi e}.
\end{align}
Now since $\eta\sim\mathcal{CN}\brac{0,\abs b^{2}}$ and $\abs a^{2}=\snr^{-\gamma_{a}},\abs b^{2}=\snr^{-\gamma_{b}},\abs c^{2}=\snr^{-\gamma_{c}}$
and Lemma \ref{fact:Jensens_gap}, we get
\begin{align}
 & h\brac{Y|X}\nonumber \\
 & \eqdof\lgbrac{\abs a^{2}\rho_{11}^{2}+\abs b^{2}\rho_{12}^{2}+\abs c^{2}\rho_{12}^{2}+\abs a^{2}\abs c^{2}\rho_{11}^{2}\rho_{12}^{2}+1}\nonumber \\
 & \quad+\lgbrac{\abs a^{2}\rho_{21}^{2}+\abs b^{2}\rho_{22}^{2}+\abs c^{2}\rho_{22}^{2}+\abs a^{2}\abs c^{2}\rho_{21}^{2}\rho_{22}^{2}+1}
\end{align}
and hence
\begin{align}
 & \underset{\snr\rightarrow\infty}{\text{lim}}\frac{h\brac{Y|X}}{\lgbrac{\snr}}\nonumber \\
 & \eqdof\max\brac{-\gamma_{a}+\gamma_{11},-\gamma_{b}+\gamma_{12},-\gamma_{c}+\gamma_{12},-\gamma_{a}-\gamma_{c}+\gamma_{11}+\gamma_{12},0}\nonumber \\
 & \quad+\max\brac{-\gamma_{a}+\gamma_{21},-\gamma_{b}+\gamma_{22},-\gamma_{c}+\gamma_{22},-\gamma_{a}-\gamma_{c}+\gamma_{21}+\gamma_{22},0}.\label{eq:dof_2x2ach2}
\end{align}
Using (\ref{eq:dof_2x2ach1}), (\ref{eq:dof_2x2ach2}) with $\gamma_{a}=0,\gamma_{c}=0,\gamma_{b}=0$,
$\gamma_{11}=\gamma_{22}=\gamma_{D}>\gamma_{CL}=\gamma_{12}=\gamma_{21}$
we get
\begin{equation}
\underset{\snr\rightarrow\infty}{\text{lim}}\frac{h\brac{GX}}{\lgbrac{\snr}}\geq2T\gamma_{D},
\end{equation}
and
\begin{equation}
\underset{\snr\rightarrow\infty}{\text{lim}}\frac{h\brac{Y|X}}{\lgbrac{\snr}}=2\brac{\gamma_{D}+\gamma_{CL}}.
\end{equation}
 Hence we have
\begin{equation}
\underset{\snr\rightarrow\infty}{\text{lim}}\frac{1}{T}\frac{I\brac{X;Y}}{\lgbrac{\snr}}\geq2\brac{\brac{1-\frac{1}{T}}\gamma_{D}-\frac{1}{T}\gamma_{CL}}
\end{equation}
 achievable. Also with $\gamma_{a}=0,\gamma_{c}=\gamma_{CL},\gamma_{b}=0$,
$\gamma_{11}=\gamma_{22}=\gamma_{D}>\gamma_{CL}=\gamma_{12}=\gamma_{21}$
in (\ref{eq:dof_2x2ach1}), (\ref{eq:dof_2x2ach2}) we get
\begin{equation}
\underset{\snr\rightarrow\infty}{\text{lim}}\frac{h\brac{GX}}{\lgbrac{\snr}}\geq2\gamma_{D}+\brac{T-1}\brac{2\gamma_{D}-\gamma_{CL}}
\end{equation}
 and
\begin{equation}
\underset{\snr\rightarrow\infty}{\text{lim}}\frac{h\brac{Y|X}}{\lgbrac{\snr}}=2\gamma_{D}.
\end{equation}
Hence for $T=2$
\begin{equation}
\underset{\snr\rightarrow\infty}{\text{lim}}\frac{1}{2}\frac{I\brac{X;Y}}{\lgbrac{\snr}}\geq\brac{\gamma_{D}-\frac{1}{2}\gamma_{CL}}
\end{equation}
is achievable. Hence the outer bounds for all regimes of $T$ from
Table \ref{tab:Sol_outerbd_2x2_sym_mimo} are achievable.

\section{Gaussian codebooks for $M\times M$ MIMO channels\label{app:Gaussian-code-noncohmimo}}

Here we prove Theorem \ref{thm:gaussian_codebooks} for an $M\times M$
MIMO channel (Figure \ref{fig:MIMO-with-direct-and-cross}) with coherence
time $T>M$ and with exponents $\gamma_{D}$ in the direct links and
$\gamma_{CL}$ in the crosslinks ($\gamma_{D}>\gamma_{CL}$). We consider
i.i.d. Gaussian codebooks across antennas and time periods and prove
that a gDoF of $M\brac{\brac{1-\frac{1}{T}}\gamma_{D}-\frac{M-1}{T}\gamma_{CL}}$
is achievable. Using Gaussian codebooks, the rate $R\geq I\brac{GX+W;X}$
is achievable, where
\begin{equation}
X=\left[\begin{array}{ccc}
X_{11} & \cdots & X_{1T}\\
\vdots &  & \vdots\\
X_{M1} & \cdots & X_{MT}
\end{array}\right]=\sbrac{\begin{array}{ccc}
\overline{X_{1}} & \ldots & \overline{X_{T}}\end{array}},\label{eq:Xstructure}
\end{equation}

\begin{equation}
\overline{X_{i}}=\tran\left[\begin{array}{ccc}
X_{1i} & \ldots & X_{Mi}\end{array}\right]
\end{equation}
 with all of the elements of the $M\times T$ matrix $X$ being i.i.d.
$\mathcal{CN}\brac{0,1}$ and $W$ being an $M\times T$ matrix with
i.i.d. $\mathcal{CN}\brac{0,1}$ noise elements. The channel matrix
\begin{equation}
G=\left[\begin{array}{cccc}
g_{11} & g_{12} & . & g_{1M}\\
g_{21} & g_{22} & . & .\\
. & . & . & .\\
g_{M1} & . & . & g_{MM}
\end{array}\right]\label{eq:Gstructure}
\end{equation}
 has independent elements with $g_{ii}\sim\mathcal{CN}\brac{0,\snr^{\gamma_{D}}}$
and rest of the elements distributed according to $\mathcal{CN}\brac{0,\snr^{\gamma_{CL}}}$.
We will show that the mutual information satisfies
\begin{equation}
I\brac{GX+W;X}\geqdof M\brac{\brac{T-1}\gamma_{D}-\brac{M-1}\gamma_{CL}}\lgbrac{\snr}.
\end{equation}
We have
\begin{align}
I\brac{GX+W;X} & =h\brac{GX+W}-h\brac{\rline{GX+W\vphantom{a^{a^{a^{a}}}}}X}\\
 & \geq h\brac{\rline{GX+W\vphantom{a^{a^{a^{a}}}}}G}\nonumber \\
 & \qquad-h\brac{\rline{GX+W\vphantom{a^{a^{a^{a}}}}}X}.
\end{align}
Now
\begin{align}
 & h\brac{\rline{GX+W\vphantom{a^{a^{a^{a}}}}}G}\nonumber \\
 & \overset{\brac i}{\geq}h\brac{\rline{GX}G,W}\\
 & =h\brac{\rline{GX}G}\\
 & \overset{\brac{ii}}{=}T\times h\brac{\rline{G\overline{X_{1}}}G}\\
 & \overset{\brac{iii}}{=}T\expect{\lgbrac{\abs{\det\brac{\pi eG}}^{2}}}\\
 & \overset{\brac{iv}}{\eqdof}TM\gamma_{D}\lgbrac{\snr},
\end{align}
where $\brac i$ is using the fact that  conditioning reduces entropy
and conditioning on $W$, $\brac{ii}$ is using the structure of $X$
from (\ref{eq:Xstructure}) and the fact that elements $X_{ij}$ are
i.i.d. Gaussian, $\brac{iii}$ is again using the fact that $X_{ij}$
are i.i.d. Gaussian and $\brac{iv}$ is by repeated application of
Lemma \vref{fact:Jensens_gap}, Tower property of expectation on Gaussian
distributed $g_{ij}$ and the structure of the determinant involved.
Now we will show that
\begin{equation}
h\brac{\rline{GX+W\vphantom{a^{a^{a^{a}}}}}X}\leqdof M\brac{\gamma_{D}\lgbrac{\snr}+\brac{M-1}\gamma_{CL}\lgbrac{\snr}}
\end{equation}
and will complete the proof.
\begin{align}
 & h\brac{\rline{GX+W\vphantom{a^{a^{a^{a}}}}}X}\nonumber \\
 & \overset{\brac i}{\leqdof}\sum_{i}h\brac{\rline{\left[\begin{array}{cccc}
g_{i1} & g_{i2} & . & g_{iM}\end{array}\right]X+\underline{W_{i}}}X}\\
 & \overset{\brac{ii}}{=}Mh\brac{\rline{\left[\begin{array}{cccc}
g_{11} & g_{12} & . & g_{1M}\end{array}\right]X+\underline{W_{1}}}X},
\end{align}
where $\brac i$ is using the fact that  conditioning reduces entropy
and $\underline{W_{i}}$ is a $1\times T$ vector with i.i.d. $\mathcal{CN}\brac{0,1}$
elements, $\brac{ii}$ is by symmetry of the channel with $g_{ii}\sim\mathcal{CN}\brac{0,\snr^{\gamma_{D}}}$
and rest of the $g_{ij}$ distributed according to $\mathcal{CN}\brac{0,\snr^{\gamma_{CL}}}$
and the i.i.d. nature of $X_{ij}$. Now we will show that
\begin{equation}
h\brac{\rline{\left[\begin{array}{cccc}
g_{11} & g_{12} & . & g_{1M}\end{array}\right]X+\underline{W_{1}}}X}\leqdof\gamma_{D}\lgbrac{\snr}+\brac{M-1}\gamma_{CL}\lgbrac{\snr}
\end{equation}
and will complete the proof. Let us denote $\sbrac{\begin{array}{cccc}
w_{11} & w_{12} & . & w_{1T}\end{array}}=\underline{W_{1}}$, $\underline{g_{1}}=\left[\begin{array}{cccc}
g_{11} & g_{12} & . & g_{1M}\end{array}\right]$. We have
\begin{align}
 & h\brac{\rline{\underline{g_{1}}X+\underline{W_{1}}}X}\nonumber \\
 & \leq h\brac{\rline{\underline{g_{1}}\overline{X_{1}}+w_{11}}X}\nonumber \\
 & \quad+\sum_{i=2}^{M}h\brac{\rline{\underline{g_{1}}\overline{X_{i}}+w_{1i}}X,\underline{g_{1}}\overline{X_{1}}+w_{11}}\nonumber \\
 & \quad+\sum_{i=M+1}^{T}h\brac{\rline{\underline{g_{1}}\overline{X_{i}}+w_{1i}}X,\cbrac{\underline{g_{1}}\overline{X_{k}}+w_{1k}}_{k=1}^{M}}.\label{eq:mimo_train1}
\end{align}
Now for the first term in (\ref{eq:mimo_train1}), we have
\begin{align}
h\brac{\rline{\underline{g_{1}}\overline{X_{1}}+w_{11}}X} & =h\brac{\rline{\sum_{j=1}^{M}g_{j1}X_{j1}+w_{11}}X_{j1}}\nonumber \\
 & =\expect{\lgbrac{\pi e\brac{1+\abs{X_{j1}}^{2}\brac{\rho_{D}^{2}+\brac{M-1}\rho_{CL}^{2}}}}}\\
 & \leqdof\gamma_{D}\lgbrac{\snr}\label{eq:mimo_train2}
\end{align}
using Lemma \ref{fact:Jensens_gap} and since $\gamma_{D}\geq\gamma_{CL}$.
Now consider the second term \sloppy in (\ref{eq:mimo_train1}),
$h\brac{\rline{\underline{g_{1}}\overline{X_{i}}+w_{1i}}X,\underline{g_{1}}\overline{X_{1}}+w_{11}}.$
In $\underline{g_{1}}=\left[\begin{array}{cccc}
g_{11} & g_{12} & . & g_{1M}\end{array}\right]$, only $g_{11}$ has $\snr$ exponent $\gamma_{D}$ and it can be
removed due to the conditioning as follows:
\begin{align}
 & h\brac{\rline{\underline{g_{1}}\overline{X_{i}}+w_{1i}}X,\underline{g_{1}}\overline{X_{1}}+w_{11}}\nonumber \\
 & \overset{\brac i}{\leq}h\brac{\rline{\underline{g_{1}}\left[\begin{array}{c}
0\\
X_{11}X_{2i}-X_{1i}X_{21}\\
.\\
X_{11}X_{Mi}-X_{1i}X_{M1}
\end{array}\right]+X_{11}w_{1i}-X_{1i}w_{11}}X}-\expect{\lgbrac{\abs{X_{11}}}}\\
 & =\expect{\lgbrac{\pi e\brac{\rho_{CL}^{2}\sum_{j=2}^{M}\abs{X_{11}X_{ji}-X_{1i}X_{j1}}^{2}+\abs{X_{11}}^{2}+\abs{X_{1i}}^{2}}}}-\expect{\lgbrac{\abs{X_{11}}}}\\
 & \overset{\brac{ii}}{\eqdof}\gamma_{CL}\lgbrac{\snr},\label{eq:mimo_train3}
\end{align}
where $\brac i$ is by multiplying $\underline{g_{1}}\overline{X_{i}}+w_{1i}$
with $X_{11}$ and subtracting $X_{1i}\brac{\underline{g_{1}}\overline{X_{1}}+w_{11}}$
from it and using the fact that  conditioning reduces entropy, and
$\brac{ii}$ is by repeated application of Lemma  \ref{fact:Jensens_gap}
and Tower property of expectation on Gaussian distributed $X_{ij}$.

Now consider the last term in (\ref{eq:mimo_train1})
\[
h\brac{\rline{\underline{g_{1}}\overline{X_{i}}+w_{1i}\vphantom{a^{a^{a^{a^{a}}}}}}X,\cbrac{\underline{g_{1}}\overline{X_{k}}+w_{1k}}_{k=1}^{M}}.
\]
 This term would not have any gDoF since all the $\snr$ exponents
from $\underline{g_{1}}=\left[\begin{array}{cccc}
g_{11} & g_{12} & . & g_{1M}\end{array}\right]$ can be canceled due to availability of $M$ linear equations in the
conditioning. Let
\[
X_{M\times M}=\left[\begin{array}{ccc}
X_{11} & \ldots & X_{1M}\\
\vdots &  & \vdots\\
X_{M1} & \ldots & X_{MM}
\end{array}\right],\ \underline{w_{1}}=\left[\begin{array}{ccc}
w_{11} & \ldots & w_{1M}\end{array}\right].
\]
In the conditioning, $\underline{g_{1}}X_{M\times M}+\underline{w_{1}}$
and $X_{M\times M}$ are available. Let $\text{Adj}\brac{X_{M\times M}}$
be the adjoint of $X_{M\times M}$ and $\text{det}\brac{X_{M\times M}}$
be the determinant of $X_{M\times M}$. Hence the term $\underline{g_{1}}\text{det}\brac{X_{M\times M}}\overline{X_{i}}+\underline{w_{1}}\text{Adj}\brac{X_{M\times M}}\overline{X_{i}}$
is available in the conditioning. The $M$ linear equations in the
conditioning can cancel off the gDoF contribution from $\underline{g_{1}}=\left[\begin{array}{cccc}
g_{11} & g_{12} & . & g_{1M}\end{array}\right]$ only if $\text{det}\brac{X_{M\times M}}$ is non-zero. Since $X$
is Gaussian i.i.d., this is true almost surely. We handle this more
precisely in the following steps:
\begin{align}
 & h\brac{\rline{\underline{g_{1}}\overline{X_{i}}+w_{1i}\vphantom{a^{a^{a^{a^{a}}}}}}X,\cbrac{\underline{g_{1}}\overline{X_{k}}+w_{1k}}_{k=1}^{M}}\\
 & \overset{\brac i}{\leq}h\brac{\rline{\underline{g_{1}}\overline{X_{i}}+w_{1i}\vphantom{a^{a^{a^{a^{a}}}}}}X,\underline{g_{1}}\text{det}\brac{X_{M\times M}}\overline{X_{i}}+\underline{w_{1}}\text{Adj}\brac{X_{M\times M}}\overline{X_{i}}}\\
 & \overset{\brac{ii}}{=}h\brac{\rline{\underline{g_{1}}\text{det}\brac{X_{M\times M}}\overline{X_{i}}+\text{det}\brac{X_{M\times M}}w_{1i}\vphantom{a^{a^{a^{a^{a}}}}}}X,\underline{g_{1}}\text{det}\brac{X_{M\times M}}\overline{X_{i}}+\underline{w_{1}}\text{Adj}\brac{X_{M\times M}}\overline{X_{i}}}\nonumber \\
 & \qquad-\expect{\lgbrac{\abs{\text{det}\brac{X_{M\times M}}}}}\\
 & \overset{\brac{iii}}{\leq}h\brac{\rline{w_{1i}\text{det}\brac{X_{M\times M}}-\underline{w_{1}}\text{Adj}\brac{X_{M\times M}}\overline{X_{i}}\vphantom{\vphantom{a^{a^{a^{a^{a}}}}}}}X}\nonumber \\
 & \qquad-\expect{\lgbrac{\text{det}\brac{X_{M\times M}}}}\\
 & \overset{\brac{iv}}{\leqdof}h\brac{\rline{w_{1i}\text{det}\brac{X_{M\times M}}-\underline{w_{1}}\text{Adj}\brac{X_{M\times M}}\overline{X_{i}}\vphantom{a^{a^{a^{a^{a}}}}}}X}\\
 & \overset{\brac v}{=}\lgbrac{\expect{\abs{\text{det}\brac{X_{M\times M}}}^{2}+\left\Vert \text{Adj}\brac{X_{M\times M}}\overline{X_{i}}\right\Vert ^{2}}}+\lgbrac{\pi e}\\
 & \overset{\brac{vi}}{\leqdof}0,\label{eq:mimo_train4}
\end{align}
where $\brac i$ is using the availability of $\underline{g_{1}}\text{det}\brac{X_{M\times M}}\overline{X_{i}}+\underline{w_{1}}\text{Adj}\brac{X_{M\times M}}\overline{X_{i}}$
in conditioning and using the fact that conditioning reduces entropy,
$\brac{ii}$ is by multiplying with $\text{det}\brac{X_{M\times M}}$
and compensating with $-\expect{\lgbrac{\abs{\text{det}\brac{X_{M\times M}}}}}$
since $\text{det}\brac{X_{M\times M}}$ is known from the values in
conditioning, $\brac{iii}$ is by subtracting the term available from
conditioning and using the fact that conditioning reduces entropy,
$\brac{iv}$ is because $\expect{\lgbrac{\abs{\text{det}\brac{X_{M\times M}}}}}$
is finite by repeated application of Lemma \ref{fact:Jensens_gap}
and the Tower property of expectation on Gaussian distributed $X_{ij}$,
$\brac v$ is because $w_{1k}\sim\mathcal{CN}\brac{0,1}$ i.i.d. and
$i>M$ and $\brac{vi}$ is because $X_{ij}\sim\mathcal{CN}\brac{0,1}$
i.i.d.

Now by substituting (\ref{eq:mimo_train4}), (\ref{eq:mimo_train3})
and (\ref{eq:mimo_train2}) in (\ref{eq:mimo_train1}), we get the
desired result.

\section{Outer bound for the MISO channel with $T<M$ \label{app:MISO_T<M}}

Here we prove the gDoF outer bound given in Theorem \ref{thm:MISO_dof}
for the $M\times1$ MISO system with $1<T<M$. The steps follow similar
to the case with $T\geq M$, given in Section \ref{subsec:MISO}.
We have the structure of input distribution as $X=LQ$ with
\begin{equation}
L=\left[\begin{array}{cccc}
x_{11} & 0 & 0\\
. & . & 0 & 0\\
. & . & . & 0\\
. & . & . & x_{TT}\\
. & . & . & .\\
x_{M1} & . & . & x_{MT}
\end{array}\right].
\end{equation}
For the channel we have, $G=\left[\begin{array}{cccc}
g_{11} & . & . & g_{1M}\end{array}\right]$, $g_{1i}\sim\mathcal{CN}\brac{0,\rho_{1i}^{2}}$, $\rho_{1i}^{2}=\snr^{\gamma_{1i}}$,
$Y=GX+W$, where $W$ is a $1\times T$ vector with i.i.d. $\mathcal{CN}\brac{0,1}$
components. We assume $\rho_{11}^{2}\geq\rho_{1i}^{2}$ without loss
of generality. Now note that $WQ$ has the same distribution as $W$
and is independent of $Q$ (using the fact that $W$ is isotropically
distributed). Hence
\begin{align}
Y & =\brac{GL+W}Q\\
 & =\left[\begin{array}{ccccc}
\brac{w_{11}+\sum_{i=1}^{M}x_{i1}g_{1i}} & \brac{w_{12}+\sum_{i=2}^{M}x_{i2}g_{1i}} & . & . & \brac{w_{1T}+\sum_{i=T}^{M}x_{i2}g_{1i}}\end{array}\right]Q.
\end{align}
Now using Lemma \ref{lem:isotropic_entropy_to_radial} from page \pageref{lem:isotropic_entropy_to_radial},
we get
\begin{align}
h\brac Y & =h\brac{\sum_{j=1}^{T}\abs{w_{1j}+\sum_{i=j}^{M}x_{ij}g_{1i}}^{2}}\nonumber \\
 & \qquad+\brac{T-1}\expect{\lgbrac{\sum_{j=1}^{T}\abs{w_{1j}+\sum_{i=j}^{M}x_{ij}g_{1i}}^{2}}}+\lgbrac{\frac{\pi^{T}}{\Gamma\brac T}}\\
 & \overset{\brac i}{\leq}h\brac{\sum_{j=1}^{T}\abs{w_{1j}+\sum_{i=j}^{M}x_{ij}g_{1i}}^{2}}\nonumber \\
 & \qquad+\brac{T-1}\expect{\lgbrac{\sum_{j=1}^{T}\brac{1+\sum_{i=j}^{M}\abs{x_{ij}}^{2}\rho_{ij}^{2}}}}+\lgbrac{\frac{\pi^{T}}{\Gamma\brac T}}\\
 & \overset{\brac{ii}}{\leq}h\brac{\sum_{j=1}^{T}\abs{w_{1j}+\sum_{i=j}^{M}x_{ij}g_{1i}}^{2}}\nonumber \\
 & \qquad+\brac{T-1}\expect{\lgbrac{\sum_{i=1}^{M}\rho_{1i}^{2}\brac{\sum_{j=1}^{\min\brac{i,T}}\abs{x_{ij}}^{2}}+T}}+\lgbrac{\frac{\pi^{T}}{\Gamma\brac T}},
\end{align}
where $\brac i$ is using the Tower property of expectation and Jensen's
inequality and $\brac{ii}$ is using $\sum_{j=1}^{T}\sum_{i=j}^{M}\abs{x_{ij}}^{2}\rho_{1i}^{2}=\sum_{i=1}^{M}\sum_{j=1}^{\min\brac{i,T}}\abs{x_{ij}}^{2}\rho_{1i}^{2}$.
Now using (\ref{eq:h(Y(n)|X)}), we have
\begin{align}
h\brac{Y|X} & =\expect{\lgbrac{\det\brac{L^{\dagger}\text{diag }\brac{\rho_{11}^{2},\ldots,\rho_{1M}^{2}}L+I_{T}}}}\nonumber \\
 & \qquad+\brac T\lgbrac{\pi e}\\
 & =\expect{\lgbrac{\prod_{i=1}^{M}\brac{1+\omega_{i}}}}+T\lgbrac{\pi e},
\end{align}
where $\omega_{i}$ are the eigenvalues of $L^{\dagger}\text{diag }\brac{\rho_{11}^{2},\ldots,\rho_{1M}^{2}}L$.
Hence
\begin{align}
h\brac{Y|X} & =\expect{\lgbrac{\prod_{i=1}^{M}\brac{1+\omega_{i}}}}+T\lgbrac{\pi e}\\
 & \geq\expect{\lgbrac{1+\sum\omega_{i}}}+T\lgbrac{\pi e}.
\end{align}
The last step is true because $\omega_{i}\geq0$. Now
\begin{align}
\sum\omega_{i} & =\text{Trace}\brac{L^{\dagger}\text{diag }\brac{\rho_{11}^{2},\ldots,\rho_{1M}^{2}}L}\nonumber \\
 & =\text{Trace}\brac{\text{diag }\brac{\rho_{11}^{2},\ldots,\rho_{1M}^{2}}LL^{\dagger}}\nonumber \\
 & =\sum_{i=1}^{M}\rho_{1i}^{2}\brac{\sum_{j=1}^{\min\brac{i,T}}\abs{x_{ij}}^{2}}.
\end{align}
Hence
\begin{align}
h\brac{Y|X} & \geq\expect{\lgbrac{1+\sum_{i=1}^{M}\rho_{1i}^{2}\brac{\sum_{j=1}^{\min\brac{i,T}}\abs{x_{ij}}^{2}}}}+T\lgbrac{\pi e}.
\end{align}
Hence
\begin{align}
I\brac{X;Y} & \overset{}{\leq}h\brac{\sum_{j=1}^{M}\abs{w_{1j}+\sum_{i=j}^{M}x_{ij}g_{1i}}^{2}+\sum_{i=M+1}^{T}\abs{w_{1i}}^{2}}\nonumber \\
 & \quad+\brac{T-2}\expect{\lgbrac{\sum_{i=1}^{M}\rho_{1i}^{2}\brac{\sum_{j=1}^{i}\abs{x_{ij}}^{2}}+T}}\nonumber \\
 & \quad+\lgbrac{\frac{\pi^{T}}{\Gamma\brac T}}-T\lgbrac{\pi e}\\
 & \overset{.}{\leq}\brac{T-1}\lgbrac{\sum_{i=1}^{M}\rho_{1i}^{2}MT+T},
\end{align}
where the last step was using Lemma \ref{lem:max_entropy} and Jensen's
inequality. Hence
\begin{align}
\underset{\snr\rightarrow\infty}{\text{limsup}}\ \frac{I\brac{X;Y}}{\lgbrac{\snr}} & \overset{}{\leq}\brac{T-1}\gamma_{11}.
\end{align}

\section{Proof of Lemma \ref{lem:dof_equivalence_abs_lin_comb_gaussian_vector}\label{app:dof_equivalence_squared_norm_gaussiann_vector}}

Here we prove that $h\brac{\abs{ag_{11}+bg_{12}+w_{11}}^{2}+\abs{cg_{12}+w_{12}}^{2}+\sum_{i=3}^{T}\abs{w_{1i}}^{2}}$
and
\[
\expect{\lgbrac{\abs a^{2}\rho_{11}^{2}+\brac{\abs b^{2}+\abs c^{2}}\rho_{12}^{2}+1}}
\]
have the same gDoF. For this, consider the point to point channel
\begin{equation}
\mathcal{C}_{1}:V=\abs{ag_{11}+bg_{12}+w_{11}}^{2}+\abs{cg_{12}+w_{12}}^{2}+\sum_{i=3}^{T}\abs{w_{1i}}^{2}
\end{equation}
with inputs $a,b,c$ and power constraint $T$. Its capacity is given
by
\begin{alignat}{1}
C_{1} & =\max_{p\brac{a,b,c};\expect{\abs a^{2}+\abs b^{2}+\abs c^{2}}\leq T}\left\{ \vphantom{a^{a^{a^{a^{a^{a^{a}}}}}}}h\brac{\brac{\abs{ag_{11}+bg_{12}+w_{11}}^{2}+\abs{cg_{12}+w_{12}}^{2}+\sum_{i=3}^{T}\abs{w_{1i}}^{2}}}\right.\nonumber \\
 & \qquad\qquad\qquad\qquad\qquad\left.-h\brac{\rline{\abs{ag_{11}+bg_{12}+w_{11}}^{2}+\abs{cg_{12}+w_{12}}^{2}+\sum_{i=3}^{T}\abs{w_{1i}}^{2}}a,b,c}\vphantom{\vphantom{a^{a^{a^{a^{a^{a^{a}}}}}}}}\right\} .
\end{alignat}
From \cite[(32)]{lapidoth2003capacity} we have
\begin{align}
I\brac{U;V} & \leq\expect{\lgbrac V}-h\brac{\rline VU}+\lgbrac{\Gamma\brac{\alpha}}\nonumber \\
 & \qquad+\alpha\brac{1+\lgbrac{\expect V}-\expect{\lgbrac V}}-\alpha\lgbrac{\alpha}\label{eq:lapidoth_outer_squarednormgaussian}
\end{align}
for any $\alpha>0$ for channels whose output $V$ takes values in
$\mathbb{R}^{+}$. We will use this result to bound $I\brac{U;V}$
for any input distribution $p\brac{a,b,c};\expect{\abs a^{2}+\abs b^{2}+\abs c^{2}}\leq T$
for the channel $\mathcal{C}_{1}$ with $U=\brac{a,b,c}$ as input.
Now
\begin{alignat}{1}
 & h\brac{\rline VU}\nonumber \\
 & =h\brac{\rline{\abs{ag_{11}+bg_{12}+w_{11}}^{2}+\abs{cg_{12}+w_{12}}^{2}+\sum_{i=3}^{T}\abs{w_{1i}}^{2}}a,b,c}\\
 & \overset{\brac i}{\leq}\expect{\lgbrac{e\expect{\rline{\abs{ag_{11}+bg_{12}+w_{11}}^{2}+\abs{cg_{12}+w_{12}}^{2}+\sum_{i=3}^{T}\abs{w_{1i}}^{2}}a,b,c}}}\\
 & \overset{\brac{ii}}{=}\expect{\lgbrac{e\brac{\brac{\rho_{11}^{2}\abs a^{2}+\rho_{12}^{2}\abs b^{2}+1}+\brac{\rho_{12}^{2}\abs c^{2}+1}+\brac{T-2}}}}\\
 & =\expect{\lgbrac{\rho_{11}^{2}\abs a^{2}+\rho_{12}^{2}\abs b^{2}+\rho_{12}^{2}\abs c^{2}+T}}+\lgbrac e,\label{eq:loc1_lapidoth_squared_norm}
\end{alignat}
where $\brac i$ was using the definition of conditional entropy and
Lemma \ref{lem:max_entropy}, $\brac{ii}$ was using the fact that
given $\brac{a,b,c}$, $ag_{11}+bg_{12}+w_{11},cg_{12}+w_{12}$ are
sums of independent Gaussians. Note that
\begin{alignat}{1}
\expect{\lgbrac V} & =\expect{\expect{\lgbrac{\rline{\abs{ag_{11}+bg_{12}+w_{11}}^{2}+\abs{cg_{12}+w_{12}}^{2}+\sum_{i=3}^{T}\abs{w_{1i}}^{2}}a,b,c}}}\nonumber \\
 & \leq\expect{\lgbrac{\rho_{11}^{2}\abs a^{2}+\rho_{12}^{2}\abs b^{2}+\rho_{12}^{2}\abs c^{2}+T}}\label{eq:loc2_lapidoth_squared_norm}
\end{alignat}
using Jensen's inequality. Also
\begin{alignat}{1}
\expect{\lgbrac V} & =\expect{\expect{\lgbrac{\rline{\abs{ag_{11}+bg_{12}+w_{11}}^{2}+\abs{cg_{12}+w_{12}}^{2}+\sum_{i=3}^{T}\abs{w_{1i}}^{2}}a,b,c}}}\nonumber \\
 & \geq\expect{\lgbrac{\rho_{11}^{2}\abs a^{2}+\rho_{12}^{2}\abs b^{2}+\rho_{12}^{2}\abs c^{2}+T}}-3\gamma_{E}\lgbrac e\label{eq:loc3_lapidoth_squared_norm}
\end{alignat}
by using Lemma \ref{fact:Jensens_gap} on page \pageref{fact:Jensens_gap}
for exponentially distributed $\abs{ag_{rd2}+bg_{rd1}+w_{d1}}^{2},$
$\abs{cg_{rd1}+w_{d2}}^{2}$ (for given $a,b,c$) and Lemma \ref{fact:Jensens_gap_chi_squared}
for chi-squared distributed $\sum_{i=3}^{T}\abs{w_{di}}^{2}$.
\begin{lem}
The term $\expect{\lgbrac V}-h\brac{\rline VU}$ is upper bounded
by $\lgbrac{3+T}+\frac{7}{2}\lgbrac e$ independent of $\snr$.\label{claim:new_squared_channel_1}
\end{lem}
\begin{IEEEproof}
It suffices to show that for any constant $\brac{a',b',c'}$, $\expect{\rline{\lgbrac V\vphantom{a^{a^{a^{a^{a^{a}}}}}}}U=\brac{a',b',c'}}-h\brac{\rline{V\vphantom{a^{a^{a^{a^{a^{a}}}}}}}U=\brac{a',b',c'}}\leq\lgbrac{3+T}+\frac{7}{2}\lgbrac e$
independent of $a',b',c'$ and $\snr$.
\[
\expect{\rline{\lgbrac V\vphantom{a^{a^{a^{a^{a^{a}}}}}}}U=\brac{a',b',c'}}-h\brac{\rline{V\vphantom{a^{a^{a^{a^{a^{a}}}}}}}U=\brac{a',b',c'}}\qquad\qquad\qquad\qquad\qquad
\]
\begin{alignat}{1}
 & =\expect{\lgbrac{\abs{a'g_{11}+b'g_{12}+w_{11}}^{2}+\abs{c'g_{12}+w_{12}}^{2}+\sum_{i=3}^{T}\abs{w_{1i}}^{2}}}\nonumber \\
 & \qquad-h\brac{\abs{a'g_{11}+b'g_{12}+w_{11}}^{2}+\abs{c'g_{12}+w_{12}}^{2}+\sum_{i=3}^{T}\abs{w_{1i}}^{2}}\\
 & \overset{\brac i}{\leq}\lgbrac{\expect{\abs{a'g_{11}+b'g_{12}+w_{11}}^{2}+\abs{c'g_{12}+w_{12}}^{2}+\sum_{i=3}^{T}\abs{w_{1i}}^{2}}}\nonumber \\
 & \qquad-h\brac{\abs{a'g_{11}+b'g_{12}+w_{11}}^{2}+\abs{c'g_{12}+w_{12}}^{2}}\\
 & \overset{\brac{ii}}{=}\lgbrac{\rho_{11}^{2}\abs{a'}^{2}+\rho_{12}^{2}\abs{b'}^{2}+\rho_{12}^{2}\abs{c'}^{2}+T}\nonumber \\
 & \qquad-h\brac{\abs{a'\rho_{11}\eta_{11}+b'\rho_{12}\eta_{12}+w_{11}}^{2}+\abs{c'\rho_{12}\eta_{12}+w_{12}}^{2}},
\end{alignat}
where $\brac i$ is using Jensen's inequality and the fact that  conditioning
reduces entropy to remove $\sum_{i=3}^{T}\abs{w_{di}}^{2}$ in the
negative term; $\brac{ii}$ is using the fact that $a'g_{11}+b'g_{12}+w_{11},c'g_{12}+w_{12}$
are sums of independent Gaussians. We also introduced $\eta_{ij}\sim\mathcal{CN}\brac{0,1}$
in $\brac{ii}$ so that $g_{ij}=\rho_{ij}\eta_{ij}$.

Consider the case when $1\leq\max\brac{\abs{a'\rho_{11}},\abs{b'\rho_{12}},\abs{c'\rho_{12}}}$.
Assume $1\leq\abs{b'\rho_{12}}=\max\brac{\abs{a'\rho_{11}},\abs{b'\rho_{12}},\abs{c'\rho_{12}}}$
\[
\expect{\rline{\lgbrac V\vphantom{a^{a^{a^{a^{a^{a}}}}}}}U=\brac{a',b',c'}}-h\brac{\rline{V\vphantom{a^{a^{a^{a^{a^{a}}}}}}}U=\brac{a',b',c'}}\qquad\qquad\qquad\qquad\qquad
\]
\begin{alignat}{1}
 & \leq\lgbrac{\rho_{11}^{2}\abs{a'}^{2}+\rho_{12}^{2}\abs{b'}^{2}+\rho_{12}^{2}\abs{c'}^{2}+T}-\lgbrac{\rho_{12}^{2}\abs{b'}^{2}}\nonumber \\
 & \qquad-h\brac{\abs{\frac{a'\rho_{11}}{b'\rho_{12}}\eta_{11}+\eta_{12}+\frac{w_{11}}{b'\rho_{12}}}^{2}+\abs{\frac{c'}{b'}\eta_{12}+\frac{w_{12}}{b'\rho_{12}}}^{2}}.
\end{alignat}
Now using the result from \ifarxiv  Appendix \ref{app:entropybound_sum_exponentials} \else  \cite[Appendix K]{Joyson_2x2_mimov5} \fi
to lower bound the entropy of sum of norm-squared of Gaussian vectors,
we have
\begin{align}
 & h\brac{\abs{\frac{a'\rho_{11}}{b'\rho_{12}}\eta_{11}+\eta_{12}+\frac{w_{11}}{b'\rho_{12}}}^{2}+\abs{\frac{c'}{b'}\eta_{12}+\frac{w_{12}}{b'\rho_{12}}}^{2}}\nonumber \\
 & \geq h\brac{\rline{\abs{\frac{a'\rho_{11}}{b'\rho_{12}}\eta_{11}+\eta_{12}+\frac{w_{11}}{b'\rho_{12}}}^{2}+\abs{\frac{c'}{b'}\eta_{12}+\frac{w_{12}}{b'\rho_{12}}}^{2}}\eta_{11}}\nonumber \\
 & \geq-\frac{7}{2}\lgbrac e.
\end{align}
Hence we get
\[
\expect{\rline{\lgbrac V\vphantom{a^{a^{a^{a^{a^{a}}}}}}}U=\brac{a',b',c'}}-h\brac{\rline{V\vphantom{a^{a^{a^{a^{a^{a}}}}}}}U=\brac{a',b',c'}}\qquad\qquad\qquad\qquad\qquad
\]
\begin{alignat}{1}
 & \overset{}{\leq}\lgbrac{\rho_{11}^{2}\abs{a'}^{2}+\rho_{12}^{2}\abs{b'}^{2}+\rho_{12}^{2}\abs{c'}^{2}+T}-\lgbrac{\rho_{12}^{2}\abs{b'}^{2}}+\frac{7}{2}\lgbrac e\\
 & \overset{\brac i}{\leq}\lgbrac{3+T}+\frac{7}{2}\lgbrac e,
\end{alignat}
where in step $\brac i$ we used $1\leq\abs{b'\rho_{12}}\leq\max\brac{\abs{a'\rho_{11}},\abs{b'\rho_{12}},\abs{c'\rho_{12}}}$.

Similarly for other cases $1\leq\abs{a'\rho_{11}}=\max\brac{\abs{a'\rho_{11}},\abs{b'\rho_{12}},\abs{c'\rho_{12}}}$
and $1\leq\abs{c'\rho_{12}}=\max\brac{\abs{a'\rho_{11}},\abs{b'\rho_{12}},\abs{c'\rho_{12}}}$,
we can show that $\expect{\rline{\lgbrac V\vphantom{a^{a^{a^{a^{a^{a}}}}}}}U=\brac{a',b',c'}}-h\brac{\rline{V\vphantom{a^{a^{a^{a^{a^{a}}}}}}}U=\brac{a',b',c'}}$
is upper bounded by $\lgbrac{3+T}+\frac{7}{2}\lgbrac e$.

Now if $1>\max\brac{\abs{a'\rho_{11}},\abs{b'\rho_{12}},\abs{c'\rho_{12}}}$
\[
\expect{\rline{\lgbrac V\vphantom{a^{a^{a^{a^{a^{a}}}}}}}U=\brac{a',b',c'}}-h\brac{\rline{V\vphantom{a^{a^{a^{a^{a^{a}}}}}}}U=\brac{a',b',c'}}\qquad\qquad\qquad\qquad\qquad\qquad\qquad\qquad\qquad\qquad\qquad\qquad
\]
\begin{alignat}{1}
 & \leq\lgbrac{\rho_{11}^{2}\abs{a'}^{2}+\rho_{12}^{2}\abs{b'}^{2}+\rho_{12}^{2}\abs{c'}^{2}+T}\nonumber \\
 & \qquad-h\brac{\abs{a'\rho_{11}\eta_{11}+b'\rho_{12}\eta_{12}+w_{11}}^{2}+\abs{c'\rho_{12}\eta_{12}+w_{12}}^{2}}\\
 & \overset{\brac i}{\leq}\lgbrac{3+T}-h\brac{\rline{\abs{a'\rho_{11}\eta_{11}+b'\rho_{12}\eta_{12}+w_{11}}^{2}+\abs{c'\rho_{12}\eta_{12}+w_{12}}^{2}\vphantom{a^{a^{a^{a^{a^{a6}}}}}}}\eta_{11},\eta_{12},w_{12}}\\
 & =\lgbrac{3+T}-h\brac{\rline{\abs{a'\rho_{11}\eta_{11}+b'\rho_{12}\eta_{12}+w_{11}}^{2}}\eta_{11},\eta_{12}}\\
 & \overset{\brac{ii}}{\leq}\lgbrac{3+T}+\frac{7}{2}\lgbrac e,
\end{alignat}
where in step $\brac i$ we used the fact $1>\max\brac{\abs{a'\rho_{11}},\abs{b'\rho_{12}},\abs{c'\rho_{12}}}$
and the fact that conditioning reduces entropy, in step $\brac{ii}$
we used the result from \ifarxiv  Appendix \ref{app:entropybound_sum_exponentials} \else  \cite[Appendix K]{Joyson_2x2_mimov5} \fi
to lower \sloppy bound $h\brac{\rline{\abs{a'\rho_{11}\eta_{11}+b'\rho_{12}\eta_{12}+w_{11}}^{2}}\eta_{11},\eta_{12}}.$
\end{IEEEproof}
Using (\ref{eq:loc1_lapidoth_squared_norm}), (\ref{eq:loc2_lapidoth_squared_norm}),
(\ref{eq:loc3_lapidoth_squared_norm}) and using Lemma \ref{claim:new_squared_channel_1},
we get
\[
\expect{\lgbrac V}\eqdof\expect{\lgbrac{\rho_{11}^{2}\abs a^{2}+\rho_{12}^{2}\abs b^{2}+\rho_{12}^{2}\abs c^{2}+T}}\qquad\qquad\qquad\qquad\qquad\qquad\qquad\qquad\qquad
\]
\begin{equation}
\eqdof h\brac{\rline{\abs{ag_{11}+bg_{12}+w_{11}}^{2}+\abs{cg_{12}+w_{12}}^{2}+\sum_{i=3}^{T}\abs{w_{1i}}^{2}}a,b,c}=h\brac{\rline{V\vphantom{\frac{q}{b}}}U}
\end{equation}
and the above approximation is tight within a constant independent
of $\snr$. Hence it follows that
\begin{alignat}{1}
C_{1} & \eqdof\max_{p\brac{a,b,c};\expect{\abs a^{2}+\abs b^{2}+\abs c^{2}}\leq T}\left\{ h\brac{\abs{ag_{11}+bg_{12}+w_{11}}^{2}+\abs{cg_{12}+w_{12}}^{2}+\sum_{i=3}^{T}\abs{w_{1i}}^{2}}\right.\nonumber \\
 & \qquad\qquad\qquad\qquad\qquad\qquad\qquad\left.-\expect{\lgbrac{\rho_{11}^{2}\abs a^{2}+\rho_{12}^{2}\abs b^{2}+\rho_{12}^{2}\abs c^{2}+T}}\right\}
\end{alignat}
and the above equality is tight within a constant independent of $\snr$.
Now we shall prove that
\begin{equation}
\underset{\snr\rightarrow\infty}{\text{limsup}}\ C_{1}\brac{\snr}-\lgbrac{\lgbrac{\snr}}<\infty
\end{equation}
 and hence it will prove our claim that \sloppy $h\brac{\abs{ag_{11}+bg_{12}+w_{11}}^{2}+\abs{cg_{12}+w_{12}}^{2}+\sum_{i=3}^{T}\abs{w_{1i}}^{2}}$
and $\expect{\lgbrac{\rho_{11}^{2}\abs a^{2}+\rho_{12}^{2}\abs b^{2}+\rho_{12}^{2}\abs c^{2}+T}}$
have the same gDoF.

Now looking at (\ref{eq:lapidoth_outer_squarednormgaussian}) again,
if the term $\lgbrac{\expect V}-\expect{\lgbrac V}$ does not approach
infinity with the $\snr$, then the result follows directly by choosing
any fixed $\alpha>0$. When $\lgbrac{\expect V}-\expect{\lgbrac V}$
does tend to infinity with $\snr$, we choose
\begin{equation}
\alpha^{*}=\brac{1+\lgbrac{\expect V}-\expect{\lgbrac V}}^{-1}\label{eq:lapidoth_alphastar}
\end{equation}
with $\alpha^{*}\downarrow0$ with the $\snr$ and we have $\lgbrac{\Gamma\brac{\alpha^{*}}}=\lgbrac{\frac{1}{\alpha^{*}}}+o\brac 1$
and $\alpha^{*}\lgbrac{\alpha^{*}}=o\brac 1$ where $o\brac 1$ tends
to zero as $\alpha^{*}$ tends to zero, following \cite[(337)]{lapidoth2003capacity}.
Hence using (\ref{eq:lapidoth_alphastar}) and Lemma \ref{claim:new_squared_channel_1}
on (\ref{eq:lapidoth_outer_squarednormgaussian}), we get
\begin{alignat}{1}
C_{1} & \leq\lgbrac{3+T}+\frac{7}{2}\lgbrac e+1+\lgbrac{\frac{1}{\alpha^{*}}}+o\brac 1\\
 & =r_{4}+\lgbrac{\frac{1}{\alpha^{*}}}+o\brac 1.
\end{alignat}
We let $r_{4}=\lgbrac{3+T}+\brac{7/2}\lgbrac e+1$ in the last step.
Now
\begin{alignat}{1}
\frac{1}{\alpha^{*}} & =1+\lgbrac{\expect V}-\expect{\lgbrac V}\\
 & =1+\lgbrac{\expect{\abs{ag_{11}+bg_{12}+w_{11}}^{2}+\abs{cg_{12}+w_{12}}^{2}+\sum_{i=3}^{T}\abs{w_{1i}}^{2}}}\nonumber \\
 & \qquad-\expect{\lgbrac{\abs{ag_{11}+bg_{12}+w_{11}}^{2}+\abs{cg_{12}+w_{12}}^{2}+\sum_{i=3}^{T}\abs{w_{1i}}^{2}}}\\
 & \overset{\brac i}{=}1+\lgbrac{\expect{\rho_{11}^{2}\abs a^{2}+\rho_{12}^{2}\abs b^{2}+\rho_{12}^{2}\abs c^{2}+T}}\nonumber \\
 & \qquad-\expect{\lgbrac{\abs{ag_{11}+bg_{12}+w_{11}}^{2}+\abs{cg_{12}+w_{12}}^{2}+\sum_{i=3}^{T}\abs{w_{1i}}^{2}}}\\
 & \overset{\brac{ii}}{\leq}1+\lgbrac{\rho_{11}^{2}T+\rho_{12}^{2}T+T}\\
 & \qquad-\expect{\lgbrac{\abs{ag_{11}+bg_{12}+w_{11}}^{2}+\abs{cg_{12}+w_{12}}^{2}+\sum_{i=3}^{T}\abs{w_{1i}}^{2}}}\\
 & \overset{\brac{iii}}{\leq}1+\lgbrac{\rho_{11}^{2}+\rho_{12}^{2}+1}+\lgbrac T-\expect{\expect{\rline{\lgbrac{\abs{ag_{11}+bg_{12}+w_{11}}^{2}+0}}a,b}}\\
 & \overset{\brac{iv}}{\leq}1+\lgbrac{\rho_{11}^{2}+\rho_{12}^{2}+1}+\lgbrac T-\expect{\lgbrac{\rho_{11}^{2}\abs a^{2}+\rho_{12}^{2}\abs b^{2}+1}}+\gamma_{E}\lgbrac e\\
 & \overset{\brac v}{\leq}1+\lgbrac{\rho_{11}^{2}+\rho_{12}^{2}+1}+\lgbrac T-0+\gamma_{E}\lgbrac e,
\end{alignat}
where $\brac i$ is using the Tower property of expectation and that
given $\brac{a,b,c}$, $ag_{11}+bg_{12}+w_{11},cg_{12}+w_{12}$ are
sums of independent Gaussians, $\brac{ii}$ is using power constraints
on $a,b,c$, $\brac{iii}$ is because $\abs{cg_{12}+w_{12}}^{2}+\sum_{i=3}^{T}\abs{w_{1i}}^{2}>0$
, $\brac{iv}$ is using Lemma \ref{fact:Jensens_gap} on page \pageref{fact:Jensens_gap}
and $\brac v$ is because $\lgbrac{\rho_{11}^{2}\abs a^{2}+\rho_{12}^{2}\abs b^{2}+1}>0$.
Hence
\begin{alignat}{1}
C_{1} & \leq r_{4}+\lgbrac{1+\lgbrac{\rho_{11}^{2}+\rho_{12}^{2}+\rho_{12}^{2}+1}+\lgbrac T+\gamma\lgbrac e}+o\brac 1
\end{alignat}
and the proof is complete.

\section{Proof of Lemma \ref{lem:dof_equivalence_h(xi22)}\label{app:Proof-of-Lemma_dof_equivalence_h(xi22)}}

Here we prove that $h\brac{\rline{\abs{\xi_{22}}^{2}\vphantom{a^{a^{a^{a}}}}}\abs{\xi_{11}}^{2}}\eqdof h\brac{\rline{\abs{\xi_{22}}^{2}\vphantom{a^{a^{a^{a}}}}}\abs{\xi_{11}}^{2},a,b,c}$
with $\abs{\xi_{11}}^{2},\abs{\xi_{22}}^{2}$ defined in (\ref{eq:xi_11_defn}),
(\ref{eq:xi_22_defn}) on page \pageref{eq:xi_11_defn}. The inequality
$h\brac{\rline{\abs{\xi_{22}}^{2}\vphantom{a^{a^{a^{a}}}}}\abs{\xi_{11}}^{2},a,b,c}\leq\expect{\lgbrac{e\expect{\rline{\abs{\xi_{22}}^{2}\vphantom{a^{a^{a^{a}}}}}a,b,c}}}$
follows using Lemma \ref{lem:max_entropy}. We now only need to show
that $I\brac{\rline{\abs{\xi_{22}}^{2};a,b,c\vphantom{a^{a^{a^{a}}}}}\abs{\xi_{11}}^{2}}$
has zero gDoF. Now
\[
I\brac{\rline{\abs{\xi_{22}}^{2};a,b,c\vphantom{a^{a^{a^{a}}}}}\abs{\xi_{11}}^{2}}\leq I\brac{\abs{\xi_{22}}^{2};a,b,c,\abs{\xi_{11}}^{2}}.
\]
We will show that $I\brac{\abs{\xi_{22}}^{2};a,b,c,\abs{\xi_{11}}^{2}}$
has no gDoF. From \cite[(32)]{lapidoth2003capacity} we have
\begin{align}
I\brac{U;V} & \leq\expect{\lgbrac V}-h\brac{\rline VU}+\lgbrac{\Gamma\brac{\alpha}}\nonumber \\
 & \qquad+\alpha\brac{1+\lgbrac{\expect V}-\expect{\lgbrac V}}-\alpha\lgbrac{\alpha}
\end{align}
for any $\alpha>0$ for channels whose output $V$ takes values in
$\mathbb{R}^{+}$. We will use this result to bound $I\brac{\abs{\xi_{22}}^{2};a,b,c,\abs{\xi_{11}}^{2}}$
with $U=\brac{a,b,c,\abs{\xi_{11}}^{2}},V=\abs{\xi_{22}}^{2}$ for
any distribution of $a,b,c$ with the power constraint $\expect{\abs a^{2}+\abs b^{2}+\abs c^{2}}\leq T$.
The result from \cite{lapidoth2003capacity} can be applied assuming
the channel induced by $p\brac{\rline{\abs{\xi_{22}}^{2}\vphantom{a^{a^{a^{a}}}}}a,b,c,\abs{\xi_{11}}^{2}}$
satisfies the Borel measurability conditions in \cite[Theorem 5.1]{lapidoth2003capacity},
\emph{i.e.,} for any given Borel set $\mathcal{B\subset\mathbb{R}}^{+}$,
$f_{\mathcal{B}}\brac v=p\brac{\rline{\mathcal{B}\vphantom{a^{a^{a^{a}}}}}v=\brac{a,b,c,\abs{\xi_{11}}^{2}}}$
is a Borel measurable function.

Recall that from (\ref{eq:xi_11_defn}) and (\ref{eq:xi_22_defn}),
we have
\begin{align*}
\abs{\xi_{11}}^{2} & =\abs{ag_{11}+bg_{12}+w_{11}}^{2}+\abs{cg_{12}+w_{12}}^{2}+\sum_{i=3}^{T}\abs{w_{1i}}^{2},\\
\abs{\xi_{22}}^{2} & =\abs{ag_{21}+bg_{22}+w_{21}}^{2}+\abs{cg_{22}+w_{22}}^{2}+\sum_{i=3}^{T}\abs{w_{2i}}^{2}\\
 & \qquad-\frac{\abs{\brac{ag_{21}+bg_{22}+w_{21}}\brac{ag_{11}+bg_{12}+w_{11}}^{*}+\brac{cg_{22}+w_{22}}\brac{cg_{12}+w_{12}}^{*}+\sum_{i=3}^{T}w_{2i}w_{1i}^{*}}^{2}}{\abs{ag_{11}+bg_{12}+w_{11}}^{2}+\abs{cg_{12}+w_{12}}^{2}+\sum_{i=3}^{T}\abs{w_{1i}}^{2}}.
\end{align*}
We first consider $\lgbrac{\expect V}-\expect{\lgbrac V}=\expect{\lgbrac{\abs{\xi_{22}}^{2}}}-h\brac{\rline{\abs{\xi_{22}}^{2}\vphantom{a^{a^{a^{a^{a}}}}}}a,b,c,\abs{\xi_{11}}^{2}}$
and show that it is bounded independent of $\snr$. Note that we can
manipulate $\abs{\xi_{22}}^{2}$ as
\begin{align}
 & \abs{\xi_{22}}^{2}\nonumber \\
 & =\brac{\abs{ag_{21}+bg_{22}+w_{21}}^{2}+\abs{cg_{22}+w_{22}}^{2}+\sum_{i=3}^{T}\abs{w_{2i}}^{2}}\nonumber \\
 & \qquad-\abs{\brac{ag_{21}+bg_{22}+w_{21}}u_{1}^{*}+\brac{cg_{22}+w_{22}}u_{2}^{*}+\sum_{i=3}^{T}w_{2i}u_{i}^{*}}^{2}\\
 & =\left\Vert \sbrac{ag_{21}+bg_{22}+w_{21},cg_{22}+w_{22},w_{23},\ldots,w_{2T}}\vphantom{a^{a^{a^{a^{a^{a^{a^{a^{a^{a^{a^{a^{a^{a}}}}}}}}}}}}}}\right.\nonumber \\
 & \qquad\left.-\sbrac{ag_{21}+bg_{22}+w_{21},cg_{22}+w_{22},w_{23},\ldots,w_{2T}}\left[\begin{array}{c}
u_{1}^{*}\\
u_{2}^{*}\\
.\\
.\\
u_{T}^{*}
\end{array}\right]\sbrac{u_{1},\ldots,u_{T}}\vphantom{a^{a^{a^{a^{a^{a^{a^{a^{a^{a^{a^{a^{a^{a}}}}}}}}}}}}}}\right\Vert ^{2},
\end{align}
where $\left\Vert \cdot\right\Vert $ indicates 2-norm for a vector
and $\brac{u_{i}}$ forms a unit norm complex vector
\begin{equation}
\sbrac{u_{1},\ldots,u_{T}}=\frac{\sbrac{ag_{11}+bg_{12}+w_{11},cg_{12}+w_{12},w_{13},\ldots,w_{1T}\vphantom{a^{a^{a^{a^{a}}}}}}}{\abs{ag_{11}+bg_{12}+w_{11}}^{2}+\abs{cg_{12}+w_{12}}^{2}+\sum_{i=3}^{T}\abs{w_{1i}}^{2}},
\end{equation}

\begin{align}
\abs{\xi_{22}}^{2} & =\abs{ag_{21}+bg_{22}+w_{21}-\brac{\brac{ag_{21}+bg_{22}+w_{21}}u_{1}^{*}+\brac{cg_{22}+w_{22}}u_{2}^{*}+\sum_{i=3}^{T}w_{2i}u_{i}^{*}}u_{1}}^{2}\nonumber \\
 & \qquad+\abs{cg_{22}+w_{22}-\brac{\brac{ag_{21}+bg_{22}+w_{21}}u_{1}^{*}+\brac{cg_{22}+w_{22}}u_{2}^{*}+\sum_{i=3}^{T}w_{2i}u_{i}^{*}}u_{2}}\nonumber \\
 & \qquad+\abs{w_{23}-\brac{\brac{ag_{21}+bg_{22}+w_{21}}u_{1}^{*}+\brac{cg_{22}+w_{22}}u_{2}^{*}+\sum_{i=3}^{T}w_{2i}u_{i}^{*}}u_{3}}^{2}\nonumber \\
 & \qquad+\ldots\nonumber \\
 & \qquad+\abs{w_{2T}-\brac{\brac{ag_{21}+bg_{22}+w_{21}}u_{1}^{*}+\brac{cg_{22}+w_{22}}u_{2}^{*}+\sum_{i=3}^{T}w_{2i}u_{i}^{*}}u_{T}}^{2}\\
 & \overset{\brac i}{=}\sum_{i=1}^{T}\abs{\eta_{21}\kappa_{1i}+\eta_{22}\kappa_{2i}+\sum_{j=1}^{T}w_{2j}\kappa_{(j+2)i}}^{2},
\end{align}
where in step $\brac i$ $\eta_{ij}$ are independent $\mathcal{CN}\brac{0,1}$
after the substitution $g_{ij}=\rho_{ij}\eta_{ij}$, also $\kappa_{ij}$
are functions of $a,b,c,\rho_{ij},u_{i}$ obtained after collecting
the coefficients of $\eta_{ij},w_{2j}$. Note that $\max_{i,j}\brac{\abs{\kappa{}_{ij}}}\geq1$.
Now
\begin{align}
 & \expect{\lgbrac{\abs{\xi_{22}}^{2}}}-h\brac{\rline{\abs{\xi_{22}}^{2}\vphantom{a^{a^{a^{a^{a}}}}}}a,b,c,\abs{\xi_{11}}^{2}}\nonumber \\
 & =\expect{\lgbrac{\sum_{i=1}^{T}\abs{\eta_{21}\kappa_{1i}+\eta_{22}\kappa_{2i}+\sum_{j=1}^{T}w_{2j}\kappa_{(j+2)i}}^{2}}}\nonumber \\
 & \qquad-h\brac{\rline{\sum_{i=1}^{T}\abs{\eta_{21}\kappa_{1i}+\eta_{22}\kappa_{2i}+\sum_{j=1}^{T}w_{2j}\kappa_{(j+2)i}}^{2}\vphantom{a^{a^{a^{a^{a}}}}}}a,b,c,\abs{\xi_{11}}^{2}}\\
 & \leq\expect{\lgbrac{\sum_{i=1}^{T}\abs{\eta_{21}\kappa_{1i}+\eta_{22}\kappa_{2i}+\sum_{j=1}^{T}w_{2j}\kappa_{(j+2)i}}^{2}}}\nonumber \\
 & \qquad-h\brac{\rline{\sum_{i=1}^{T}\abs{\eta_{21}\kappa_{1i}+\eta_{22}\kappa_{2i}+\sum_{j=1}^{T}w_{2j}\kappa_{(j+2)i}}^{2}\vphantom{a^{a^{a^{a^{a}}}}}}\cbrac{\kappa_{ij}}},
\end{align}
where the last step uses the fact that conditioning reduces entropy
and Markovity $\brac{a,b,c,\abs{\xi_{11}}^{2}}$ \xdash[1em] $\brac{\cbrac{\kappa_{ij}}}$\xdash[1em]$\brac{\sum_{i=1}^{T}\abs{\eta_{21}\kappa_{1i}+\eta_{22}\kappa_{2i}+\sum_{j=1}^{T}w_{2j}\kappa_{(j+2)i}}^{2}}$.
Note that $\eta_{21},\eta_{22},w_{2j}$ are independent of $\kappa_{ij}$.
Now it suffices to show that for any given set of \textit{constant}
$\kappa'_{ij}$ the difference
\[
\expect{\lgbrac{\sum_{i=1}^{T}\abs{\eta_{21}\kappa'_{1i}+\eta_{22}\kappa'_{2i}+\sum_{j=1}^{T}w_{2j}\kappa'_{(j+2)i}}^{2}}}-h\brac{\sum_{i=1}^{T}\abs{\eta_{21}\kappa'_{1i}+\eta_{22}\kappa'_{2i}+\sum_{j=1}^{T}w_{2j}\kappa'_{(j+2)i}}^{2}}
\]
 is uniformly bounded independent of $\kappa'_{ij}$. We will show
this by assuming $\abs{\kappa'_{11}}=\max_{i,j}\brac{\abs{\kappa'_{ij}}}$.
This is without loss of generality since $\eta_{ij},w_{ij}$ are all
i.i.d. $\mathcal{CN}\brac{0,1}$. Now
\begin{align}
 & \expect{\lgbrac{\sum_{i=1}^{T}\abs{\eta_{21}\kappa'_{1i}+\eta_{22}\kappa'_{2i}+\sum_{j=1}^{T}w_{2j}\kappa'_{(j+2)i}}^{2}}}-h\brac{\sum_{i=1}^{T}\abs{\eta_{21}\kappa'_{1i}+\eta_{22}\kappa'_{2i}+\sum_{j=1}^{T}w_{2j}\kappa'_{(j+2)i}}^{2}}\nonumber \\
 & =\expect{\lgbrac{\sum_{i=1}^{T}\abs{\eta_{21}\frac{\kappa'_{1i}}{\kappa'_{11}}+\eta_{22}\frac{\kappa'_{2i}}{\kappa'_{11}}+\sum_{j=1}^{T}w_{2j}\frac{\kappa'_{(j+2)i}}{\kappa'_{11}}}^{2}}}\nonumber \\
 & \qquad-h\brac{\sum_{i=1}^{T}\abs{\eta_{21}\frac{\kappa'_{1i}}{\kappa'_{11}}+\eta_{22}\frac{\kappa'_{2i}}{\kappa'_{11}}+\sum_{j=1}^{T}w_{2j}\frac{\kappa'_{(j+2)i}}{\kappa'_{11}}}^{2}}\\
 & \overset{\brac i}{\leq}\lgbrac{\sum_{i=1}^{T}\expect{\abs{\eta_{21}\frac{\kappa'_{1i}}{\kappa'_{11}}+\eta_{22}\frac{\kappa'_{2i}}{\kappa'_{11}}+\sum_{j=1}^{T}w_{2j}\frac{\kappa'_{(j+2)i}}{\kappa'_{11}}}^{2}}}\nonumber \\
 & \qquad-h\brac{\sum_{i=1}^{T}\abs{\eta_{21}\frac{\kappa'_{1i}}{\kappa'_{11}}+\eta_{22}\frac{\kappa'_{2i}}{\kappa'_{11}}+\sum_{j=1}^{T}w_{2j}\frac{\kappa'_{(j+2)i}}{\kappa'_{11}}}^{2}}\\
 & \overset{\brac{ii}}{=}\lgbrac{\sum_{i=1}^{T}\brac{\abs{\frac{\kappa'_{1i}}{\kappa'_{11}}}^{2}+\abs{\frac{\kappa'_{2i}}{\kappa'_{11}}}^{2}+\sum_{j=1}^{T}\abs{\frac{\kappa'_{(j+2)i}}{\kappa'_{11}}}^{2}}}\nonumber \\
 & \qquad-h\brac{\sum_{i=1}^{T}\abs{\eta_{21}\frac{\kappa'_{1i}}{\kappa'_{11}}+\eta_{22}\frac{\kappa'_{2i}}{\kappa'_{11}}+\sum_{j=1}^{T}w_{2j}\frac{\kappa'_{(j+2)i}}{\kappa'_{11}}}^{2}}\\
 & \overset{\brac{iii}}{\leq}\lgbrac{T\brac{T+2}}-h\brac{\sum_{i=1}^{T}\abs{\eta_{21}\frac{\kappa'_{1i}}{\kappa'_{11}}+\eta_{22}\frac{\kappa'_{2i}}{\kappa'_{11}}+\sum_{j=1}^{T}w_{2j}\frac{\kappa'_{(j+2)i}}{\kappa'_{11}}}^{2}}\\
 & \overset{\brac{iv}}{\leq}\lgbrac{T\brac{T+2}}-h\brac{\rline{\sum_{i=1}^{T}\abs{\eta_{21}\frac{\kappa'_{1i}}{\kappa'_{11}}+\eta_{22}\frac{\kappa'_{2i}}{\kappa'_{11}}+\sum_{j=1}^{T}w_{2j}\frac{\kappa'_{(j+2)i}}{\kappa'_{11}}}^{2}}\eta_{22},w_{2j}}\\
 & \overset{\brac v}{\leq}\lgbrac{T\brac{T+2}}+\frac{7}{2}\lgbrac e,
\end{align}
where $\brac i$ is using Jensen's inequality, $\brac{ii}$ is using
the fact that $\eta_{21}\frac{\kappa'_{1i}}{\kappa'_{11}}+\eta_{22}\frac{\kappa'_{2i}}{\kappa'_{11}}+\sum_{j=1}^{T}w_{2j}\frac{\kappa'_{(j+2)i}}{\kappa'_{11}}$
is Complex Gaussian, $\brac{iii}$ is because $\frac{\abs{\kappa'_{ij}}}{\abs{\kappa'_{11}}}\leq1$
since $\abs{\kappa'_{11}}=\max_{i,j}\brac{\abs{\kappa'_{ij}}}$ (note
that $\max_{i,j}\brac{\abs{\kappa'{}_{ij}}}\geq1$ for a valid set
of $\kappa'{}_{ij}$, due to the way $\kappa{}_{ij}$ is defined),
$\brac{iv}$ is because conditioning reduces entropy and $\brac v$
is by invoking the result from \ifarxiv  Appendix \ref{app:entropybound_sum_exponentials}\else  \cite[Appendix K]{Joyson_2x2_mimov5}\fi.

Now if the term $\lgbrac{\expect V}-\expect{\lgbrac V}$ does not
approach infinity with the $\snr$ then the desired result follows
directly by choosing any fixed $\alpha>0$. When $\lgbrac{\expect V}-\expect{\lgbrac V}$
does tend to infinity with $\snr$, following \cite[(336)]{lapidoth2003capacity}
we choose
\begin{equation}
\alpha^{*}=\brac{1+\lgbrac{\expect V}-\expect{\lgbrac V}}^{-1}
\end{equation}
with $\alpha^{*}\downarrow0$ with the $\snr$ and we have $\lgbrac{\Gamma\brac{\alpha^{*}}}=\lgbrac{\frac{1}{\alpha^{*}}}+o\brac 1$
and $\alpha^{*}\lgbrac{\alpha^{*}}=o\brac 1$ where $o\brac 1$ tends
to zero as $\alpha^{*}$ tends to zero, following \cite[(337)]{lapidoth2003capacity}.
Hence we have
\begin{align}
I\brac{\abs{\xi_{22}}^{2};a,b,c,\abs{\xi_{11}}^{2}} & \leq\brac{\lgbrac{T\brac{T+2}}+\frac{7}{2}\lgbrac e}+1+\lgbrac{\frac{1}{\alpha^{*}}}+o\brac 1,\label{eq:loc1_xi22_lapidoth}
\end{align}
\begin{align}
\frac{1}{\alpha^{*}} & =1+\lgbrac{\expect{\abs{\xi_{22}}^{2}}}-\expect{\lgbrac{\abs{\xi_{22}}^{2}}}.
\end{align}
Now
\begin{align}
\abs{\xi_{22}}^{2} & \leq\abs{ag_{21}+bg_{22}+w_{21}}^{2}+\abs{cg_{22}+w_{22}}^{2}+\sum_{i=3}^{T}\abs{w_{2i}}^{2}
\end{align}
due to the LQ transformation on (\ref{eq:2x2_preLQ}). Hence
\begin{align}
\expect{\abs{\xi_{22}}^{2}} & \overset{\brac i}{\leq}\expect{\brac{\rho_{21}^{2}\abs a^{2}+\rho_{22}^{2}\brac{\abs b^{2}+\abs c^{2}}+T}},\\
\lgbrac{\expect{\abs{\xi_{22}}^{2}}} & \overset{\brac{ii}}{\leq}\lgbrac{\rho_{21}^{2}+\rho_{22}^{2}+1}+\lgbrac T,
\end{align}
where $\brac i$ is using the fact that given $\brac{a,b,c}$, $ag_{21}+bg_{22}+w_{21}$,$cg_{22}+w_{22}$
are sums of independent Gaussians and $\brac{ii}$ is using the power
constraint on $a,b,c$. Hence we have
\begin{align}
\frac{1}{\alpha^{*}} & \leq1+\lgbrac{\rho_{21}^{2}+\rho_{22}^{2}+1}+\lgbrac T-\expect{\lgbrac{\abs{\xi_{22}}^{2}}}.\label{eq:loc2_xi22_lapidoth}
\end{align}
Now we lower bound $\expect{\lgbrac{\abs{\xi_{22}}^{2}}}$. Note that
\begin{align}
\abs{\xi_{22}}^{2}= & \abs{ag_{21}+bg_{22}+w_{21}}^{2}+\abs{cg_{22}+w_{22}}^{2}+\sum_{i=3}^{T}\abs{w_{2i}}^{2}\\
 & -\frac{\abs{\brac{ag_{21}+bg_{22}+w_{21}}\brac{ag_{11}+bg_{12}+w_{11}}^{*}+\brac{cg_{22}+w_{22}}\brac{cg_{12}+w_{12}}^{*}+\sum_{i=3}^{T}w_{2i}w_{1i}^{*}}^{2}}{\abs{ag_{11}+bg_{12}+w_{11}}^{2}+\abs{cg_{12}+w_{12}}^{2}+\sum_{i=3}^{T}\abs{w_{1i}}^{2}}
\end{align}
 is the magnitude squared of the projection of the Complex vector
\[
\sbrac{\brac{ag_{21}+bg_{22}+w_{21}},\brac{cg_{22}+w_{22}},w_{23},\ldots,w_{2T}\vphantom{a^{a^{a^{a^{a}}}}}}
\]
onto the subspace orthogonal to the Complex vector
\[
\sbrac{\brac{ag_{11}+bg_{12}+w_{11}},\brac{cg_{12}+w_{12}},w_{13},\ldots,w_{1T}\vphantom{a^{a^{a^{a^{a}}}}}}.
\]
Note that $\sbrac{\brac{cg_{12}+w_{12}}^{*},-\brac{ag_{11}+bg_{12}+w_{11}}^{*},0,\ldots,0}$
is orthogonal to
\[
\sbrac{\brac{ag_{11}+bg_{12}+w_{11}},\brac{cg_{12}+w_{12}},w_{13},\ldots,w_{1T}\vphantom{a^{a^{a^{a^{a}}}}}}.
\]
 Hence
\begin{align}
\abs{\xi_{22}}^{2} & \geq\frac{\abs{\sbrac{\begin{array}{ccccc}
ag_{21}+bg_{22}+w_{21}, & cg_{22}+w_{22}, & w_{23}, & \ldots, & w_{2T}\end{array}}\sbrac{\begin{array}{c}
cg_{12}+w_{12}\\
-\brac{ag_{11}+bg_{12}+w_{11}}\\
0\\
\vdots\\
0
\end{array}}}^{2}}{\abs{ag_{11}+bg_{12}+w_{11}}^{2}+\abs{cg_{12}+w_{12}}^{2}}\\
= & \frac{\abs{\brac{ag_{21}+bg_{22}+w_{21}}\brac{cg_{12}+w_{12}}-\brac{cg_{22}+w_{22}}\brac{ag_{11}+bg_{12}+w_{11}}}^{2}}{\abs{ag_{11}+bg_{12}+w_{11}}^{2}+\abs{cg_{12}+w_{12}}^{2}}
\end{align}
and hence
\begin{align}
 & \expect{\lgbrac{\abs{\xi_{22}}^{2}}}\\
 & \geq\expect{\lgbrac{\frac{\abs{\brac{ag_{21}+bg_{22}+w_{21}}\brac{cg_{12}+w_{12}}-\brac{cg_{22}+w_{22}}\brac{ag_{11}+bg_{12}+w_{11}}}^{2}}{\abs{ag_{11}+bg_{12}+w_{11}}^{2}+\abs{cg_{12}+w_{12}}^{2}}}}\nonumber \\
 & =\expect{\lgbrac{\abs{\brac{ag_{21}+bg_{22}+w_{21}}u'_{1}-\brac{cg_{22}+w_{22}}u'_{2}}^{2}}},
\end{align}
where $\brac{u'_{1},u'_{2}}$ is a unit norm complex vector independent
of $g_{2i},w_{2i}$. Hence
\begin{align}
 & \expect{\lgbrac{\abs{\xi_{22}}^{2}}}\nonumber \\
 & \overset{\brac i}{\geq}\expect{\lgbrac{\abs{au'_{1}}^{2}\rho_{21}^{2}+\abs{bu'_{1}-cu'_{2}}^{2}\rho_{22}^{2}+\abs{u'_{1}}^{2}+\abs{u'_{2}}^{2}}}-\gamma_{E}\lgbrac e\\
 & \overset{\brac{ii}}{\geq}\expect{\lgbrac{\abs{au'_{1}}^{2}\rho_{21}^{2}+\abs{bu'_{1}-cu'_{2}}^{2}\rho_{22}^{2}+1}}-\gamma_{E}\lgbrac e\\
 & \geq-\gamma\lgbrac{_{E}e},\label{eq:loc3_xi22_lapidoth}
\end{align}
where $\brac i$ is using the fact that given $\brac{a,b,c,u'_{1},u'_{2}}$,
$\brac{ag_{21}+bg_{22}+w_{21}}u'_{1}-\brac{cg_{22}+w_{22}}u'_{2}$
is Complex Gaussian distributed with variance $\abs{au'_{1}}^{2}\rho_{21}^{2}+\abs{bu'_{1}-cu'_{2}}^{2}\rho_{22}^{2}+\abs{u'_{1}}^{2}+\abs{u'_{2}}^{2}$
and applying Lemma \ref{fact:Jensens_gap} on page \pageref{fact:Jensens_gap}
together with Tower property of expectation. The step $\brac{ii}$
is because $\brac{u'_{1},u'_{2}}$ is a unit norm vector.

Substituting (\ref{eq:loc3_xi22_lapidoth}) in (\ref{eq:loc2_xi22_lapidoth})
we get
\begin{align}
\frac{1}{\alpha^{*}} & \leq\lgbrac{\rho_{21}^{2}+\rho_{22}^{2}+1}+1+\lgbrac T+\gamma_{E}\lgbrac e\\
 & =\lgbrac{\rho_{21}^{2}+\rho_{22}^{2}+1}+r_{2}\brac T
\end{align}
and hence by substituting the above in (\ref{eq:loc1_xi22_lapidoth}),
we get
\begin{align}
I\brac{\abs{\xi_{22}}^{2};a,b,c,\abs{\xi_{11}}^{2}} & \leq\brac{\lgbrac{T\brac{T+2}}+\frac{7}{2}\lgbrac e}+1\nonumber \\
 & \quad+\lgbrac{\lgbrac{\rho_{21}^{2}+\rho_{22}^{2}+1}+r_{2}\brac T}+o\brac 1,
\end{align}
where $r_{2}\brac T=1+\lgbrac T+\gamma_{E}\lgbrac e$ is a function
of $T$ alone. Hence $I\brac{\abs{\xi_{22}}^{2};a,b,c,\abs{\xi_{11}}^{2}}$
has zero gDoF. Now since
\[
I\brac{\rline{\abs{\xi_{22}}^{2};a,b,c\vphantom{a^{a^{a^{a^{a^{a}}}}}}}\abs{\xi_{11}}^{2}}\leq I\brac{\abs{\xi_{22}}^{2};a,b,c,\abs{\xi_{11}}^{2}},
\]
 it follows that $h\brac{\rline{\abs{\xi_{22}}^{2}\vphantom{a^{a^{a^{a^{a^{a}}}}}}}\abs{\xi_{11}}^{2}}\eqdof h\brac{\rline{\abs{\xi_{22}}^{2}\vphantom{a^{a^{a^{a^{a^{a}}}}}}}\abs{\xi_{11}}^{2},a,b,c}$.

\section{A lower bound on entropy of squared 2-norm of a Gaussian vector\label{app:entropybound_sum_exponentials}}

For complex $l_{i},k_{i},l$\textbf{ }for finite number of $i$'s\textbf{
}with $\abs{k_{i}}\leq1$ and $\eta\sim\mathcal{CN}\brac{0,1}$ we
will show that
\begin{equation}
h\brac{\abs{\eta+l}^{2}+\sum_{i}\abs{k_{i}\eta+l_{i}}^{2}}\geq-\frac{7}{2}\lgbrac e.
\end{equation}
We have
\begin{align}
 & h\brac{\abs{\eta+l}^{2}+\sum_{i}\abs{k_{i}\eta+l_{i}}^{2}}\nonumber \\
 & =h\brac{\abs l^{2}+2\text{Re}\brac{\brac{l^{*}+\sum_{i}l_{i}^{*}k_{i}}\eta}+\abs{\eta}^{2}\brac{1+\sum_{i}\abs{k_{i}}^{2}}}\\
 & =h\brac{\abs{\eta\sqrt{1+\sum_{i}\abs{k_{i}}^{2}}+\frac{l+\sum_{i}l_{i}k_{i}^{*}}{\sqrt{1+\sum_{i}\abs{k_{i}}^{2}}}}^{2}}.\label{eq:2norm_gaussianvector_eq1}
\end{align}
Now it suffices to show that $h\brac{\abs{\eta k'+l'}^{2}}>-\brac{7/2}\lgbrac e$
for $\abs{k'}\geq1$. Now,
\begin{align}
h\brac{\abs{\eta k'+l'}^{2}} & =h\brac{\abs{\eta k'}^{2}+2\abs{\eta}\abs{k'}\abs{l'}\cos\theta+\abs{l'}^{2}},
\end{align}
where $\theta$ is uniformly distributed in $\sbrac{0,2\pi}$ and
is independent of $\abs{\eta}$ since $\eta$ is circularly symmetric
Gaussian.
\begin{align}
h\brac{\abs{\eta k'+l'}^{2}} & \geq h\brac{\rline{\abs{\eta k'}^{2}+2\abs{\eta}\abs{k'}\abs{l'}\cos\theta+\abs{l'}^{2}\vphantom{\frac{a}{b}}}\theta}\\
 & =h\brac{\rline{\brac{\abs{\eta k'}+\abs{l'}\cos\theta\vphantom{\frac{a}{b}}}^{2}}\theta}
\end{align}

Consider $S=\abs{\abs{\eta}\abs{k'}+\abs{l'}\cos\theta'}$ for a constant
$\theta'$. It suffices to show that $h\brac{S^{2}}\geq-\brac{7/2}\lgbrac e$
to complete the proof. Now $\eta'=\abs{\eta}\abs{k'}$ is Rayleigh
distributed with probability density function $p_{\eta'}\brac x=\brac{x/\abs{k'}^{2}}\exp\brac{-x^{2}/\brac{2\abs{k'}^{2}}}$
and it easily follows that $p_{\eta'}\brac x\leq\brac{1/\abs{k'}}\exp\brac{-1/2}\leq\exp\brac{-1/2}$
since $\abs{k'}\geq1$. Hence the probability density function of
$S$ has $p_{s}\brac x\leq2\exp\brac{-1/2}$. Hence
\begin{align}
h\brac S & =-\expect{\lgbrac{p_{s}\brac S}}\\
 & \geq-\lgbrac{2e^{-\frac{1}{2}}}
\end{align}
Using \cite[(316)]{lapidoth2003capacity} for rates in bits, we have
\begin{align}
h\brac{S^{2}} & =h\brac S+\expect{\lgbrac S}+1\\
 & \geq-\lgbrac{2e^{-\frac{1}{2}}}+\expect{\lgbrac{\abs{\abs{\eta}\abs{k'}+\abs{l'}\cos\theta'\vphantom{\frac{a}{\frac{b}{c}}}}}}+1\\
 & =\frac{1}{2}\lgbrac e+\expect{\lgbrac{\abs{\abs{\eta}\abs{k'}+\abs{l'}\cos\theta'\vphantom{\frac{a}{\frac{b}{c}}}}}}\label{eq:2norm_gaussianvector_eq2}
\end{align}
Now it suffices to show that {\scriptsize{}$\expect{\lgbrac{\abs{\abs{\eta}\abs{k'}+\abs{l'}\cos\theta'\vphantom{\frac{a}{\frac{b}{c}}}}}}$}
is lower bounded by $-4\lgbrac e$ to complete the proof. For a random
variable $X$ we define $h^{-}\brac X=\int_{p(x)>1}p\brac x\lgbrac{p\brac x}dx$.
We have
\begin{align}
h^{-}\brac{\abs{\eta}\abs{k'}} & =\int_{p_{\eta'}\brac x>1}p_{\eta'}\brac x\lgbrac{p_{\eta'}\brac x}dx\\
 & =0\label{eq:2norm_gaussianvector_eq3}
\end{align}
since $p_{\eta'}\brac x\leq\brac{1/\abs{k'}}\exp\brac{-1/2}\leq\exp\brac{-1/2}$.
Using \cite[(257)]{lapidoth2003capacity} to bound the expected logarithm
($\expect{\lgbrac{\abs X}}\geq-\frac{1}{\brac{1-\alpha}^{2}}\lgbrac e-\frac{1}{\alpha}h^{-}\brac X$
with $h^{-}\brac X=\int_{p(x)>1}p\brac x\lgbrac{p\brac x}dx$ for
any $0<\alpha<1$ ), we have
\begin{equation}
\expect{\lgbrac{\abs X}}\geq-\frac{1}{\brac{1-\alpha}^{2}}\lgbrac e-\frac{1}{\alpha}h^{-}\brac X,\ 0<\alpha<1,
\end{equation}
\begin{align}
\expect{\lgbrac{\abs{\abs{\eta}\abs{k'}+\abs{l'}\cos\theta'\vphantom{\frac{a}{\frac{b}{c}}}}}} & \overset{\brac i}{\geq}-\frac{1}{\brac{1-\frac{1}{2}}^{2}}\lgbrac e-2h^{-}\brac{\abs{\eta}\abs{k'}+\abs{l'}\cos\theta'}\\
 & =-2h^{-}\brac{\abs{\eta}\abs{k'}}-4\lgbrac e\\
 & \overset{\brac{ii}}{=}0-4\lgbrac e,\label{eq:2norm_gaussianvector_eq4}
\end{align}
where $\brac i$ is using \cite[(257)]{lapidoth2003capacity} with
$\alpha=\frac{1}{2}$ and $\brac{ii}$ is using  (\ref{eq:2norm_gaussianvector_eq3}).
Now using   (\ref{eq:2norm_gaussianvector_eq4}) in   (\ref{eq:2norm_gaussianvector_eq2})
the proof is complete.

\section{Numerical Calculation of inner bound for $T=2$ \label{app:simulation_calculation}}

Here we provide the calculations required for numerically evaluating
the achievable rates given in Table \ref{tab:Comparison-of-rates}.
We consider the case with $T=2$. In the calculations below, the channels
are scaled, so that\emph{ }the average power per transmit symbol from
each antenna is unity. Also, Gaussian codebooks are used in the training
based schemes.

\subsection{Training Scheme Using Only One Antenna}

For a training-based scheme using only one antenna (reducing to a
SISO case), we use one symbol (of value 1) to train the channel to
obtain $Y_{1,\text{train}}=g_{11}+w$ at the receiver. The minimum
mean squared error (MMSE) estimate for the channel is
\begin{align*}
\hat{g}_{11} & =\frac{\expect{\abs{g_{11}}^{2}}}{1+\expect{\abs{g_{11}}^{2}}}Y_{1,\text{train}}\\
 & \expect{\abs{g_{11}}^{2}}\frac{g_{11}+w}{1+\expect{\abs{g_{11}}^{2}}}.
\end{align*}
The total noise including MMSE is
\begin{align*}
N_{\text{SISO}}= & \text{\ensuremath{\expect{\brac{g_{11}-\expect{\abs{g_{11}}^{2}}\frac{g_{11}+w}{1+\expect{\abs{g_{11}}^{2}}}}^{2}}}+1}\\
= & \expect{\abs{\frac{g_{11}}{1+\expect{\abs{g_{11}}^{2}}}}^{2}}+\frac{\expect{\abs{g_{11}}^{2}}^{2}}{\abs{1+\expect{\abs{g_{11}}^{2}}}^{2}}+1\\
 & =\frac{\expect{\abs{g_{11}}^{2}}}{1+\expect{\abs{g_{11}}^{2}}}+1
\end{align*}
 and after scaling with $T=2$, the achievable rate is calculated
as
\begin{align*}
R_{\text{SISO}}=\frac{1}{2} & \expect{\lgbrac{1+\frac{\abs{\hat{g}_{11}}^{2}}{N_{\text{SISO}}}}}.
\end{align*}

\subsection{Training Scheme Using Both Antennas}

If we treat the system as a parallel antenna system, treating the
crosslinks as noise and use one symbol to train the channel, then
we get $Y_{1,\text{train}}=g_{11}+g_{12}+w$ at the first receiver
antenna. The MMSE estimate for the channel to the first antenna is
\begin{align*}
\hat{g}_{11} & =\frac{\expect{\abs{g_{11}}^{2}}}{1+\expect{\abs{g_{11}}^{2}}+\expect{\abs{g_{12}}^{2}}}Y_{1,\text{train}}\\
= & \expect{\abs{g_{11}}^{2}}\frac{g_{11}+g_{12}+w}{1+\expect{\abs{g_{11}}^{2}}+\expect{\abs{g_{12}}^{2}}}.
\end{align*}
The total noise including MMSE is
\begin{align*}
N_{\text{Parallel}}= & \text{\ensuremath{\expect{\brac{g_{11}-\expect{\abs{g_{11}}^{2}}\frac{g_{11}+g_{12}+w}{1+\expect{\abs{g_{11}}^{2}}+\expect{\abs{g_{12}}^{2}}}}^{2}}}+1}+\expect{\abs{g_{12}}^{2}}\\
= & \expect{\abs{\frac{g_{11}}{1+\expect{\abs{g_{11}}^{2}}+\expect{\abs{g_{12}}^{2}}}}^{2}\brac{1+\expect{\abs{g_{12}}^{2}}}^{2}}+\frac{\expect{\abs{g_{11}}^{2}}^{2}\brac{1+\expect{\abs{g_{12}}^{2}}}}{\abs{1+\expect{\abs{g_{11}}^{2}}+\expect{\abs{g_{12}}^{2}}}^{2}}\\
 & +1+\expect{\abs{g_{12}}^{2}}\\
= & \frac{\expect{\abs{g_{11}}^{2}}\brac{1+\expect{\abs{g_{12}}^{2}}}^{2}+\expect{\abs{g_{11}}^{2}}^{2}\brac{1+\expect{\abs{g_{12}}^{2}}}}{\abs{1+\expect{\abs{g_{11}}^{2}}+\expect{\abs{g_{12}}^{2}}}^{2}}+1+\expect{\abs{g_{12}}^{2}}\\
= & \frac{\expect{\abs{g_{11}}^{2}}\brac{1+\expect{\abs{g_{12}}^{2}}}}{1+\expect{\abs{g_{11}}^{2}}+\expect{\abs{g_{12}}^{2}}}+1+\expect{\abs{g_{12}}^{2}}
\end{align*}
 and using symmetry, the achievable rate (after scaling with $T=2$)
using both antennas is calculated as
\begin{align*}
R_{\text{Parallel}}= & \expect{\lgbrac{1+\frac{\abs{\hat{g}_{11}}^{2}}{N_{\text{Parallel}}}}}.
\end{align*}

\subsection{Noncoherent Scheme}

We evaluate the mutual information carefully for $T=2$ case for numerically
calculating it. Using the input distribution as given in Theorem \ref{thm:2x2-sym-mimo},
we have
\begin{align}
I\brac{X;Y} & =h\brac Y-h\brac{\rline YX}\\
h\brac Y & =h\brac{GX+W}\\
 & =h\brac{\left[\begin{array}{cc}
g_{11} & g_{12}\\
g_{21} & g_{22}
\end{array}\right]\left[\begin{array}{cc}
a & 0\\
\eta & c
\end{array}\right]Q+W}\\
 & =h\brac{\left[\begin{array}{cc}
ag_{11}+\eta g_{12} & cg_{12}\\
ag_{21}+\eta g_{22} & cg_{22}
\end{array}\right]Q+W}\\
 & \overset{\brac i}{=}h\brac{\brac{\left[\begin{array}{cc}
ag_{11}+\eta g_{12} & cg_{12}\\
ag_{21}+\eta g_{22} & cg_{22}
\end{array}\right]+W}Q},
\end{align}
where $\brac i$ is using the fact that $W$ and $WQ$ have the same
distribution. Now
\begin{align}
h\brac Y & =h\brac{\brac{\left[\begin{array}{cc}
ag_{11}+\eta g_{12} & cg_{12}\\
ag_{21}+\eta g_{22} & cg_{22}
\end{array}\right]+W}Q}\nonumber \\
 & \overset{\brac i}{=}h\brac{\left[\begin{array}{cc}
ag_{11}+\eta g_{12}+w_{1} & cg_{12}+w_{2}\\
ag_{21}+\eta g_{22}+w_{3} & cg_{22}+w_{4}
\end{array}\right]Q}\\
 & \overset{\brac{ii}}{\geq}h\brac{\left[\begin{array}{cc}
ag_{11}+\eta g_{12}+w_{1}, & cg_{12}+w_{2}\end{array}\right]Q}\\
 & \quad+h\brac{\rline{\left[\begin{array}{cc}
ag_{21}+\eta g_{22}+w_{3}, & cg_{22}+w_{4}\end{array}\right]Q}Q,ag_{11}+\eta g_{12}+w_{1},cg_{12}+w_{2}}\\
 & \overset{\brac{iii}}{=}h\brac{\abs{ag_{11}+\eta g_{12}}^{2}+\abs{cg_{12}}^{2}}+\expect{\lgbrac{\abs{ag_{11}+\eta g_{12}}^{2}+\abs{cg_{12}}^{2}}}+\lgbrac{\frac{\pi^{T}}{\Gamma\brac T}}\\
 & \quad+h\brac{\rline{\left[\begin{array}{cc}
ag_{21}+\eta g_{22}+w_{3}, & cg_{22}+w_{4}\end{array}\right]Q}Q,ag_{11}+\eta g_{12}+w_{1},cg_{12}+w_{2}},
\end{align}
where in step $\brac i$, $w_{i}$'s are i.i.d $\mathcal{CN}\brac{0,1}$
and the step $\brac{ii}$ is using the fact that conditioning reduces
entropy. The step $\brac{iii}$ is using Lemma \ref{lem:isotropic_entropy_to_radial}.
Now
\begin{align}
 & h\brac{\rline{\left[\begin{array}{cc}
ag_{21}+\eta g_{22}+w_{3}, & cg_{22}+w_{4}\end{array}\right]Q}Q,ag_{11}+\eta g_{12}+w_{1},cg_{12}+w_{2}}\nonumber \\
 & =h\brac{\rline{\left[\begin{array}{cc}
ag_{21}+\eta g_{22}+w_{3}, & cg_{22}+w_{4}\end{array}\right]}ag_{11}+\eta g_{12}+w_{1},cg_{12}+w_{2}}\\
 & \overset{\brac i}{=}h\brac{cg_{22}+w_{4}}+h\brac{\rline{ag_{21}+\eta g_{22}+w_{3}}ag_{11}+\eta g_{12}+w_{1},cg_{12}+w_{2}}\\
 & \overset{\brac{ii}}{\geq}h\brac{cg_{22}+w_{4}}+h\brac{\rline{ag_{21}+\eta g_{22}}cg_{22},ag_{11}+\eta g_{12},cg_{12}},
\end{align}
where $\brac i$ is using the fact that $c$ is a given constant and
$cg_{22}+w_{4}$ is independent of $ag_{11}+\eta g_{12}+w_{1},cg_{12}+w_{2}$.
The step $\brac{ii}$ is by providing $w_{1},w_{2},w_{3}$ in the
conditioning and using the fact that conditioning reduces entropy.
Hence
\begin{align}
h\brac Y & \geq h\brac{\abs{ag_{11}+\eta g_{12}}^{2}+\abs{cg_{12}}^{2}}+\expect{\lgbrac{\abs{ag_{11}+\eta g_{12}}^{2}+\abs{cg_{12}}^{2}}}+\lgbrac{\frac{\pi^{T}}{\Gamma\brac T}}\nonumber \\
 & \qquad h\brac{cg_{22}+w_{4}}+h\brac{\rline{ag_{21}+\eta g_{22}}cg_{22},ag_{11}+\eta g_{12},cg_{12}}.\label{eq:simul_beg}
\end{align}
Now for numerically evaluating the terms in the above expression,
we use the following:
\begin{align}
h\brac{\abs{ag_{11}+\eta g_{12}}^{2}+\abs{cg_{12}}^{2}} & \geq h\brac{\rline{\abs{ag_{11}+\eta g_{12}}^{2}}g_{12}}
\end{align}
and
\begin{align*}
 & \expect{\lgbrac{\abs{ag_{11}+\eta g_{12}}^{2}+\abs{cg_{12}}^{2}}}\\
 & \overset{\brac i}{=}\expect{\lgbrac{\abs a^{2}\abs{g_{11}}^{2}+\abs{\eta}^{2}\abs{g_{12}}^{2}+\abs c^{2}\abs{g_{12}}^{2}+2\abs{a\eta g_{12}g_{11}}\cos\theta}}\\
 & \overset{\brac{ii}}{=}\expect{\lgbrac{\frac{\abs a^{2}\abs{g_{11}}^{2}+\abs{\eta}^{2}\abs{g_{12}}^{2}+\abs c^{2}\abs{g_{12}}^{2}+\sqrt{\abs{\abs a^{2}\abs{g_{11}}^{2}+\abs{\eta}^{2}\abs{g_{12}}^{2}+\abs c^{2}\abs{g_{12}}^{2}}^{2}-4\abs{a\eta g_{12}g_{11}}}^{2}}{2}}},
\end{align*}
where $\brac i$ is using a $\theta$ uniformly distributed in $\sbrac{0,2\pi}$
and the fact that $g_{12},g_{11},\eta$ are independent circularly
symmetric Gaussians. The step $\brac{ii}$ is by evaluating the expectation
over $\theta$. Also $h\brac{\rline{ag_{21}+\eta g_{22}}cg_{22},ag_{11}+\eta g_{12},cg_{12}}$
is evaluated using (\ref{eq:innerbound_2x2_simplification_simulation})
on page \pageref{eq:innerbound_2x2_simplification_simulation}. Also,
using (\ref{eq:h(Y|X)}), (\ref{eq:h(Y(n)|X)}) we have
\begin{align}
 & h\brac{Y|X}\nonumber \\
 & =\expect{\lgbrac{\abs a^{2}\rho_{11}^{2}+\abs{\eta}^{2}\rho_{12}^{2}+\abs c^{2}\rho_{12}^{2}+\abs a^{2}\abs c^{2}\rho_{11}^{2}\rho_{12}^{2}+1}}\nonumber \\
 & \quad+\expect{\lgbrac{\abs a^{2}\rho_{21}^{2}+\abs{\eta}^{2}\rho_{22}^{2}+\abs c^{2}\rho_{22}^{2}+\abs a^{2}\abs c^{2}\rho_{21}^{2}\rho_{22}^{2}+1}}\nonumber \\
 & \quad+2T\lgbrac{\pi e}.\label{eq:simul_end}
\end{align}
Now following the choice in Theorem \ref{thm:2x2-sym-mimo}, we choose
\[
\abs a^{2}=2,\text{\ensuremath{\eta}}\sim\mathcal{CN}\brac{0,\abs b^{2}},\abs b^{2}=1,\abs c^{2}=1/\rho_{12}^{2}
\]
 and evaluate the achievable rate using (\ref{eq:simul_beg}) to (\ref{eq:simul_end})
and scaling with $T=2$. Note that the expressions given here assume
that the transmit SNR is scaled into the link strengths $\rho_{ij}^{2}$
and the power at the antennas are unity after the scaling.
\end{document}